\documentclass[compress,final,1p,times]{elsarticle}
\usepackage[T1]{fontenc}
\usepackage[utf8]{inputenc}
\usepackage{lineno}
\usepackage{graphicx}
\usepackage{subfigure}
\usepackage{amssymb}
\usepackage{epstopdf}
\usepackage{booktabs}
\usepackage[table]{xcolor}
\usepackage{wrapfig}
\usepackage{makecell}
\usepackage{amsmath}
\usepackage{amsthm}
\usepackage{caption}
\usepackage[noend]{algpseudocode}
\usepackage{algorithmicx,algorithm}
\usepackage[font=small,labelsep=none]{caption}
\usepackage{stfloats}
\usepackage{multirow}
\usepackage{bm}
\usepackage{indentfirst}
\usepackage{latexsym}
\usepackage[wby]{callouts}
\usepackage[colorlinks,linkcolor=black,anchorcolor=black,citecolor=black]{hyperref}
\usepackage{academicons}
\newcommand{\orcid}[1]{\href{https://orcid.org/#1}{\textcolor[HTML]{A6CE39}{\aiOrcid}}}
\begin{document}

\begin{frontmatter}

\title{Unsupervised Classification for Polarimetric SAR Data Using Variational Bayesian Wishart Mixture Model with Inverse Gamma-Gamma Prior}
\tnotetext[t1]{This work was supported in part by National Natural Science Foundation of China under Grant 61801347, Grant 62001350, Grant 61801344, and Grant 61631019, in part by the Project funded by China Postdoctoral Science Foundation under Grant 2017M613076, 2020M673346, and 2016M602775, in part by the Postdoctoral Science Research Projects of Shaanxi Province, in part by the project supported by Natural Science Basic Research Plan in Shaanxi Province of China (Program No. 2020JQ-312), in part by the Aeronautical Science Foundation of China under Grant 20180181003, in part by the Joint Fund of Ministry of Education under Grant 6141A02022367, in part by the Fundamental Research Funds for the Central Universities.}
\author{Shijie Ren}
\ead{sjren@stu.xidian.edu.cn}
\author{Feng Zhou\corref{cor1}}
\cortext[cor1]{Corresponding author}
\ead{fzhou@mail.xidian.edu.cn}
\author{Changlong Wang}
\ead{clw_xjtu@163.com}
\address{Key Laboratory of Electronic Information Counter Measure and Simulation Technology of Ministry of Education, Xidian University, 2 South Taibai Road, Xi'an 710071, Shaanxi, P.R. China}

\begin{abstract}
Although various clustering methods have been successfully applied to polarimetric synthetic aperture radar (PolSAR) image clustering tasks, most of the available approaches fail to realize automatic determination of cluster number, nor have they derived an exact distribution for the number of looks. To overcome these limitations and achieve robust unsupervised classification of PolSAR images, this paper proposes the variational Bayesian Wishart mixture model (VBWMM), where variational Bayesian expectation maximization (VBEM) technique is applied to estimate the variational posterior distribution of model parameters iteratively. Besides, covariance matrix similarity and geometric similarity are combined to incorporate spatial information of PolSAR images. Furthermore, we derive a new distribution named inverse gamma-gamma (IGG) prior that originates from the log-likelihood function of proposed model to enable efficient handling of number of looks. As a result, we obtain a closed-form variational lower bound, which can be used to evaluate the convergence of proposed model. We validate the superiority of proposed method in clustering performance on four real-measured datasets and demonstrate significant improvements towards conventional methods. As a by-product, the experiments show that our proposed IGG prior is effective in estimating the number of looks.
\end{abstract}
\end{frontmatter}
\section{Introduction}
\subsection{Problem Statement and Previous Works}
Synthetic Aperture Radar (SAR) has the advantages of full-time, all-weather, long range and strong penetrating ability \cite{OliverSAR}. Compared with single-polarization SAR, the application of polarimetric SAR (PolSAR) is quite extensive due to its capability to record the backscattering information of targets under different combinations of receiving and transmitting polarization, which greatly enhances its ability to discriminate land covers \cite{OliverSAR}. With the wide application of PolSAR in civil areas such as topographic mapping, resource exploration, disaster monitoring and military fields such as battlefield reconnaissance, situation surveillance and precise guidance \cite{Usage_1,Usage_2,Usage_3,Usage_4}, the research of PolSAR image interpretation has attracted extensive attention in the past few decades.

Many researchers have conducted in-depth research on PolSAR image classification, ranging from supervised to unsupervised methods. A great deal of supervised methods have been developed for PolSAR image classification, including Bayesian methods \cite{Bayes_Supervised}, kernel methods such as support vector machine (SVM) \cite{SVM_1,SVM_2}, metric-based methods such as k-nearest neighbor \cite{TensorICA}, and neural networks \cite{NN_1,NN_2}. Recently, deep learning is emerging as a powerful leading component in image analysis. So far, restricted Boltzmann machines (RBMs) \cite{WishartRBM,EnsemRBM}, deep belief networks (DBNs) \cite{WishartDBN,WishartDSN}, autoencoders (AEs) \cite{MDPLSAE,WCAE}, convolutional neural networks (CNNs) \cite{HierFCN,CV-CNN} and recurrent neural networks (RNNs) \cite{LSTM} have found their applications in PolSAR image classification. However, their potential remains constrained to the quantities and quality of labeled data. While accurate labeled data have been proven to be difficult to obtain in practical scenarios, unsupervised classifiers can help to alleviate the dependence on labels and enhance the interpretability of PolSAR images to some extent. Therefore, unsupervised methods have become an important trend in PolSAR image classification.

For unsupervised classification methods, most of the existing approaches are based on scattering characteristics \cite{Scattering}, polarimetric decomposition \cite{OliverSAR,Unsuper_Wishart,Unsuper_Cloude,Unsuper_Dec_Lee,Unsuper_Entro}, statistical distributions \cite{APD11,relaxed,DoulgerisUDis}, as well as regional information \cite{SVWMM,region_1,region_2}. For methods derived from target decomposition, clustering results can be greatly improved \cite{Unsuper_Wishart,APD11,CloudeSPAN,Wis_1,Wis_ChernoffDis} by integrating complex Wishart classifier with polarimetric features. Besides, methods based on target decomposition can also be enhanced by introducing additional information, such as anisotropy and total polarimetric power (SPAN) \cite{CloudeSPAN} to more comprehensively characterize land covers from different perspectives, thus obtaining better clustering performance. However, how to exclude redundant information while preserving discriminative features remains to be an open problem. 

Recently, iterative Bayesian methods like expectation maximization (EM) and Markov random field (MRF) \cite{APD11,DoulgerisUDis} have emerged as powerful tools for PolSAR unsupervised image classification. Most of these methods are based on the assumption that polarimetric coherence/covariance matrices follow a series of matrix variate distributions, including complex Wishart distribution \cite{WishartDis}, $\mathcal{K}_{d}$-distribution \cite{APD11,KDis}, $\mathcal{G}_{d}^{0}$-distribution \cite{G0} and $\mathcal{U}_{d}$-distribution \cite{DoulgerisUDis}. These methods are more advantageous in modeling the scattering characteristics of PolSAR data, while avoiding feature redundancy and information loss caused by polarimetric decomposition. In this case, the main concern should be the determination of cluster number, which usually needs to be specified in advance. To deal with this problem, Doulgeris \textit{et al.} \cite{APD11} proposed to use goodness-of-fit (GoF) test based on Mellin transform \cite{GoF_Mellin} to realize automated clustering of PolSAR data. Apart from this split-and-merge mechanism, variational Bayesian (VB) inference is also a powerful tool naturally designed for the automatic determination of cluster number \cite{VB}. Liu \textit{et al.} \cite{SVWMM} proposed a spatial variant Wishart mixture model (SVWMM) with double constraints, which utilizes spatial information by introducing covariance matrix similarity and geometric similarity. However, it fails to derive a prior distribution for the equivalent number of looks (ENL), which is set to a constant for the entire dataset. In view of the weak convergence performance in SVWMM \cite{SVWMM}, the key problems lie in providing a closed-form variational lower bound and deriving an analytic prior distribution for ENL.

As an essential parameter of Wishart distribution, ENL can have a great influence on PolSAR image classification \cite{APD11,DoulgerisUDis,Frery07}. Most classifiers derived from Wishart distribution are based on the assumption that the number of looks $L$ can be roughly estimated by calculating the ENL of manually selected homogeneous regions, where speckle pattern is fully developed and the sources of heterogeneity, e.g. texture and mixture of classes, can be ignored \cite{OliverSAR}. However, this assumption does not hold in heterogeneous environments and high-resolution settings. Therefore, an accurate modeling of ENL and automatic determination of cluster number have been the problems demanding prompt solution for PolSAR image classification tasks.

Assuming that estimates computed in windows only contain variation due to fully developed speckle, existing ENL estimators \cite{LeeNoise,LeeUnsup,MLwin, ENLphase,ENLSVD,ENLSAR} are generally evaluated in sliding windows over the entire PolSAR images. Lee \textit{et al.} \cite{LeeNoise,LeeUnsup} proposed to produce a scatter plot of mean versus standard deviation of the intensity data in each sliding window, where the ENL can be inferred from its slope. Foucher \textit{et al.} \cite{MLwin} proposed a maximum likelihood method, namely ML estimator, to produce a 1-D distribution of ENL estimates, where the final ENL estimate is obtained according to their mean value. Joughin \textit{et al.} \cite{ENLphase} proposed an ENL estimator, namely TM estimator, that originates from the second-order trace moment. Ren \textit{et al.} \cite{ENLSVD} proposed an ENL estimator for SAR imageries based on singular value decomposition. However, these unsupervised approaches are proposed assuming that an arbitrary window can share the same ENL value, which does not hold in complex scenarios due to presence of multiple classes in the same windows sampled from complex scenes. In this case, the global ENL values could be severely underestimated. This is because it is difficult to statistically model the non-uniform windows containing a mixture of classes, which can lead to inconsistent ENL estimates in different polarimetric channels \cite{VarENL}. 

To overcome the aforementioned defects, Hu \cite{MixENL} proposed to detect and remove these non-uniform windows for more accurate ENL estimation. In this case, when an adequate share of sliding windows is homogeneous, the median value of ENL estimates is usually taken as the final result. Both the following approaches benefit from this strategy. Anfinsen et al. \cite{EstENL} proposed one estimator based on the second-order trace moments (TM) and another maximum likelihood estimator derived from the log-determinant matrix moment. Although the latter one proves to have lower variance than other known estimators \cite{EstENL,DTM}, there still exists a common weakness that these methods underestimate the ENL in heterogeneous regions. To alleviate the influence of textural variation on ENL estimation, Liu et al. \cite{DTM} proposed an estimator named development of trace moments (DTM), which is claimed to be robust to any distribution of texture under the product model. 

Despite the fact that the estimators above can provide an accurate ENL value for each pixel, it is hard to give a closed-from iterative solution for ENL for finite mixture models with matrix-variate distributions of PolSAR data, e.g. Wishart mixture model (WMM). To fully exploit the variance of ENL in different regions, Anfinsen \textit{et al.} \cite{relaxed} proposed a relax Wishart model, which allows ENL to vary between classes in PolSAR image classification tasks. Based on this idea, Doulgeris \textit{et al.} \cite{APD11} proposed an automated clustering method based on the relaxed-Wishart mixture model, where the numerical solution derived from the first matrix log-cumulant expression of Wishart distribution, namely WMLC \cite{GoF_Mellin}, is taken as the ENL estimate for each class. Nevertheless, it relies heavily on down-sampling strategy to adjust the number of clusters automatically, which constraints its potential to incorporate spatial information from neighboring pixels. 
\subsection{Contributions}
Although non-Gaussian mixture models like WMM have made tremendous progress in PolSAR image classification tasks, most current approaches are naturally based on the Mellin transform of matrix-variate distributions, where the main disadvantage is obviously its poor performance on large PolSAR datasets. While stochastic expectation maximization and down-sampling operation can alleviate this drawback, the sampling rate and confidence level, which are strongly related to cluster number, have to be tuned on each dataset. Moreover, excessive down-sampling on large-datasets can inevitably lead to the loss of spatial information. 

In view of the weak convergence performance in SVWMM \cite{SVWMM}, the key problems lie in providing a closed-form variational lower bound and deriving an analytic prior distribution for ENL. As a result, we take the advantage of variational Bayesian expectation maximization (VBEM) technique in dealing with automatic determination of cluster number and propose a variational Bayesian Wishart mixture model, namely WMMVBEM, to achieve robust clustering of PolSAR data. To the best of our knowledge, there has not been an appropriate distribution for ENL, thus our proposed inverse gamma-gamma (IGG) distribution should be the first attempt in the field of PolSAR image classification. Compared with the available Bayesian techniques for unsupervised PolSAR image classification, the main contributions of this paper are as follows:

(1) Considering the impact of ENL variance between different clusters, our proposed method allows each individual cluster to have its own independent ENL estimate, or to be more exact, each class can have a different number of looks that follows the derived IGG prior distribution with very simple numerical evaluations. 

(2) A closed-form variational lower bound has been derived, which helps to determine the convergence of proposed model. Moreover, instead of tuning the magnitude of mean matrix on each dataset, the hyperparameter initialization is more reasonable compared to that of SVWMM \cite{SVWMM}, resulting in more robust classification results. 

(3) Euclidean distance and Wishart distance are combined to incorporate spatial information embedded in neighboring pixels, which allows for a moderate degree of smoothness and robustness to speckles. 

The rest of this paper is organized as follows. Section \ref{sec:Method} presents the data formats, distributions and derivation process of proposed model. Section \ref{sec:Exps} demonstrates the performance of our method in experiments with real-measured airborne and spaceborne PolSAR datasets, with an aim to provide qualitative and quantitative analysis to verify the validity of proposed model. The classification results show that our proposed method can produce more accurate clustering results than $H/\alpha$-Wishart \cite{HalphaWis}, WMM \cite{APD11} and SVWMM \cite{SVWMM}. Besides, the effectiveness of inverse gamma-gamma (IGG) prior has been validated by comparing its expectation value of each class with four ENL estimation methods. Section \ref{sec:Conclusion} summarizes the main conclusions and outlines the improvement directions according to the deficiencies of existing clustering approaches. 

\section{Methodology}
\label{sec:Method}
\subsection{Overview of PolSAR Data and Complex Wishart Distribution}
\label{sec:overview}
Compared with single-channel synthetic aperture radar (SAR), Polarimetric SAR measures coherent microwave backscatter from land covers in dual or quadratic polarizations, which usually comes in the form of a complex
scattering matrix, which consists of four backscattering coefficients under different combinations of transmit-receive polarization \cite{OliverSAR}, as shown in \eqref{eq.Smat}.
\begin{equation}
\label{eq.Smat}
S \!=\! \left[ {\begin{array}{*{20}{c}}
{{S_{hh}}}&{{S_{hv}}}\\
{{S_{vh}}}&{{S_{vv}}}
\end{array}} \right]
\end{equation}
where $S_{hh}$, $S_{hv}$, $S_{vh}$ and $S_{vv}$  are the four complex scattering coefficients, providing both the amplitude and phase information.

Under the assumption of monostatic reciprocity, the scattering matrix can be reshaped to the scattering vector in (2).
\begin{equation}
s \!=\! {\left[ {\begin{array}{*{20}{c}}
		{{S_{hh}}}&{\sqrt 2 {S_{hv}}}&{{S_{vv}}}
		\end{array}} \right]^T}
\end{equation}

For PolSAR, only single-look complex data is presented by scattering matrix, whose mean value cannot effectively eliminate the influence of speckle noise. Therefore, the multi-look polarimetric SAR data does not directly average the scattering matrix, but converts the scattering matrix into a polarization covariance matrix or a polarization coherence matrix. Besides, multi-look sub-sampling operation is applied to reduce the influence of non-Gaussian noise, and thereby derive a multi-look covariance matrix as follows.

\begin{equation}
C \!=\! \frac{1}{L}\sum\limits_{i \!=\! 1}^L {{S_i}S_i^H} 
\end{equation}
where \textit{L} is number of looks and $\left(  \cdot  \right)^H$ is the conjugate transpose operation.

The complex Wishart distribution \cite{WishartDis} for multi-look covariance matrix \textit{C} is given by \eqref{eq.WisDistr}.
\begin{equation}
\label{eq.WisDistr}
{p_W}\left( {C;L,\Sigma } \right) \!=\! \frac{{{L^{Ld}}{{\left| C \right|}^{L \!-\! d}}}}{{I\left( {L,d} \right){{\left| \Sigma  \right|}^L}}}{e^{ \!-\! Ltr\left( {{\Sigma ^{-1}}C} \right)}}
\end{equation}
where $I\left( {L,d} \right) \!=\! {\pi ^{{{d\left( {d \!-\! 1} \right)} \mathord{\left/{\vphantom {{d\left( {d \!-\! 1} \right)} 2}} \right.\kern-\nulldelimiterspace} 2}}}\prod\limits_{i \!=\! 1}^d {\Gamma \left( {L \!-\! i + 1} \right)}$, \textit{L} is the number of looks, $\Sigma$ denotes the mean covariance matrix, \textit{d} is the dimension of polarimetric scattering vector, $\Gamma \left(  \cdot  \right)$ denotes standard gamma function, $\left|  \cdot  \right|$ is the determinant of a matrix, $tr\left(  \cdot  \right)$ denotes trace operator.

\subsection{Variational Bayesian Framework}
\label{sec:VBframe}
The starting point of variational Bayesian framework is to approximate the distributions of hidden variables and hyperparameters with appropriate conjugate prior distributions, assuming that the prior distributions of hidden variables are given and the data samples are independent and identically distributed. According to Jensen's inequality, the exact lower bound can be written as
\begin{equation}
	\begin{aligned}
		\log P(X) &=\ln \iint p(\theta, Z, X) d \theta d Z \\
		&=\ln \iint q(\theta, Z) \frac{p(\theta, Z, X)}{q(\theta, X)} d \theta d Z \\
		& \geq \iint q(\theta, Z) \ln \left(\frac{p(\theta, Z, X)}{q(\theta, Z)}\right) d \theta d Z=L(q)
	\end{aligned}
\end{equation}
where $p\left( {\theta ,Z,X} \right)$ is the joint distribution, $q(\theta ,Z)$ is the variational posterior distribution introduced to approximate the true posterior $p\left( {\theta ,Z|X} \right)$, which cannot be solved directly for the estimation of joint distribution requires the marginal likelihood function and normalized constant.

The variational distribution $q\left( {\theta ,Z} \right)$ assumes that the hidden variables and hyperparameters are independent, so as to factorize the distributions of various components in the form of their product, thus obtaining the simplified lower bound as follows.
\begin{equation}
	L\left(q_{\theta}, q_{z}\right)=\iint q_{\theta}(\theta) q_{z}(Z) \ln \left(\frac{p(\theta, Z, X)}{q_{\theta}(\theta) q_{z}(Z)}\right) d \theta d Z
\end{equation}

Variational inference seeks an approximate posterior that is close to the true posterior in terms of KL divergence. The posterior is typically restricted to some tractable family of distributions, and an optimization problem is formed by minimizing the KL divergence from an approximate posterior to the true posterior, as is shown in \eqref{eq.KLdist}.
\begin{equation}
	\label{eq.KLdist}
	\begin{aligned}
		\ln p\left( X \right) \!-\! L\left( {{q_\theta },{q_Z}} \right) &\!=\!  \!-\! \iint {{q_\theta }\left( \theta  \right){q_Z}\left( Z \right)\ln \left( {\frac{{p\left( {\theta ,Z,X} \right)}}{{{q_\theta }\left( \theta  \right){q_Z}\left( Z \right)}}} \right)d\theta dZ} \cr 
		&\!=\! KL\left[ {{q_\theta }\left( \theta  \right){q_Z}\left( Z \right)||p\left( {\theta ,Z|X} \right)} \right] \geqslant 0 \cr
	\end{aligned}
\end{equation}

Therefore, maximizing the lower bound is equivalent to minimizing the relative entropy between ${q_\theta }\left( \theta  \right){q_Z}\left( Z \right)$ and $p\left( {\theta ,Z|X} \right)$. The relative entropy is zero if and only if they are equivalent.

Under independent irrelevant assumption and mean field approximation \cite{VB,VG,PRML}, the posterior distribution can be approximated as ${q_Z}\left( Z \right) \!=\! \prod\limits_{n \!=\! 1}^N {{q_{_{{Z_n}}}}\left( {{z_n}} \right)} $, so as to reduce the complexity of joint probability estimation. Thus maximizing the log-likelihood function of incomplete data requires the optimization of each factor, which gives out the iterative solutions of the hyperparameters for the variational posterior by matching the form of prior distributions with the corresponding posteriors.

Apart from most other techniques derived from Mellin statistics and goodness-of-fit tests \cite{GoF_Mellin}, variational Bayesian inference also provides an alternative to achieve automated clustering. SVWMM \cite{SVWMM} made an attempt to combine Wishart mixture model with variational inference. However, this approach fails to treat ENL as a distinguishing feature, which can have a significant impact on PolSAR image classification with matrix-variate distributions. Up to now, none of the existing methods has derived an appropriate distribution for ENL. Therefore, it is of great importance to find a distribution used to obtain an accurate ENL estimation for each class. In response to this issue, an inverse gamma-gamma (IGG) prior is put forward in Section \ref{sec:IGGPrior}. 

\subsection{Proposed Inverse-Gamma Gamma Prior}
\label{sec:IGGPrior}
For VB methods, the variation lower bound, also named evidence lower bound (ELBO), requires a conjugate relationship between latent variables, which takes the product of two exponential family distributions. In allusion to this, the WMMVBEM method proposed in this paper assumes that the mean covariance matrix and number of looks are conjugated, with a closed-form variational lower bound used to determine the convergence of model. 

For complex Wishart distribution, it is common practice to assume the conjugate prior of mean covariance matrix as inverse Wishart distribution. However, one may find it hard to statistically model an analytical probability distribution for ENL, which is mainly caused by the cross-terms derived from complex Wishart distributed posterior. Thus how to integrate automatic estimation of ENL in variational Bayesian framework poses a great challenge.

To cope with this problem, here the proposed model first utilizes the terms containing mean covariance matrix in the posterior to obtain an iterative solution for its hyperparameters. After that, the prior of \textit{L} is derived by comparing the remaining terms of variational posterior and real posterior, as is shown in \eqref{eq.IGG}.
\begin{equation}
\label{eq.IGG}
p_{L}(l)=\frac{1}{2 b^{-a / 2} c^{a / 2} K_{a}\left(2 \sqrt{b c}\right)} l^{a-1} e^{-b l-c / l}
\end{equation}
where $K_a(\cdot)$ is a modified Bessel function of the second kind with order $a$. Note that the probability density function (PDF) of $L$ is a three-parameter continuous distribution parameterized by shape parameter $a$ and scale parameters $b$ and $c$. From the iterative solutions in \eqref{eq.ak}\verb|-|\eqref{eq.ck}, it is known that these parameters are always greater than zero, thus the proposed prior can be regarded as the product of an inverse-gamma distribution and a gamma distribution, namely inverse gamma-gamma (IGG) distribution. Note that the denominator in \eqref{eq.IGG} serves as the normalization term that makes sure the distribution integrates to one and satisfies the definition of PDF.
\subsubsection*{Approximation for the Logarithm of PDF}
\label{subsubsec:AppLogPL}	
For any two positive real numbers $a$ and $z$, the following asymptotic approximation holds.
\begin{equation}
\label{eq.K_a}
{K_a }\left( z \right) \sim \frac{{\Gamma \left( a  \right)}}{2}{\left( {\frac{2}{z}} \right)^a }, a \in {\cal R}^{+}, z \in {\cal R}^{+}.
\end{equation}

Take the logarithm on both sides of \eqref{eq.K_a}, it is easy to get that
\begin{equation}
\label{eq.lnK_a}
\ln {K_a }\left( z \right) \sim \ln \Gamma \left( a  \right) \!-\! a \ln z + \left( {a  \!-\! 1} \right)\ln 2, a \in \mathbb{R}, z \in \mathbb{R}.
\end{equation}

By substituting $z$ with $2\!\sqrt {{b}{c}}$ in \eqref{eq.K_a}, the logarithm of PDF in \eqref{eq.IGG} can be simplified into \eqref{eq.logpL}.
\begin{equation}
\label{eq.logpL}
\ln {p_L}\left( l \right) = \left( {a - 1} \right)\ln l - {b_1}l - {{{b_2}} \mathord{\left/
		{\vphantom {{{b_2}} l}} \right.
		\kern-\nulldelimiterspace} l} - \ln \Gamma \left( a \right) + a\ln {b_1}
\end{equation}
\subsubsection*{Expectation Terms Related to \textit{L}}
\label{subsubsec:ExpsL}
By definition, an expectation term related to $L$ can be obtained via an convergent improper integral, e.g. $\left\langle L^{n} \right\rangle=\int_{0}^{\infty} l^{n} p(l) d l, n \in \mathbb{R}$, where $p(l)$ denotes the PDF of IGG prior in \eqref{eq.IGG}. The expectations used in the proposed model are given in \eqref{eq.EL_ti}\verb|-|\eqref{eq.EL_2_ti}. 
\begin{equation}
	\label{eq.EL_ti}
	\left\langle L \right\rangle  \!=\! \sqrt {\frac{{{c}}}{{{b}}}} \frac{{{K_{a + 1}}\left(2\!\sqrt {{b}{c}}\right)}}{{{K_a}\left(2\!\sqrt {{b}{c}}\right)}}
\end{equation}
\begin{equation}
	\label{eq.ElnL_ti}
	\left\langle {\ln L} \right\rangle  \!=\! \frac{1}{2}\ln \frac{{{c}}}{{{b}}} \!-\! \frac{{{{K}_a^{'}}\left(2\!\sqrt {{b}{c}}\right)}}{{{K_a}\left(2\!\sqrt {{b}{c}}\right)}}
\end{equation}
\begin{equation}
	\label{eq.EL_1_ti}
	\left\langle {{L^{-1}}} \right\rangle  \!=\! \sqrt {\frac{{{b}}}{{{c}}}} \frac{{{K_{a \!-\! 1}\left(2\!\sqrt {{b}{c}}\right)}}}{{{K_a}\left(2\!\sqrt {{b}{c}}\right)}}
\end{equation}
\begin{equation}
	\label{eq.EL_2_ti}
	\left\langle {{L^{ \!-\! 2}}} \right\rangle \!=\! \frac{{{b}}}{{{c}}}\frac{{{K_{a \!-\! 2}}\left(2\!\sqrt {{b}{c}}\right)}}{{{K_{a \!-\! 1}}\left(2\!\sqrt {{b}{c}}\right)}}\frac{{{K_{a \!-\! 1}\left(2\!\sqrt {{b}{c}}\right)}}}{{{K_a}\left(2\!\sqrt {{b}{c}}\right)}}
\end{equation}

\subsection{Variational Bayesian Expectation Maximization for PolSAR Data}
\label{sec:VBEM4PolSAR}
The proposed VBWMM performs unsupervised classification for PolSAR data based on complex Wishart mixture distribution, where the conditional probability of WMM with $K$ components can be written as
\begin{equation}
\label{eq.WMM_K}
p\left( {C|z,L,\Sigma } \right) \!=\! \prod\limits_{n \!=\! 1}^N {\prod\limits_{k \!=\! 1}^K {W{{\left( {{C_n}|{L_k},{\Sigma _k}} \right)}^{{z_{n,k}}}}} } 
\end{equation}
\begin{figure}[htbp]
\centering
\includegraphics[width=.35\textwidth]{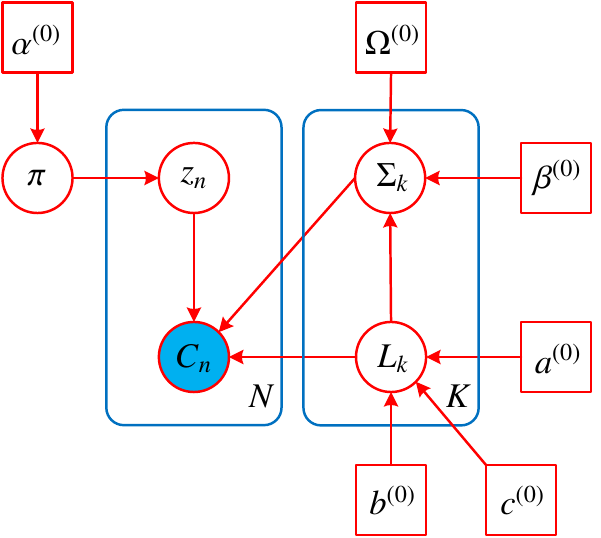}
\caption{Graph model of the proposed WMMVBEM.}
\label{fig.GraphMod}
\end{figure}

The graph model of proposed VBWMM can be illustrated as Fig.~\ref{fig.GraphMod}, with the distributions summarized as below.
\begin{equation}
\begin{aligned}
\pi &\sim Dir(\alpha) \\
z_{n} &\sim Cat(\pi) \\
L_{k} &\sim IGG\left(a_{k}, b_{k}, c_{k}\right) \\
\Sigma_{k}^{-1} &\sim W\left(\beta_{k} L_{k}, \Omega_{k}\right) \\
C_{n} &\sim W\left(L^{z_{n}}, \Sigma^{z_{n}}\right)
\end{aligned}
\end{equation}
where $z_{n}$ is a binary vector that follows categorical distribution, whose parameter $\pi$ satisfies the Dirichlet distribution. For VBEM methods, the variational lower bound exists only when the latent variables are conjugate distributed \cite{PRML}. Therefore, it is reasonable to assume that $\Sigma $ and \textit{L} are conjugated when both variables are unknown, which can be written as $p\left( {\Sigma ,L} \right) \!=\! p\left( {\Sigma |L} \right)p\left( L \right)$. To make the posterior tractable, the prior of $\Sigma _k^{-1}$ is set as complex Wishart distribution with two parameters, namely ${\beta _k}{L_k}$ and ${\Omega _k}$. Meanwhile, the prior of \textit{L} is formulated as inverse-gamma gamma distribution, as defined in \eqref{eq.IGG}.

The latent variables $\Phi  \!=\! \left[ {z,\pi ,\Sigma ,L} \right]$ and hyperparameters $H \!=\! \left[ {{\alpha^{(0)}},{\beta^{(0)}},{\Omega^{(0)}},{a^{(0)}},{b^{(0)}},{c^{(0)}}} \right]$ make up the parameter set of proposed model. Under the assumption that mean covariance matrix $\Sigma$ and number of looks \textit{L} are conjugate distributed, the joint probability distribution between observed and latent variables can be factorized as

\begin{equation}
p\left( {C,z,\pi ,L,\Sigma } \right) \!=\! p\left( {C|z,L,\Sigma } \right) \cdot p\left( {z|\pi } \right) \cdot p\left( \pi  \right)  
\cdot p\left( {\Sigma |L} \right) \cdot p\left( L \right)
\end{equation}

Thus the log-likelihood function of proposed model can be written as
\begin{equation}
\label{eq.LogLike}
\begin{aligned}
L(\Phi, H) &=\sum_{k=1}^{K}\left[\ln p\left(\Sigma_{k} \mid \beta_{k}^{(0)} L_{k}, \left(\Omega_{k}^{(0)}\right)^{-1}\right)+\ln p\left(L_{k} \mid a_{k}^{(0)}, b_{k}^{(0)}, c_{k}^{(0)}\right)\right] \\
&+\sum_{n=1}^{N} \sum_{k=1}^{K} z_{n,k} \ln W\left(C_{n} \mid L_{k}, \Sigma_{k}\right)+\ln p(z \mid \pi)+\ln p\left(\pi \mid \alpha^{(0)}\right)
\end{aligned}
\end{equation}

The VE-step and VM-step make up the iterative process of VBEM algorithm. In the VE-step, the lower bound is maximized with respect to the posterior $q\left( z \right)$ over latent variables, which is the standard procedure using the log-likelihood of  $q\left( z \right)$ by retaining the terms containing $z_{n,k}$ to calculate the posterior class probabilities for each pixel. The logarithm of $q\left( z \right)$ is expanded as
\begin{equation}
\begin{aligned}
\ln q\left( z \right) &\!=\! \sum\limits_{n \!=\! 1}^N {\sum\limits_{k \!=\! 1}^K {{z_{n,k}}\ln { r_{n,k}}} }  + const\\
&\!=\! \sum\limits_{n \!=\! 1}^N {\sum\limits_{k \!=\! 1}^K {{z_{n,k}}\left\{ {d\left\langle {{L_k}\ln {L_k}} \right\rangle  + \left( {\left\langle {{L_k}} \right\rangle  \!-\! d} \right)\ln \left| {{C_n}} \right|} \right.} } 
- \left\langle {\sum\limits_{i \!=\! 1}^d {\ln \Gamma \left( {{L_k} \!-\! i + 1} \right)} } \right\rangle + \left\langle {{L_k}} \right\rangle \left\langle {\ln \left| {\Sigma _k^{-1}} \right|} \right\rangle \\
&\left. { \!-\! \left\langle {{L_k}} \right\rangle tr\left( {\left\langle {\Sigma _k^{-1}} \right\rangle {C_n}} \right) + \left\langle {\ln {\pi _k}} \right\rangle  \!-\! \frac{{d\left( {d \!-\! 1} \right)}}{2}\ln \pi } \right\} + const 
\end{aligned}
\end{equation}
where $r_{n,k}$ denotes the posterior probability that the $n$th sample belongs to the $k$th component, ${\pi _k}$ is the prior probability of $K$th component, ${C_n}$ denotes the covariance matrix of \textit{n}th pixel, $\left\langle  \cdot  \right\rangle $ is equivalent to the expectation operator $E\left[  \cdot  \right]$.

\subsubsection*{Approximation for Log-gamma Functions}
The iterative functional equation in \eqref{eq.loggamma_iter1} allows one to determine log-gamma function values in one strip of width 1 in $z \in \mathbb{R}$ from the neighboring strip \cite{Psi}. Starting with an accurate approximation for large real $z$, one may go step by step down to the desired $z$. 
\begin{equation}
\label{eq.loggamma_iter1}
\ln \Gamma(z)=\ln \Gamma(z+1)-\ln z
\end{equation}

One can easily generalize \eqref{eq.loggamma_iter1} to the case of multi-step iterative sequences, as is illustrated in \eqref{eq.loggamma_iterm}.
\begin{equation}	
\label{eq.loggamma_iterm}
\ln \Gamma(z-m)=\ln \Gamma(z)-\sum_{k=1}^{m} \ln (z-k), m \in \mathbb{N}^+
\end{equation}

Using \eqref{eq.loggamma_iterm} and McLaughlin formula, the expectation of the summation of log-gamma functions can be expanded as \eqref{eq.ElnSumGamma}.
\begin{equation}
\label{eq.ElnSumGamma}
\begin{aligned}
\left\langle {\sum\limits_{i \!=\! 1}^d {\ln \Gamma \left( {{L_k} \!-\! i + 1} \right)} } \right\rangle &= \left\langle {3\ln \Gamma \left( {{L_k}} \right) \!-\! 2\ln \left( {{L_k} \!-\! 1} \right) \!-\! \ln \left( {{L_k} \!-\! 2} \right)} \right\rangle \\
&= 3\left\langle {{L_k}\ln {L_k}} \right\rangle  \!-\! 3\left\langle {{L_k}} \right\rangle  + \frac{3}{2}\ln 2\pi \!-\! \left[ {\frac{3}{2}\left\langle {\ln {L_k}} \right\rangle  + 2\left\langle {\ln \left( {{L_k} \!-\! 1} \right)} \right\rangle  + \left\langle {\ln \left( {{L_k} \!-\! 2} \right)} \right\rangle } \right] \\
&= 3\left\langle {{L_k}\ln {L_k}} \right\rangle  \!-\! \frac{9}{2}\left\langle {\ln {L_k}} \right\rangle  \!-\! 3\left\langle {{L_k}} \right\rangle + 4\left\langle {\frac{1}{{{L_k}}}} \right\rangle  + \frac{3}{2}\ln 2\pi
\end{aligned}
\end{equation}

As $L > 2$ gets larger, $L^{L}$ and $\Gamma(L)$ can grow much faster than an exponential function, which can lead to numerical overflow and thus result in failure to estimate $L$. An alternative approach is to eliminate both terms in the log-likelihood function with the following Stirling's formula in \eqref{eq.Stirling}, and thus resulting in \eqref{eq.logGammaAppro}. 

Asymptotically as $L \rightarrow+\infty$, the magnitude of $\Gamma(L)$ is given by Stirling's formula, as defined in \eqref{eq.Stirling}.
\begin{equation}
\label{eq.Stirling}
\Gamma(z+1) \sim \sqrt{2 \pi z}\left(\frac{z}{e}\right)^{z}
\end{equation}
where the symbol $\sim$ implies asymptotic convergence. In another word, the ratio of two sides converges to 1 as $z \rightarrow+\infty$.

Following \eqref{eq.loggamma_iter1} and \eqref{eq.Stirling}, it is easy to get an approximation for log-gamma function to prevent numerical overflow, as is shown in \eqref{eq.logGammaAppro}.
\begin{equation}
\label{eq.logGammaAppro}
\ln \Gamma \left( z \right) \sim \left( {z \!-\! \frac{1}{2}} \right)\ln z \!-\! z + \frac{1}{2}\ln 2\pi 
\end{equation}

Without loss of generality, let us assume that ${\tilde r}_{n,k} \!=\! \ln { r_{n,k}}$. Therefore, the logarithm of posterior probability ${\tilde r_{n,k}}$ for the $K$th component over \textit{n}th pixel can be written as
\begin{equation}
	\label{eq.rtnk}
	\begin{aligned}
		\tilde{r}_{n k}=& d\left\langle L_{k} \ln L_{k}\right\rangle+\left(\left\langle L_{k}\right\rangle-d\right) \ln \left|C_{n}\right|-\frac{d(d-1)}{2} \ln \pi-\left\langle\sum_{i=1}^{d} \ln \Gamma\left(L_{k}-i+1\right)\right\rangle \\
		&-\left\langle L_{k}\right\rangle\left\langle\ln \left|\Sigma_{k}\right|\right\rangle-\left\langle L_{k}\right\rangle \operatorname{tr}\left(\left\langle\Sigma_{k}^{-1}\right\rangle C_{n}\right)+\left\langle\ln \pi_{k}\right\rangle \\
		=& \frac{9}{2}\left\langle\ln L_{k}\right\rangle+\left(3+\ln \left|C_{n}\right|+\left\langle\ln \left|\Sigma_{k}^{-1}\right|\right\rangle-\operatorname{tr}\left(\Omega_{k} C_{n}\right)\right)\left\langle L_{k}\right\rangle-4\left\langle\frac{1}{L_{k}}\right\rangle \\
		&+\left\langle\ln \pi_{k}\right\rangle-3 \ln \left|C_{n}\right|-3 \ln \pi-\frac{3}{2} \ln 2 \pi
	\end{aligned}
\end{equation}
where
\begin{equation}
	\label{eq.ElogSigma_1}
	\begin{aligned}
		\left\langle {\ln \left| {\Sigma _k^{-1}} \right|} \right\rangle  &= \ln \left| {{\Omega _k}} \right| \!-\! d\left\langle {\ln {\beta _k}{L_k}} \right\rangle  + \left\langle {\sum\limits_{i \!=\! 1}^d {\psi \left( {{\beta _k}{L_k} \!-\! i + 1} \right)} } \right\rangle \\
		&= \ln \left| {{\Omega _k}} \right| \!-\! \frac{9}{{2{\beta _k}}}\left\langle {\frac{1}{{{L_k}}}} \right\rangle  \!-\! \frac{4}{{\beta _k^2}}\left\langle {\frac{1}{{L_k^2}}} \right\rangle
	\end{aligned}
\end{equation}

It should be noted that The ${L_k}\ln {L_k}$ term in \eqref{eq.ElogSigma_1} is eliminated using \eqref{eq.loggamma_iterm} and \eqref{eq.logGammaAppro}. Besides, the expectation of digamma functions in \eqref{eq.ElogSigma_1} is obtained by taking the derivative with respect to $L$ on both sides of \eqref{eq.ElnSumGamma}. 

By completing the square, the posterior expectation of ${z_{n,k}}$ is obtained in \eqref{eq.rnk}.
\begin{equation}
\label{eq.rnk}
{r_{n,k}} \!=\! {{\exp \left( {{{\tilde r}_{n,k}}} \right)} \mathord{\left/
		{\vphantom {{\exp \left( {{{\tilde r}_{n,k}}} \right)} {\sum\limits_{k \!=\! 1}^K {\exp \left( {{{\tilde r}_{n,k}}} \right)} }}} \right.
		\kern-\nulldelimiterspace} {\sum\limits_{k \!=\! 1}^K {\exp \left( {{{\tilde r}_{n,k}}} \right)} }}
\end{equation}

The VM-step determines the iterative solutions of hyperparameters by modifying the variational distribution towards true posterior. From the graph model in Fig. \ref{fig.GraphMod}, it is observed that the parameters $\alpha _k$, $\beta _k$, $\Omega _k$, $a_k$, $b_{k}$, $c_{k}$ need to be estimated.

To facilitate the introduction of spatial information, the iterative solution of ${\alpha _k}$ can be written as
\begin{equation}
{\alpha _{n,k}} \!=\! {\alpha _{n,k}^{(0)}} + {r_{n,k}} 
\end{equation}
\begin{equation}
{\alpha _k} \!=\! \sum\limits_{n \!=\! 1}^N {{\alpha _{n,k}}} 
\end{equation}

The logarithms of $p\left( {{\Sigma _k}|{L_k}} \right)$ an $q\left( {{\Sigma _k}|{L_k}} \right)$ can be obtained by retaining the terms containing $\Sigma $ from \eqref{eq.LogLike}.
\begin{equation}
\label{eq.logpSigma_L}
\begin{aligned}
\log p\left( {{\Sigma _k}|{L_k}} \right) &= \left( {{\beta _{k}^{(0)}}{L_k} \!-\! d} \right)\ln \left| {\Sigma _k^{-1}} \right| \!-\! {\beta _{k}^{(0)}}{L_k}tr\left( {\left(\Omega _{k}^{(0)}\right)^{-1}\Sigma _k^{-1}} \right) \\ 
&+ {L_k}\sum\limits_{n \!=\! 1}^N {\left\{ {{r_{n,k}}\left[ {\ln \left| {\Sigma _k^{-1}} \right| \!-\! tr\left( {\Sigma _k^{-1}{C_n}} \right)} \right]} \right\} + const}  
\end{aligned}
\end{equation}
\begin{equation}
\label{eq.logqSigma_L}
\log q\left( {{\Sigma _k}|{L_k}} \right) \!=\! \left( {{\beta _k}{L_k} \!-\! d} \right)\ln \left| {\Sigma _k^{-1}} \right| \!-\! {\beta _k}{L_k}tr\left( {\Omega _k^{-1}\Sigma _k^{-1}} \right) + const
\end{equation}

By comparing the log-likelihoods in \eqref{eq.logpSigma_L} and \eqref{eq.logqSigma_L}, the iterative solutions of ${\beta _k}$ and $\Omega _k^{-1}$ can be obtained in \eqref{eq.betak} and \eqref{eq.Omegak_1}, respectively.
\begin{equation}
\label{eq.betak}
{\beta _k} \!=\! {\beta _{k}^{(0)}} + {N_k}
\end{equation}
\begin{equation}
\label{eq.Omegak_1}
\Omega _k^{-1} \!=\! \frac{1}{{{\beta _k}}}\left( {{\beta _{k}^{(0)}}\left(\Omega _{k}^{(0)}\right)^{-1} + \sum\limits_{n \!=\! 1}^N {{r_{n,k}}{C_n}} } \right)
\end{equation}
where $N_{k} \!=\! \sum\limits_{n \!=\! 1}^N {{r_{n,k}}}$.

From \eqref{eq.Omegak_1}, it is known that $\beta$ can be regarded as a regularization term that measures the relative importance between prior distribution and observation data. In other words, it determines the relative weight of initial clusters and iterative updates, thus $\beta_{k}^{(0)}$ should be set according to the dataset size and richness of land covers.

Considering the conjugate relationship between mean covariance matrix and number of looks, together with \eqref{eq.loggamma_iterm} and \eqref{eq.logGammaAppro}, the logarithms of $p\left( L \right)$ and $q\left( L \right)$ are derived based on the terms containing \textit{L} in their log-likelihood functions, as shown in \eqref{eq.logpLk} and \eqref{eq.logqLk}.
\begin{equation}
\label{eq.logpLk}
\begin{aligned}
\ln p\left( {{L_k}} \right) &= \ln p\left( {{\Sigma _k},{L_k}} \right) \!-\! \ln p\left( {{\Sigma _k}|{L_k}} \right)\\
&= \frac{9}{2}\left( {{N_k} + 1} \right)\ln {L_k} + \left[ {3\left( {{N_k} + {\beta _{k}^{(0)}}\ln {\beta _{k}^{(0)}}} \right) + \sum\limits_{n \!=\! 1}^N {{r_{n,k}}\ln \left| {{C_n}} \right|}  \!-\! {\beta _{k}^{(0)}}\ln \left| {{\Omega _{k}^{(0)}}} \right|} \right]{L_k} \\
&- 4\left( {{N_k} + \frac{1}{{{\beta _{k}^{(0)}}}}} \right)\frac{1}{{{L_k}}} + \log IGG\left( {{L_k}|{a_{0k}},{b_{k}^{(0)}},{c_{k}^{(0)}}} \right) + const
\end{aligned}
\end{equation}

\begin{equation}
	\label{eq.logqLk}
	\begin{aligned}
		\log q\left( {{L_k}} \right) &= \log q\left( {{\Sigma _k},{L_k}} \right) \!-\! \log q\left( {{\Sigma _k}|{L_k}} \right)\\
		&= \frac{9}{2}\ln {\beta _k}{L_k} \!-\! \frac{4}{{{\beta _k}}}\frac{1}{{{L_k}}} \!-\! {\beta _k}{L_k} {  \ln \left| {{\Omega _k}} \right|} + \log IGG\left( {{L_k}|{a_k},{b_{k}},{c_{k}}} \right) + const
	\end{aligned}
\end{equation}

By comparing the logarithms of $p\left( {{L_k}} \right)$ and $q\left( {{L_k}} \right)$, the statistical distribution of \textit{L} can be assumed as \eqref{eq.IGG}, thus obtaining the iterative solutions of $a_k$, $b_{k}$ and $c_{k}$, as illustrated in \eqref{eq.ak}\verb|-|\eqref{eq.ck}.
\begin{equation}
\label{eq.ak}
{a_k} \!=\! {a_{k}^{(0)}} + \frac{9}{2}{N_k}
\end{equation}
\begin{equation}
\label{eq.bk}
\begin{aligned}
{b_{k}} \!=\! {b_{k}^{(0)}} + {\beta _{k}^{(0)}}\ln \left| {{\Omega _{k}^{(0)}}} \right| \!-\! {\beta _k}\ln \left| {{\Omega _k}} \right| \!-\! \sum\limits_{n \!=\! 1}^N {{r_{n,k}}\ln \left| {{C_n}} \right|}  \!-\! 3\left( {{N_k} + {\beta _{k}^{(0)}}\ln {\beta _{k}^{(0)}}} \right)
\end{aligned}
\end{equation}
\begin{equation}
\label{eq.ck}
\begin{aligned}
{c_{k}} \!=\! {c_{k}^{(0)}} + 4\left( {{N_k} + \frac{1}{{{\beta _{k}^{(0)}}}} \!-\! \frac{1}{{{\beta _k}}}} \right)
\end{aligned}
\end{equation}

The number of looks for each cluster is roughly estimated through the intensity variance test in a single channel \cite{APD11}, which is used to initialize the hyperparameters of IGG prior. From \eqref{eq.ak}\verb|-|\eqref{eq.ck}, one can observe that these parameters are highly related with $N_k$. Besides, the spatial information in Section \ref{sec:SpaSim} is introduced in a patch-wise manner. Therefore, the uncertainty of $L_k$ increases as the sample size in the $K$th cluster decreases, especially in the case of sub-sampling or small datasets. The numerical integration techniques related to \textit{L} can be derived from Tur{\'a}n-type inequalities \cite{BoundsBes}, as shown in \eqref{eq.EL_ti}\verb|-|\eqref{eq.EL_2_ti} and \eqref{eq.seineq1}\verb|-|\eqref{eq.clo_2}.

The variational lower bound of proposed model is given as follows.
\begin{equation}
\label{eq.ELBO}
\begin{aligned}
L(C,\theta ) &= \sum\limits_z  \iiint q\left( {\Sigma ,L,z,\pi } \right)\ln \left\{ {\frac{{p\left( {C,\Sigma ,L,z,\pi } \right)}}{{q\left( {\Sigma ,L,z,\pi } \right)}}} \right\}d\pi  dL d\Sigma  \\
&= \left\langle {\ln p\left( {C,\Sigma ,L,z,\pi } \right)} \right\rangle  \!-\! \left\langle {\ln q\left( {\Sigma ,L,z,\pi } \right)} \right\rangle \\
&= \left\langle {\ln p\left( {C|\Sigma ,L,z,\pi } \right)} \right\rangle  + \left\langle {\ln p\left( {\Sigma ,L} \right)} \right\rangle  + \left\langle {\ln p\left( {z|\pi } \right)} \right\rangle \\
&+ \left\langle {\ln p\left( \pi  \right)} \right\rangle  \!-\! \left\langle {\ln q\left( {\Sigma ,L} \right)} \right\rangle  \!-\! \left\langle {\ln q\left( z \right)} \right\rangle  \!-\! \left\langle {\ln q\left( \pi  \right)} \right\rangle
\end{aligned}
\end{equation}
where the expansion terms of \eqref{eq.ELBO} are given in \eqref{eq.ElogpC}\verb|-|\eqref{eq.ElogqSigmaL}. To avoid duplication, the expressions of $\left\langle {\ln p\left( {z|\pi } \right)} \right\rangle$, $\left\langle {\ln p\left( \pi  \right)} \right\rangle$, $\left\langle {\ln q\left( \pi  \right)} \right\rangle$ have been omitted in this section, which can be found in \cite{PRML}.

To summarize, this section develops and evaluates a variational Bayesian expectation maximization (VBEM) method for the proposed VBWMM, which performs unsupervised classification for PolSAR data based on complex Wishart mixture distribution. The proposed method can be divided into four steps, as presented in Algorithm \ref{alg.WMMVBEM}.

\begin{algorithm}
	\caption{Detailed implementation of WMMVBEM.}
	\label{alg.WMMVBEM}
	\hspace*{0.02in} {\bf Input:}
	PolSAR covariance matrices $\{C_n\}_{n=1}^{N}$, maximum iterations $T_{m}$, tolerance $\tau$, window size $win$, initial cluster number $K$, hyperparameters $\alpha^{(0)}$, $\pi_0$, $\beta^{(0)}$ $b^{(0)}$, $c^{(0)}$.\\
	\hspace*{0.02in} {\bf Output:}
	Classification result, lower bound and ENL estimate for each class.
	\begin{algorithmic}[1]		
		\State To avoid exponential overflow, add the diagonal elements of $\left\{C_n\right\}_{n=1}^{N}$ with $eps=10^{-15}$.
		\State Prepare the normalized Euclidean distance and Wishart distance between the central pixel and others from the same patch.
		\State Use K-means clustering to cluster normalized diagonal elements of $\left\{C_n\right\}_{n=1}^{N}$ into $K$ classes. Initialize $r_{n,k}$ ($r_{n,k}\!=\!1$ when the $n$th pixel belongs to $k$th class, otherwise $r_{n,k}\!=\!0$). 
		\State Set the hyperparameters of each component, e.g. $b_{k}^{(0)}$ and $c_{k}^{(0)}$. Besides, $\Omega_{k}^{(0)}$ is set to ${{I_{d}\!\cdot\!tr(\bar C_{k}^{(0)})} \mathord{\left/{\vphantom{{I_{d}\!\cdot\!tr(\bar C_{k}^{(0)})} {d}}} \right.	\kern-\nulldelimiterspace} {d}}$, where $\bar C_{k}^{(0)}$ is the mean covariance matrix of each cluster.
		\State Set the ENL of each cluster to the nominal number of looks and calculate $a_{k}^{(0)}$ according to the numerical solution of \eqref{eq.EL_ti}.
		\For{$iter \!=\! 1\!:\!T_{m}$}\Comment{Start iteration.}
		\State Update $r_{n,k}$ with \eqref{eq.rnk}.	\Comment{VE-step.}
		\State Update $\alpha_{n,k}$, $\beta_{n,k}$, $\Omega_{k}^{-1}$, $a_{n,k}$, $b_{nk}$, $c_{nk}$ with \eqref{eq.alpha_k_spa}\verb|-|\eqref{eq.ck_spa}.\Comment{VM-step.}
		\State Update ELBO $\mathcal{L}_{iter}$ with \eqref{eq.ELBO}. 
		\If{$\left(\mathcal{L}_{iter}-\mathcal{L}_{iter-1}\right)/\mathcal{L}_{iter-1} \!\le\! \tau$}\Comment{Converged?}
		\State break\Comment{Early stop.}
		\EndIf
		\EndFor
		\State \Return Clustering result, lower bound $\mathcal{L}$ and expectation values of $L_{k}$.
	\end{algorithmic}
\end{algorithm}

\subsection{Spatial Information based on Similarities}
\label{sec:SpaSim}
In the initialization stage, the proposed model can make full use of the spatial information embedded in neighboring pixels by introducing geometric similarity based on Euclidean distance and covariance matrix similarity based on Wishart distance \cite{WisDis}. The geometric similarity decreases with the increase of Euclidean distance from the central pixel, and the covariance matrix similarity measures the Wishart distance between mean covariance matrix and each pixel in the same patch. 

Both similarities have found their modifications in the iterative solution of each hyperparameter, as shown in \eqref{eq.alpha_k_spa}\verb|-|\eqref{eq.ck_spa}. These predefined constants can be computationally efficient with the use of tensor multiplication. By introducing geometric similarity, $\alpha _k$ is updated to
\begin{equation}
\label{eq.alpha_k_spa}
\alpha _k \!=\! \sum\limits_{n \!=\! 1}^N {\sum\limits_{m \in {P_n}} {{\gamma _{n,m}}{\alpha _{n,k}}} } 
\end{equation}
where ${\gamma _{n,m}} \!=\! {{{\sigma _{n,m}}} \mathord{\left/{\vphantom {{{\sigma _{n,m}}}{\sum\limits_{m \in {P_n}} {{\sigma _{n,m}}} }}} \right.\kern-\nulldelimiterspace} {\sum\limits_{m \in {P_n}} {{\sigma _{n,m}}} }}$, ${\sigma _{n,m}} \!=\! {1 \mathord{\left/{\vphantom {1 {\left( {1 + d_{n,m}^2} \right)}}} \right.\kern-\nulldelimiterspace} {\left( {1 + d_{n,m}^2} \right)}}$, $d_{n,m}$ is the Euclidean distance between the center pixel and the \textit{m}th covariance matrix in the \textit{n}th batch.

The covariance matrix similarity is introduced by summing up the multiplication of $\omega _{n,m}$ by the terms containing $r_{n,k}$, thus the corresponding iterative solutions are updated to \eqref{eq.betak_spa}\verb|-|\eqref{eq.ck_spa}.

\begin{equation}
\label{eq.betak_spa}
\beta _k \!=\! {\beta _{k}^{(0)}} + \sum\limits_{n \!=\! 1}^N {\sum\limits_{m \in {P_n}} {{r_{n,k}}{\omega _{n,m}}} }
\end{equation}

\begin{equation}
\label{eq.omegak_spa}
\Omega _k^{-1} \!=\! \frac{1}{{{\beta _k}}}\left( {{\beta _{k}^{(0)}}\left(\Omega _{k}^{(0)}\right)^{-1} + \sum\limits_{n \!=\! 1}^N {\sum\limits_{m \in {P_n}} {{r_{n,k}}{\omega _{n,m}}{C_m}} } } \right)
\end{equation}

\begin{equation}
\label{eq.ak_spa}
a_k \!=\! {a_{k}^{(0)}} + \frac{9}{2}{\sum\limits_{n \!=\! 1}^N {\sum\limits_{m \in {P_n}} {{r_{n,k}}{\omega _{n,m}}} }}    
\end{equation}

\begin{equation}
	\label{eq.bk_spa}
	\begin{aligned}
		{b_{k}} \!=\! {b_{k}^{(0)}} + {\beta _{k}^{(0)}}\ln \left| {{\Omega _{k}^{(0)}}} \right| \!-\! {\beta _k}\ln \left| {{\Omega _k}} \right| \!-\! \sum\limits_{n \!=\! 1}^N {{r_{n,k}}\ln \left| {{C_n}} \right|}  \!-\! 3\left( {\sum\limits_{n \!=\! 1}^N {\sum\limits_{m \in {P_n}} {{r_{n,k}}{\omega _{n,m}}} }  + {\beta _{k}^{(0)}}\ln {\beta _{k}^{(0)}}} \right)
	\end{aligned}
\end{equation}

\begin{equation}
	\label{eq.ck_spa}
	\begin{aligned}
		{c_{k}} \!=\! {c_{k}^{(0)}} + 4\left( {\sum\limits_{n \!=\! 1}^N {\sum\limits_{m \in {P_n}} {{r_{n,k}}{\omega _{n,m}}} }  + \frac{1}{{{\beta _{k}^{(0)}}}} \!-\! \frac{1}{{{\beta _k}}}} \right)
	\end{aligned}
\end{equation}

$\omega _{n,m}$ is defined as the normalized Wishart distance in \eqref{eq.normWisDis}.

\begin{equation}
\label{eq.normWisDis}
\omega _{n,m} \!=\! {{\exp \left( { \!-\! {D_W}\left( {{C_m},{{\bar C}_n}} \right)} \right)} \mathord{\left/
		{\vphantom {{\exp \left( { \!-\! {D_W}\left( {{C_m},{{\bar C}_n}} \right)} \right)} {\sum\limits_{m \in {P_n}} {\exp \left( { \!-\! {D_W}\left( {{C_m},{{\bar C}_n}} \right)} \right)} }}} \right.
		\kern-\nulldelimiterspace} {\sum\limits_{m \in {P_n}} {\exp \left( { \!-\! {D_W}\left( {{C_m},{{\bar C}_n}} \right)} \right)} }}
\end{equation}

where ${D_W}\left(  \cdot  \right)$ is the Wishart distance between two matrices \cite{WisDis}, ${C_m}$ and ${\bar C_n}$ are the \textit{m}th pixel and the mean value of covariance matrices in the \textit{n}th patch, respectively.

\section{Experimental Results}
\label{sec:Exps}
In this section, the proposed method is evaluated on four real-measured PolSAR datasets recorded by EMISAR, AIRSAR and RADARSAT-2. Section~\ref{subsec:DataPara} first introduces the experimental data, comparative methods, experimental arrangement and parameter settings. Then, ablation study on the proposed method is conducted in Section \ref{subsec:Exp_Abl}, which validates the effectiveness of proposed method in automated clustering and ENL estimation. From Section~\ref{subsec:Exp_Abl} to Section~\ref{subsec:ExpQuebec}, the superiority of proposed method is demonstrated by comparison with $H/\alpha$-Wishart classifier as the baseline, and four other state-of-the-art methods, where standard-Wishart and relaxed-Wishart \cite{APD11} are based on Mellin-kind statistics, WMM and SVWMM \cite{SVWMM} are derived from VB framework. Besides, the 1-D ENL distribution and estimation results of three ENL estimators, including ML \cite{MLwin}, TM \cite{EstENL} and WMLC \cite{GoF_Mellin}, have validated the effectiveness of proposed method in ENL estimation.
\subsection{Experimental Data and Settings}
\label{subsec:DataPara}
The Foulum dataset in \ref{subsec:Exp_Abl} is utilized to validate the effectiveness of proposed VBWMM in automatic determination of cluster number and qualitatively analyze the impact of hyperparameters on the clustering results. The San Francisco dataset in \ref{subsec:ExpSanP} is used to validate the effectiveness of proposed model under natural and man-made scenes, e.g. sea and urban. The Flevoland dataset in \ref{subsec:ExpAIRFlev} is applied to quantitatively evaluate the performance of proposed method in agricultural areas. The Quebec dataset in \ref{subsec:ExpQuebec} is adopted to validate the performance of proposed method on spaceborne PolSAR platforms and in heterogeneous areas. 

To avoid inconsistent ENL estimates in different polarimetric channels \cite{VarENL}, the non-uniform windows containing a mixture of classes have been excluded when conducting ENL estimation. To obtain more reasonable clustering results and accurate estimation of ENL values, the sliding window should be neither large nor small. This is because tiny sliding windows can make it hard for the estimators to capture the variance of ENL, resulting in invalid ENL estimates. On the other hand, large non-uniform windows can generally contain a mixture of classes, resulting in inconsistent estimates between channels. Therefore, the 1-D ENL distribution and estimation results of ML \cite{MLwin} and TM \cite{EstENL} are obtained with a sliding estimation window of size $5 \!\times\! 5$ to cover the whole image. The WMLC estimator \cite{GoF_Mellin} is first evaluated on 25 samples randomly chosen from the same class, to obtain the ENL values over the same cluster, where the final result is given by the median value of these estimates. For better visualization, the 1-D distribution of ENL values together with the estimations of ML \cite{MLwin}, TM \cite{EstENL} and WMLC \cite{GoF_Mellin} are illustrated to validate the effectiveness of proposed method in ENL estimation. 

For the initialization of hyperparameters, $\left(\Omega _{k}^{(0)}\right)^{-1}$ is set as ${I_d} \cdot {{tr\left( {{{\bar C}_k}} \right)} \mathord{\left/{\vphantom {{tr\left( {{{\bar C}_k}} \right)} d}} \right.\kern-\nulldelimiterspace} d}$, where $\bar C_k$ is the average covariance matrix of each cluster initialized by K-means, $I_d$ is the $d \!\times\! d$ unit diagonal matrix in size. To avoid possible discrimination on any component, the parameters $\lbrace \alpha_{n,k}^{(0)}, a_{n,k}^{(0)}, b_{nk10}, b_{nk20}, \beta_{k}^{(0)}\rbrace$ are set equal for each cluster. $\alpha_{n,k}^{(0)} \!=\! 1 \!\times\! 10^{-1} \!\sim\! 1 \!\times\! 10^{0}$, $b_{nk10} \!=\! 1 \!\sim\! 5$ and $b_{nk20} \!=\! 1 \!\sim\! 5$ are the general choice, with $a_{n,k}^{(0)}$ derived from the expectation value of $L$, which should equal to the nominal number of looks. $\beta_{k}^{(0)}$ plays an important role in determining the number of clusters, where a large $\beta_{k}^{(0)}$ is recommended if $K$ is significantly greater than the expected cluster number. If K is set properly, $\beta_{k}^{(0)}$ is commonly set as $5 \!\times\! 10^{0} \!\sim\! 5 \!\times\! 10^{1}$. 

The experiments are implemented with MATLAB R2020b and C++ on Windows 10 operation system with 64GB RAM and Intel Core i9-9900K CPU.  As a matter of experience, the hyperparameters are set equal to avoid bias against any cluster. In the initialization stage, K-means is used to automatically determine the initial mean covariance matrix of each component $C_{0k}$, with an aim to fasten the convergence process and obtain more robust clustering results. The GoF methods \cite{APD11} realize automatic determination of components by sub-sampling and tuning confidence level. In the experiments, many attempts have been made to find the most suitable parameter combinations for standard-Wishart, relaxed-Wishart \cite{APD11,AGKQ}, WMM and SVWMM \cite{SVWMM}. The experimental results show that the proposed model is superior to the above mentioned comparative methods in clustering performance and ENL estimation. Another advantage of proposed method lies in the conjugation relationship between the mean covariance matrix $\Sigma$ and number of looks \textit{L}, which results in a closed-form variational lower bound, which serves as the stopping criteria for the iterative process. When the clustering results get relatively stable, the changing rate of variational lower bound generally falls between $10^{-9}$ and $10^{-8}$, so the convergence tolerance is generally set to $10^{-8}$ in the experiments. 

\subsection{Ablation Study on Hyperparameters}
\label{subsec:Exp_Abl}
\begin{figure}[htbp]
	\centering
	\includegraphics[width=0.45\textwidth]{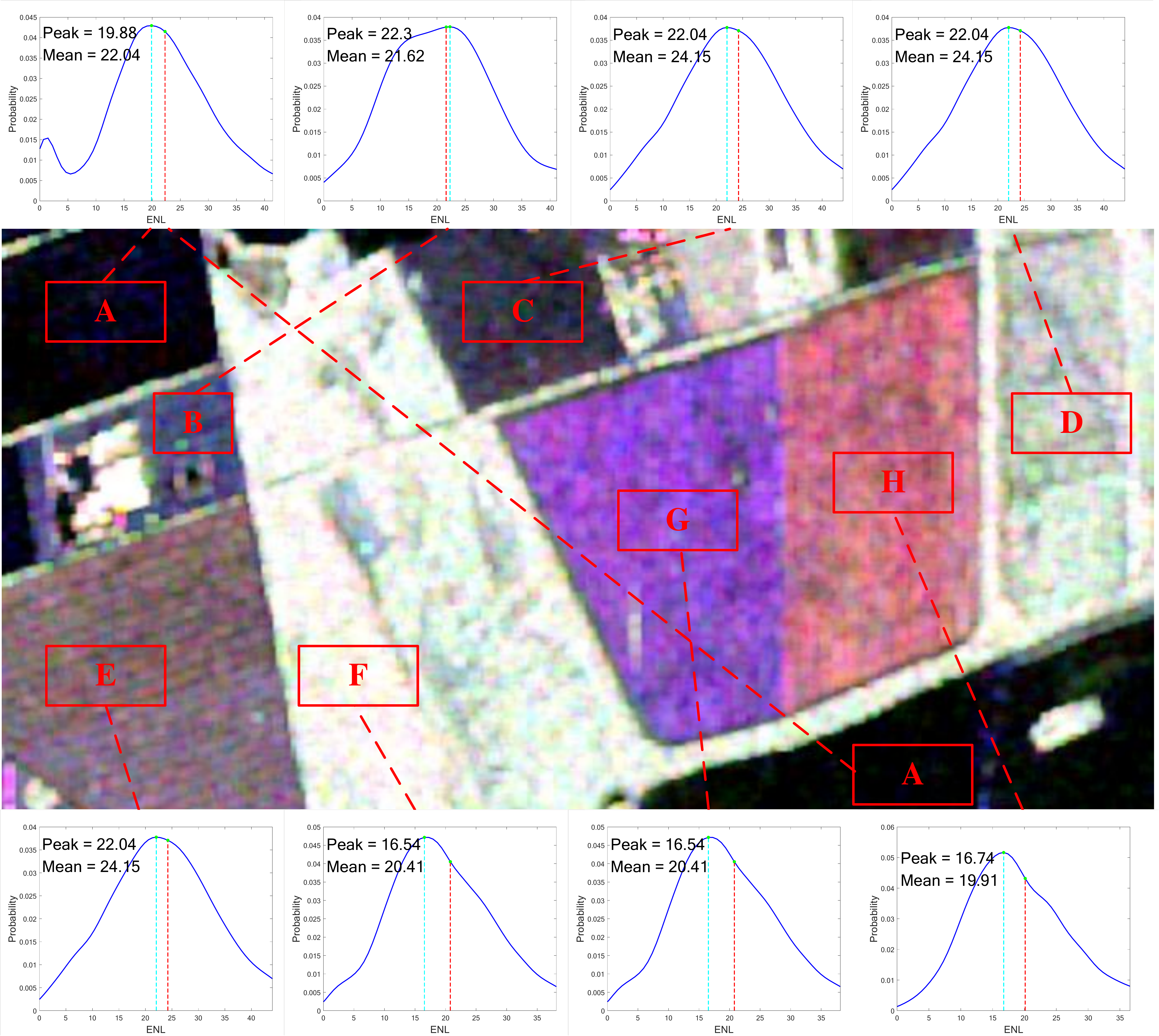}
	\caption{Foulum dataset and ENL histograms with distribution shape (single-look estimated, 8 classes).}
	\label{fig.ENL_Distributions}
\end{figure}

The PolSAR data used in the following experiments is a sub area selected from EMISAR Foulum dataset, which contains forests and a variety of crops, as illustrated in Fig. \ref{fig.ENL_Distributions}. From Fig. \ref{fig.ENL_Distributions}, it is observed that the ENL estimate of each class can be quite different, where the ENL distributions have rejected the top 5\% largest values in the histograms for better visualization. Therefore, the variance of ENL between clusters cannot be neglected. In our proposed method, the number of looks for each cluster is assumed as a variable rather than a constant. To alleviate the impact of speckle noise, the $6 \!\times\! 3$ window box-car filter \cite{DoulgerisUDis} is utilized to perform multi-look preprocess for the following tests. The purpose of this section is to analyze the influence of three hyperparameters, namely $\beta_ {0}$, $\alpha_ {n,k}^{(0)}$ and spatial sliding window size $win$, on the clustering performance and variational lower bound of proposed model. To demonstrate the effectiveness of proposed VBWMM in adaptive determination of cluster number, the initial cluster number $K$ is set to 50. Moreover, to confirm the validity of proposed IGG prior in ENL estimation, $\langle L \rangle$ is set to 15 for each cluster, which is slightly lower than the nominal number of looks 18. 

Theoretically speaking, it can be concluded from \eqref{eq.betak} that the product of $\beta ^{(0)}$ and inversion of $\Omega ^{(0)}$ acts as the regularization term in the iterative solution of $\beta$. Therefore, $\beta ^{(0)}$ should be neither large nor small, otherwise the components with less data may be ignored or excessively stressed, which can make it hard to converge to the true posterior of observation data. To analyze the effect of $\beta ^{(0)}$, $\alpha _{n,k}^{(0)}$ is set to $1 \!\times\! 10^{-1}$, $\beta ^{(0)}$ picks its value from $5 \!\times\! 10^0$, $5 \!\times\! 10^1$, $5 \!\times\! 10^2$ and $3 \!\times\! 10^3$.

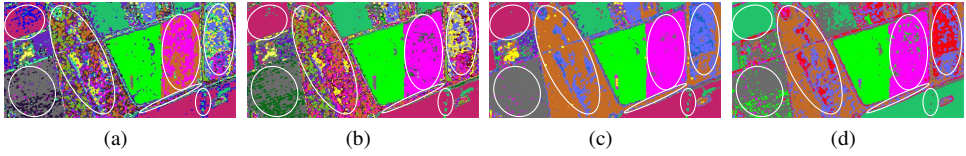
\begin{figure*}[htbp]
	\centering
	\subfigtopskip=2pt
	\subfigbottomskip=2pt
	\subfigcapskip=-5pt
	\subfigure[]{
		\label{fig.beta5}
		\begin{annotate}{\includegraphics[width=.225\textwidth]{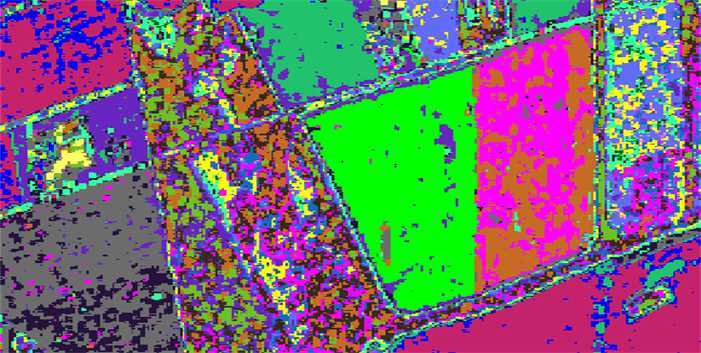}}{0.2}
			\draw[color=white,rotate around={25:(-2.5, 0)}] (-2.5, 0) ellipse (1.5 and 3.9);
			\draw[color=white,rotate around={25:(-5.8, -2.1)}] (-5.8, -2.1) ellipse (1.5 and 1.7);	
			\draw[color=white,rotate around={25:(-6.2, 2.3)}] (-6.2, 2.5) ellipse (1.35 and 1.1);
			\draw[color=white] (6.5, 1.35) ellipse (1.0 and 2.35);
			\draw[color=white] (5.45, -2.8) ellipse (0.5 and 0.9);
			\draw[color=white] (4, 0.5) ellipse (1.25 and 2.5);			
			\draw[color=white,rotate around={25:(3.3, -2.55)}] (3.3, -2.55) ellipse (2.25 and 0.2);				
		\end{annotate}	
	}
	\hspace{-.045\textwidth}
	\subfigure[]{
		\label{fig.beta50}
		\begin{annotate}{\includegraphics[width=.225\textwidth]{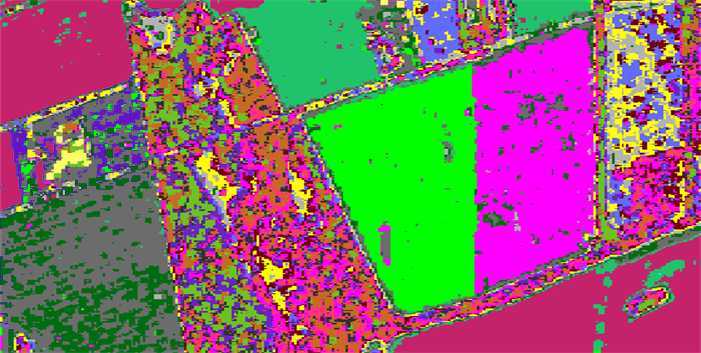}}{0.2}
			\draw[color=white,rotate around={25:(-2.5, 0)}] (-2.5, 0) ellipse (1.5 and 3.9);
			\draw[color=white,rotate around={25:(-5.8, -2.1)}] (-5.8, -2.1) ellipse (1.5 and 1.7);	
			\draw[color=white,rotate around={25:(-6.2, 2.3)}] (-6.2, 2.5) ellipse (1.35 and 1.1);
			\draw[color=white] (6.5, 1.35) ellipse (1.0 and 2.35);
			\draw[color=white] (5.45, -2.8) ellipse (0.5 and 0.9);
			\draw[color=white] (4, 0.5) ellipse (1.25 and 2.5);			
			\draw[color=white,rotate around={25:(3.3, -2.55)}] (3.3, -2.55) ellipse (2.25 and 0.2);			
		\end{annotate}	
	}
	\hspace{-.045\textwidth}
	\subfigure[]{
		\label{fig.beta500}
		\begin{annotate}{\includegraphics[width=.225\textwidth]{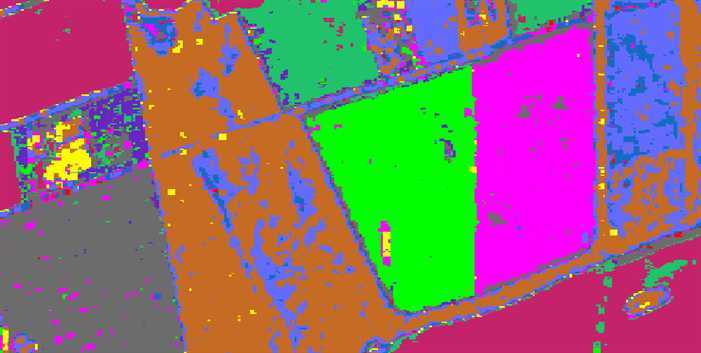}}{0.2}
			\draw[color=white,rotate around={25:(-2.5, 0)}] (-2.5, 0) ellipse (1.5 and 3.9);
			\draw[color=white,rotate around={25:(-5.8, -2.1)}] (-5.8, -2.1) ellipse (1.5 and 1.7);	
			\draw[color=white,rotate around={25:(-6.2, 2.3)}] (-6.2, 2.5) ellipse (1.35 and 1.1);
			\draw[color=white] (6.5, 1.35) ellipse (1.0 and 2.35);
			\draw[color=white] (5.45, -2.8) ellipse (0.5 and 0.9);
			\draw[color=white] (4, 0.5) ellipse (1.25 and 2.5);			
			\draw[color=white,rotate around={25:(3.3, -2.55)}] (3.3, -2.55) ellipse (2.25 and 0.2);		
		\end{annotate}	
	}
	\hspace{-.045\textwidth}
	\subfigure[]{
		\label{fig.beta3000}
		\begin{annotate}{\includegraphics[width=.225\textwidth]{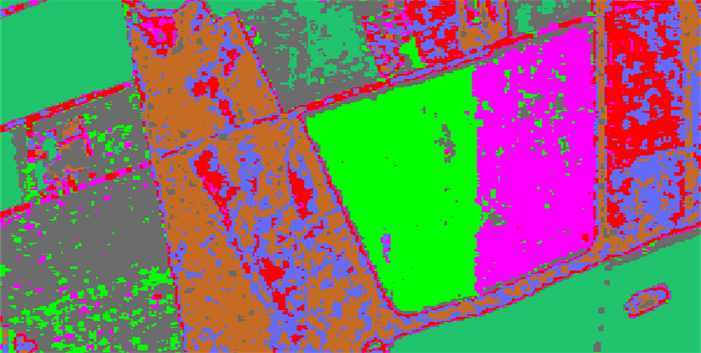}}{0.2}
			\draw[color=white,rotate around={25:(-2.5, 0)}] (-2.5, 0) ellipse (1.5 and 3.9);
			\draw[color=white,rotate around={25:(-5.8, -2.1)}] (-5.8, -2.1) ellipse (1.5 and 1.7);	
			\draw[color=white,rotate around={25:(-6.2, 2.3)}] (-6.2, 2.5) ellipse (1.35 and 1.1);
			\draw[color=white] (6.5, 1.35) ellipse (1.0 and 2.35);
			\draw[color=white] (5.45, -2.8) ellipse (0.5 and 0.9);
			\draw[color=white] (4, 0.5) ellipse (1.25 and 2.5);			
			\draw[color=white,rotate around={25:(3.3, -2.55)}] (3.3, -2.55) ellipse (2.25 and 0.2);		
		\end{annotate}	
	}
	\caption{Ablation analysis on $\beta^{(0)}$. (a) $\beta^{(0)}\!=\!5\!\times\!10^{0}$, (b) $\beta^{(0)}\!=\!5\!\times\!10^{1}$, (c) $\beta^{(0)}\!=\!5\!\times\!10^{2}$, (d) $\beta^{(0)}\!=\!3\!\times\!10^{3}$.}
	\label{fig.Foulum_Beta}
\end{figure*}	

The classification maps for various values of $\beta^{(0)}$ are demonstrated in Fig. \ref{fig.Foulum_Beta}, where the resulting cluster numbers are 44, 31, 15, and 9, respectively. As such, there is an evidence that the resulting number of clusters can be less than the initial 50 components, and the resulting cluster number gradually decreases with the increase of $\beta^{(0)}$. Therefore, the proposed method can automatically determine the cluster number according to the PolSAR data if the initial number of components $K$ is large enough. 

However, it is also found that too large $\beta ^{(0)}$ values tend to merge different clusters into the same land cover, such as when $\beta ^{(0)}$ is set as $3 \!\times\! 10^3$, the proposed model cannot distinguish the two agricultural areas in the white eclipses in the middle and bottom left corner. This indicates that the model with a large regularization item $\beta ^{(0)}$ is more likely to make a rough estimate to approximate the true posterior and is more inclined to merge the clusters with less samples or similar features, thus cannot overstep the local optimum by iteration. Therefore, large $\beta ^{(0)}$ values are recommended only when the expected cluster number is far less than the initial number of clusters $K$. 

On the contrary, smaller $K$s help to distinguish similar land covers and preserve more texture details. For example, when $\beta ^{(0)} \!=\! 5 \!\times\! 10^0$, more details can be found compared to Fig. \ref{fig.beta50}, Fig. \ref{fig.beta500} and Fig. \ref{fig.beta3000}, with the land covers in black eclipses separated into two or multiple clusters. However, the proposed model fails to remove the clusters with few samples when $\beta ^{(0)} \!=\! 5 \!\times\! 10^0$, thus resulting in redundant noise-like clusters, especially in the white eclipses of Fig. \ref{fig.beta5}, which makes the clustering result hard to interpret. Although $\beta^{(0)}$ could result in noise-like clusters and over-smoothing results, the good classification map can be achieved by appropriately selecting the values of $\beta^{(0)}$. 

\begin{figure*}[ht]
	\centering
	\subfigtopskip=2pt
	\subfigbottomskip=2pt
	\subfigcapskip=-5pt
	\subfigure[]{
		\label{fig.elbo_beta}
		\includegraphics[width=.3\textwidth]{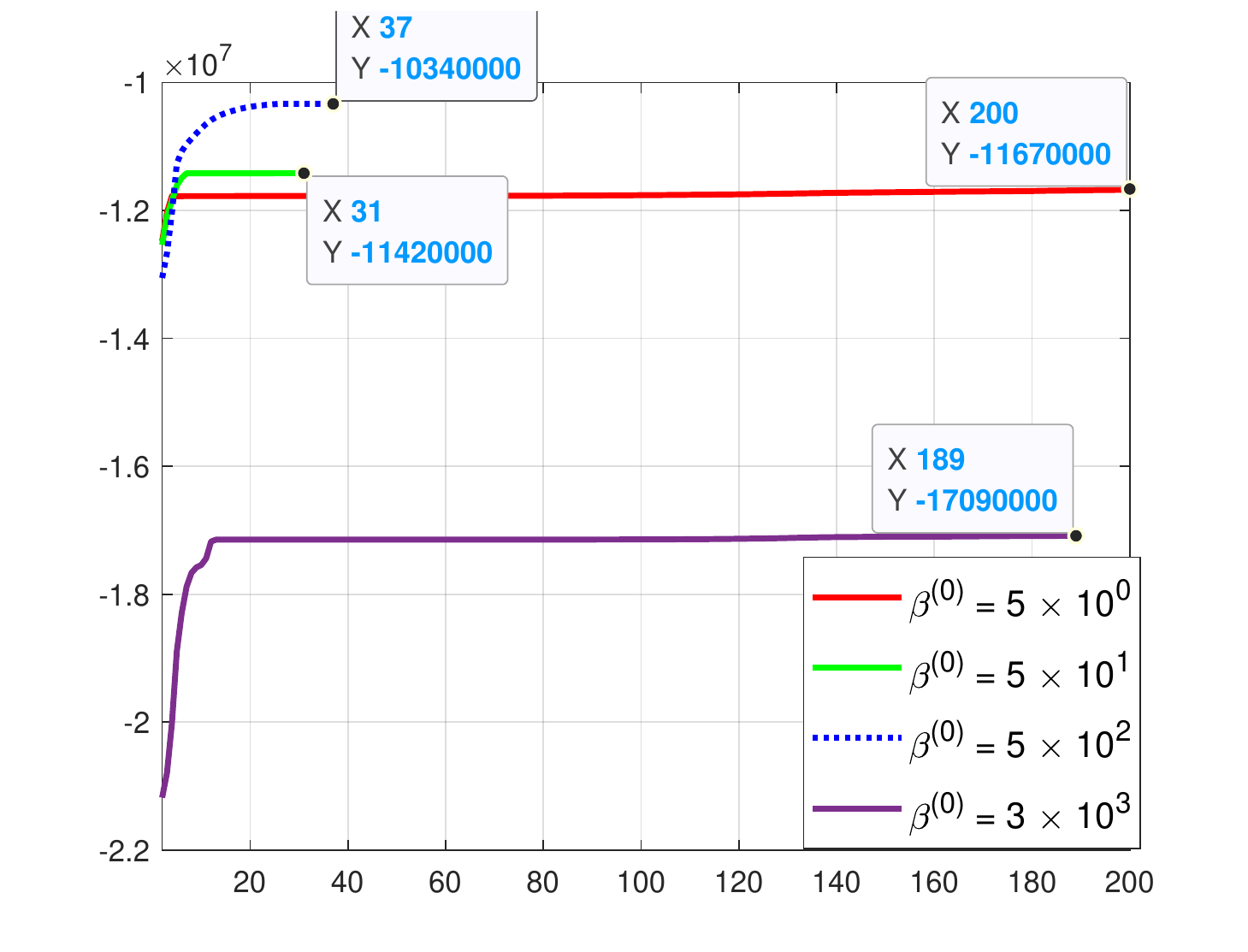}
	}
	\subfigure[]{
		\label{fig.elbo_alp}
		\includegraphics[width=.3\textwidth]{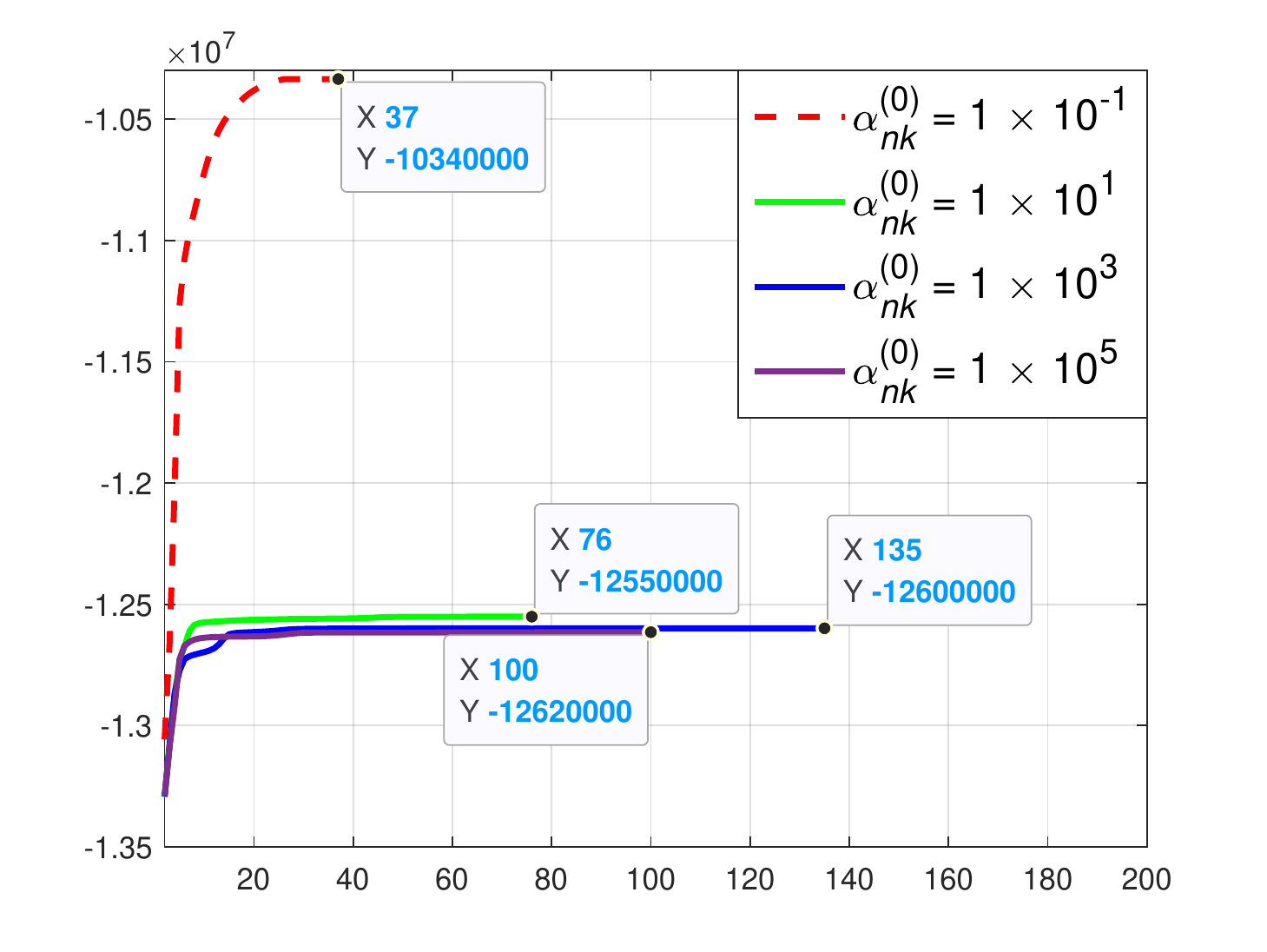}
	}
	\subfigure[]{
		\label{fig.elbo_win}
		\includegraphics[width=.3\textwidth]{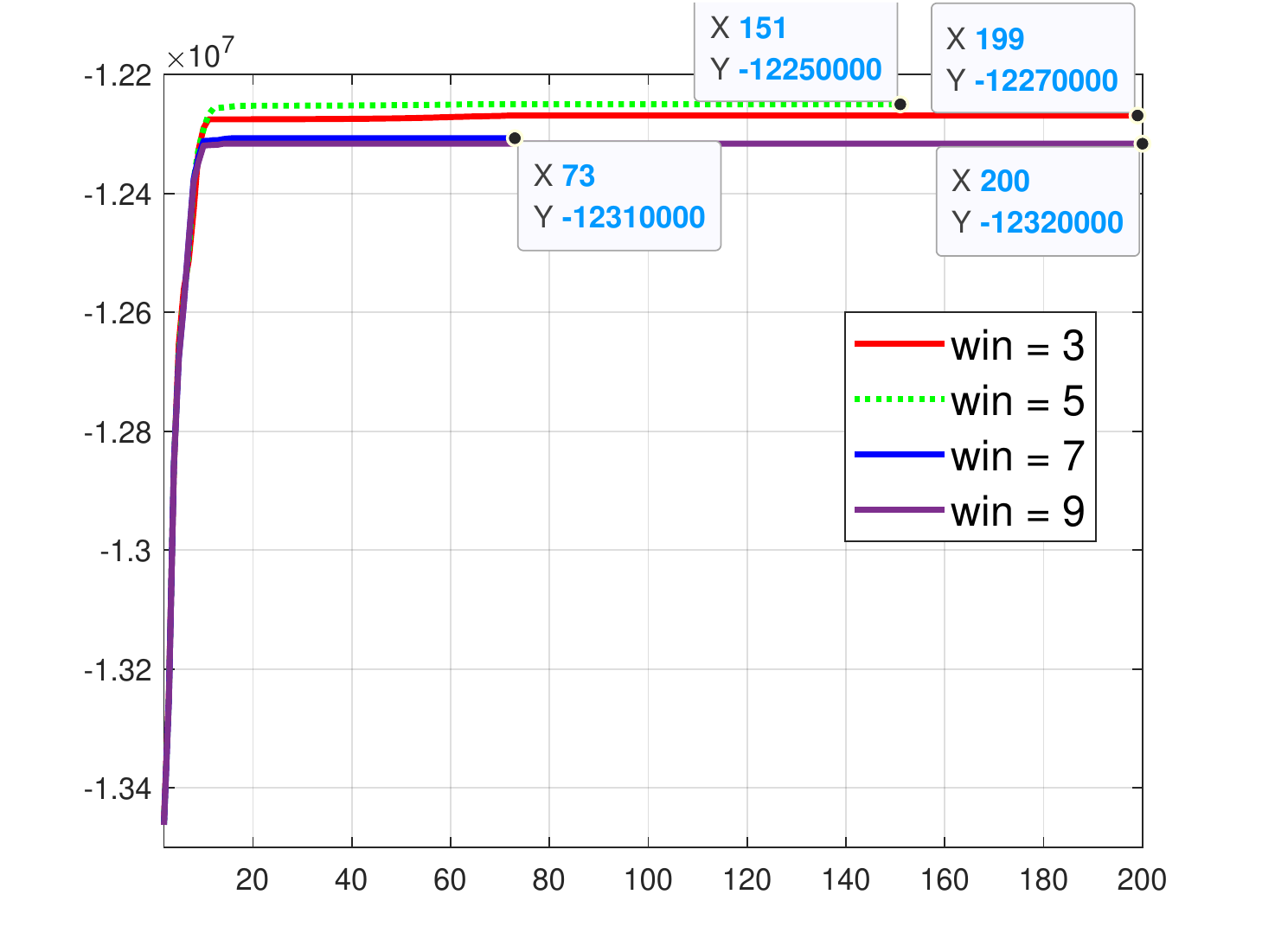}
	}
	\caption{The impact of different factors on ELBO. (a) $\beta^{(0)}$, (b) $\alpha^{(0)}$, (c) window size $win$.}
	\label{fig.Foulum_ELBO}
\end{figure*}

When $\beta^{(0)}$ is taken as $5 \!\times\! 10^0$, $5 \!\times\! 10^1$, $5 \!\times\! 10^2$ and $3 \!\times\! 10^3$, the lower bound converges to $-1.167 \!\times\!10^7$, $-1.142 \!\times\!10^7$, $-1.034 \!\times\!10^7$ and  $-1.709 \!\times\!10^7$ respectively, where it is observed that the lower bound first increases and then decreases with the value of $\beta^{(0)}$, as illustrated in Fig. \ref{fig.elbo_beta}. It is worth noticing that the classification map is closest to the corresponding Pauli RGB image when $\beta^{(0)} \!=\! 5 \!\times\!10^2$, as shown in Fig. \ref{fig.beta500}. Besides, one can observe that the lower bound tends to converge faster with the increase of $\beta ^{(0)}$, this is because the observation data gradually make less contribution to the clustering task. However, if $\beta ^{(0)}$ is too small, as is illustrated by the red curve when $\beta ^{(0)} \!=\! 5 \!\times\! 10^{0}$, it would be hard for the lower bound to converge to a steady state because of the lack of constraints on prior and redundancy in cluster centers, especially for large $K$s. Therefore, the hyperparameter value $\beta^{(0)}$ should depend on the initial cluster number $K$ and expected number of clusters.

\begin{figure*}[ht]
	\centering
	\subfigtopskip=2pt
	\subfigbottomskip=2pt
	\subfigcapskip=-5pt
	\subfigure[]{
		\label{fig.alp-1}
		\begin{annotate}{\includegraphics[width=.225\textwidth]{Foulum_VBWMM.png}}{0.2}
			\draw[color=white,rotate around={25:(-2.5, 0)}] (-2.5, 0) ellipse (1.5 and 3.9);
			\draw[color=white,rotate around={25:(-5.8, -2.1)}] (-5.8, -2.1) ellipse (1.5 and 1.7);	
			\draw[color=white,rotate around={25:(-6.2, 2.3)}] (-6.2, 2.5) ellipse (1.35 and 1.1);
			\draw[color=white] (6.5, 1.35) ellipse (1.0 and 2.35);
			\draw[color=white] (5.45, -2.8) ellipse (0.5 and 0.9);
			\draw[color=white] (4, 0.5) ellipse (1.25 and 2.5);			
			\draw[color=white,rotate around={25:(3.3, -2.55)}] (3.3, -2.55) ellipse (2.25 and 0.2);		
		\end{annotate}	
	}
	\hspace{-.045\textwidth}
	\subfigure[]{
		\label{fig.alp1}
		\begin{annotate}{\includegraphics[width=.225\textwidth]{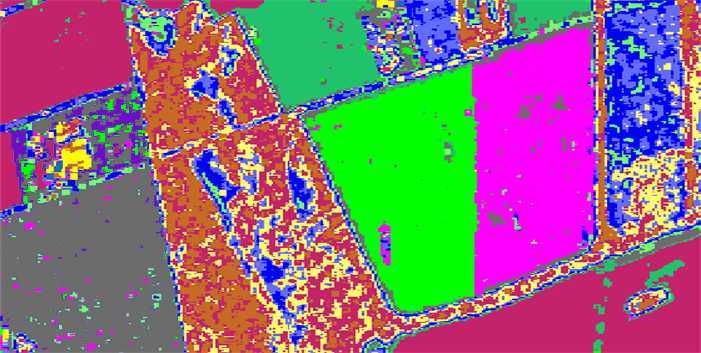}}{0.2}
			\draw[color=white,rotate around={25:(-2.5, 0)}] (-2.5, 0) ellipse (1.5 and 3.9);
			\draw[color=white,rotate around={25:(-5.8, -2.1)}] (-5.8, -2.1) ellipse (1.5 and 1.7);	
			\draw[color=white,rotate around={25:(-6.2, 2.3)}] (-6.2, 2.5) ellipse (1.35 and 1.1);
			\draw[color=white] (6.5, 1.35) ellipse (1.0 and 2.35);
			\draw[color=white] (5.45, -2.8) ellipse (0.5 and 0.9);
			\draw[color=white] (4, 0.5) ellipse (1.25 and 2.5);			
			\draw[color=white,rotate around={25:(3.3, -2.55)}] (3.3, -2.55) ellipse (2.25 and 0.2);		
		\end{annotate}	
	}
	\hspace{-.045\textwidth}
	\subfigure[]{
		\label{fig.alp3}
		\begin{annotate}{\includegraphics[width=.225\textwidth]{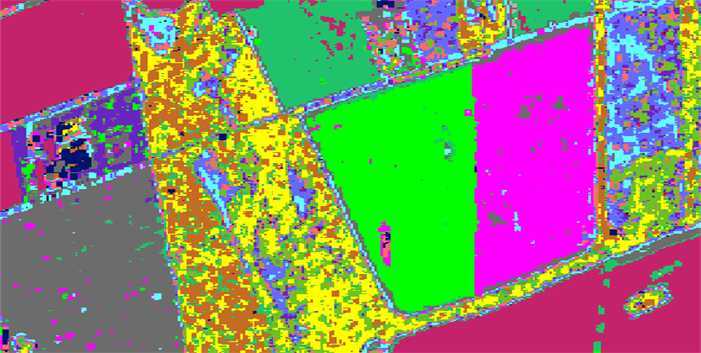}}{0.2}
			\draw[color=white,rotate around={25:(-2.5, 0)}] (-2.5, 0) ellipse (1.5 and 3.9);
			\draw[color=white,rotate around={25:(-5.8, -2.1)}] (-5.8, -2.1) ellipse (1.5 and 1.7);	
			\draw[color=white,rotate around={25:(-6.2, 2.3)}] (-6.2, 2.5) ellipse (1.35 and 1.1);
			\draw[color=white] (6.5, 1.35) ellipse (1.0 and 2.35);
			\draw[color=white] (5.45, -2.8) ellipse (0.5 and 0.9);
			\draw[color=white] (4, 0.5) ellipse (1.25 and 2.5);			
			\draw[color=white,rotate around={25:(3.3, -2.55)}] (3.3, -2.55) ellipse (2.25 and 0.2);		
		\end{annotate}	
	}
	\hspace{-.045\textwidth}
	\subfigure[]{
		\label{fig.alp5}
		\begin{annotate}{\includegraphics[width=.225\textwidth]{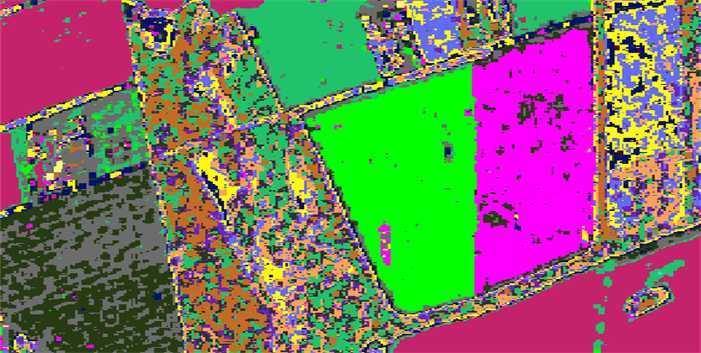}}{0.2}
			\draw[color=white,rotate around={25:(-2.5, 0)}] (-2.5, 0) ellipse (1.5 and 3.9);
			\draw[color=white,rotate around={25:(-5.8, -2.1)}] (-5.8, -2.1) ellipse (1.5 and 1.7);	
			\draw[color=white,rotate around={25:(-6.2, 2.3)}] (-6.2, 2.5) ellipse (1.35 and 1.1);
			\draw[color=white] (6.5, 1.35) ellipse (1.0 and 2.35);
			\draw[color=white] (5.45, -2.8) ellipse (0.5 and 0.9);
			\draw[color=white] (4, 0.5) ellipse (1.25 and 2.5);			
			\draw[color=white,rotate around={25:(3.3, -2.55)}] (3.3, -2.55) ellipse (2.25 and 0.2);		
		\end{annotate}	
	}
	\caption{Ablation analysis on $\alpha^{(0)}$. (a) $\alpha^{(0)}\!=\!10^{-1}$, (b) $\alpha^{(0)}\!=\!10^{1}$, (c) $\alpha^{(0)}\!=\!10^{3}$, (d) $\alpha^{(0)}\!=\!10^{5}$.}
	\label{fig.Foulum_Alpha}
\end{figure*}	

To analyze the impact of $\alpha_{n,k}^{(0)}$, $\beta ^{(0)}$ is set to the optimal value $5 \!\times\! {10^2}$. For better comparison, $\alpha _{n,k}^{(0)}$ picks its value from $1 \!\times\! 10^{-1}$, $1 \!\times\! 10^{1}$, $1 \!\times\! 10^{3}$ and $1 \!\times\! 10^{5}$. By comparing the clustering results in Fig. \ref{fig.Foulum_Alpha}, it is easy to find that the number of clusters shows no direct correlation with $\alpha _{n,k}^{(0)}$. In VB framework, $\alpha _{n,k}^{(0)}$ plays a different role in controlling the smoothness of resulting classification maps. Although the clustering results in field A (red pixels) are similar for diversifying values of $\alpha _{n,k}^{(0)}$, it is quite different in forest D and forest F (right and middle parts of the classification maps), as shown in the white eclipses in Fig. \ref{fig.Foulum_Alpha}. In another word, large $\alpha _{n,k}^{(0)}$ values (e.g. $\alpha_{n,k}^{(0)} \!=\! 1 \!\times\! 10^{3}$ in Fig. \ref{fig.alp3} and $1 \!\times\! 10^{5}$ in Fig. \ref{fig.alp5}) can lead to a complex structure in heterogeneous areas, such as forests. 

On the contrary, simple structures are often recognized for relatively small $\alpha _{n,k}^{(0)}$ values (e.g. Fig. \ref{fig.alp-1} and Fig. \ref{fig.alp1}), which are usually recommended for better visual interpretation. It is observed from Fig. \ref{fig.alp-1} that the structure and details in red eclipses are explicitly exhibited, which is in accordance with those in the Pauli RGB image in Fig. \ref{fig.ENL_Distributions}. Besides, as $\alpha _{n,k}^{(0)}$ increases from $1 \!\times\! 10^{-1}$ to $1 \!\times\! 10^{5}$, the lower bound gradually converges to a lower state, as is shown in Fig. \ref{fig.elbo_alp}. However, the lower bound may decrease if $\alpha _{n,k}^{(0)}$ is too small, indicating that the convergence cannot be ensured for inappropriate $\alpha _{n,k}^{(0)}$ values. 

\begin{figure*}[ht]
	\centering
	\subfigtopskip=2pt
	\subfigbottomskip=2pt
	\subfigcapskip=-5pt
	\subfigure[]{
		\label{fig.win3}
		\begin{annotate}{\includegraphics[width=.225\textwidth]{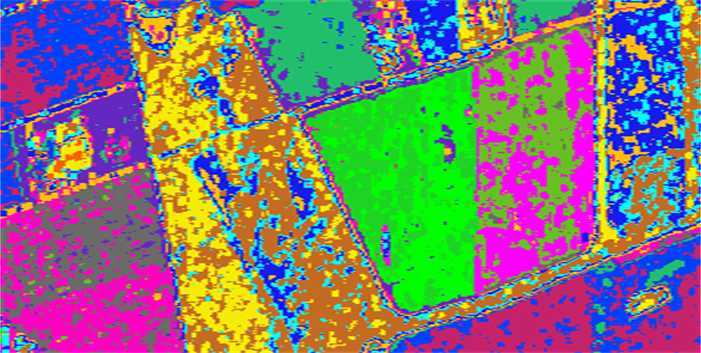}}{0.2}
			\draw[color=white,rotate around={25:(-2.5, 0)}] (-2.5, 0) ellipse (1.5 and 3.9);
			\draw[color=white,rotate around={25:(-5.8, -2.1)}] (-5.8, -2.1) ellipse (1.5 and 1.7);	
			\draw[color=white,rotate around={25:(-6.2, 2.3)}] (-6.2, 2.5) ellipse (1.35 and 1.1);
			\draw[color=white] (6.5, 1.35) ellipse (1.0 and 2.35);
			\draw[color=white] (5.45, -2.8) ellipse (0.5 and 0.9);
			\draw[color=white] (4, 0.5) ellipse (1.25 and 2.5);			
			\draw[color=white,rotate around={25:(3.3, -2.55)}] (3.3, -2.55) ellipse (2.25 and 0.2);		
		\end{annotate}	
	}
	\hspace{-.045\textwidth}
	\subfigure[]{
		\label{fig.win5}
		\begin{annotate}{\includegraphics[width=.225\textwidth]{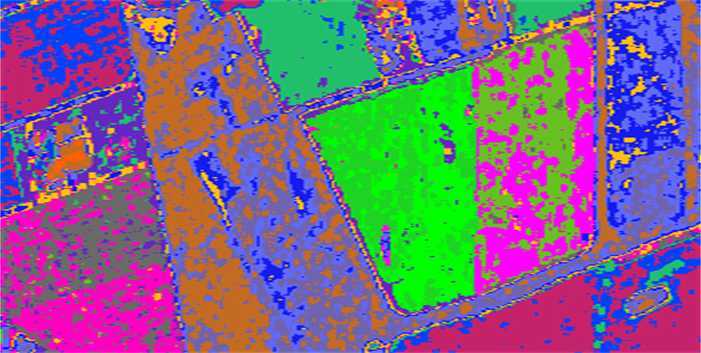}}{0.2}
			\draw[color=white,rotate around={25:(-2.5, 0)}] (-2.5, 0) ellipse (1.5 and 3.9);
			\draw[color=white,rotate around={25:(-5.8, -2.1)}] (-5.8, -2.1) ellipse (1.5 and 1.7);	
			\draw[color=white,rotate around={25:(-6.2, 2.3)}] (-6.2, 2.5) ellipse (1.35 and 1.1);
			\draw[color=white] (6.5, 1.35) ellipse (1.0 and 2.35);
			\draw[color=white] (5.45, -2.8) ellipse (0.5 and 0.9);
			\draw[color=white] (4, 0.5) ellipse (1.25 and 2.5);			
			\draw[color=white,rotate around={25:(3.3, -2.55)}] (3.3, -2.55) ellipse (2.25 and 0.2);		
		\end{annotate}	
	}
	\hspace{-.045\textwidth}
	\subfigure[]{
		\label{fig.win7}
		\begin{annotate}{\includegraphics[width=.225\textwidth]{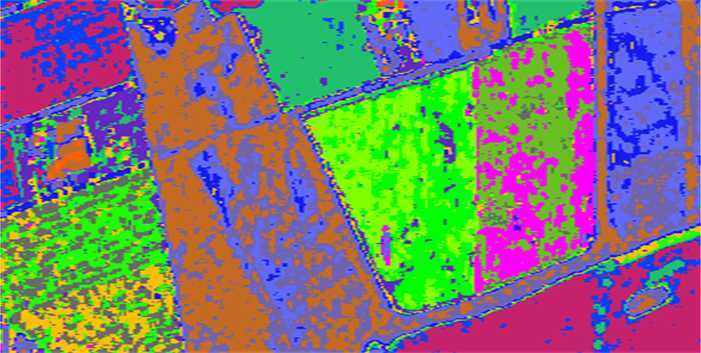}}{0.2}
			\draw[color=white,rotate around={25:(-2.5, 0)}] (-2.5, 0) ellipse (1.5 and 3.9);
			\draw[color=white,rotate around={25:(-5.8, -2.1)}] (-5.8, -2.1) ellipse (1.5 and 1.7);	
			\draw[color=white,rotate around={25:(-6.2, 2.3)}] (-6.2, 2.5) ellipse (1.35 and 1.1);
			\draw[color=white] (6.5, 1.35) ellipse (1.0 and 2.35);
			\draw[color=white] (5.45, -2.8) ellipse (0.5 and 0.9);
			\draw[color=white] (4, 0.5) ellipse (1.25 and 2.5);			
			\draw[color=white,rotate around={25:(3.3, -2.55)}] (3.3, -2.55) ellipse (2.25 and 0.2);		
		\end{annotate}	
	}
	\hspace{-.045\textwidth}
	\subfigure[]{
		\label{fig.win9}
		\begin{annotate}{\includegraphics[width=.225\textwidth]{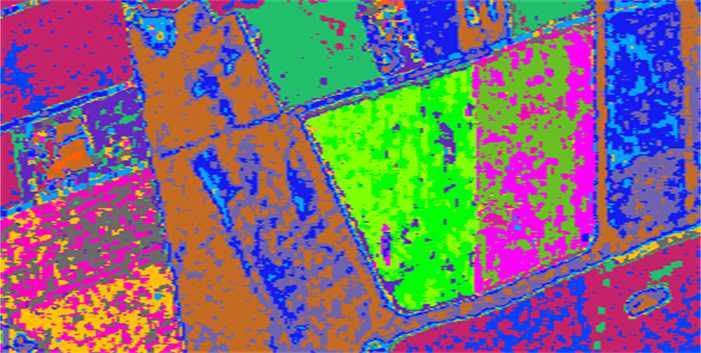}}{0.2}
			\draw[color=white,rotate around={25:(-2.5, 0)}] (-2.5, 0) ellipse (1.5 and 3.9);
			\draw[color=white,rotate around={25:(-5.8, -2.1)}] (-5.8, -2.1) ellipse (1.5 and 1.7);	
			\draw[color=white,rotate around={25:(-6.2, 2.3)}] (-6.2, 2.5) ellipse (1.35 and 1.1);
			\draw[color=white] (6.5, 1.35) ellipse (1.0 and 2.35);
			\draw[color=white] (5.45, -2.8) ellipse (0.5 and 0.9);
			\draw[color=white] (4, 0.5) ellipse (1.25 and 2.5);			
			\draw[color=white,rotate around={25:(3.3, -2.55)}] (3.3, -2.55) ellipse (2.25 and 0.2);		
		\end{annotate}	
	}
	\caption{Ablation analysis on window size $win$. (a) $win\!=\!3$, (b) $win\!=\!5$, (c) $win\!=\!7$, (d) $win\!=\!9$.}
	\label{fig.Foulum_Win}
\end{figure*}
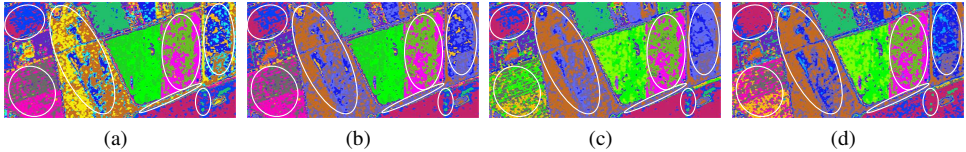	

To better visualize the impact of spatial information on the classification performance, $\beta^{(0)}$ is set to $5 \!\times\! 10^{1}$, which is smaller than the optimal value $5 \!\times\! 10^{2}$. As can be seen in Fig. \ref{fig.beta50} and Fig. \ref{fig.Foulum_Win}, the classification map becomes smoother with the increase of spatial window size $win$, especially in the white eclipses. This is because the model can capture more spatial information as the spatial window size $win$ grows from $3 \!\times\! 3$ to $9 \!\times\! 9$. For small window sizes (e.g. $3 \!\times\! 3$ in Fig. \ref{fig.win3} and $5 \!\times\! 5$ in Fig. \ref{fig.win5}), the forest F area in the middle white eclipses is classified into two clusters, which is obviously misclassification. With larger window sizes, this misclassification can be corrected by incorporating more spatial information from the neighboring pixels, and less miscellaneous pixels are observed (e.g. $win \!=\! 7 \!\times\! 7$ in Fig. \ref{fig.win7} and $win \!=\! 9 \!\times\! 9$ in Fig. \ref{fig.win9}). However, this does not mean large $win$s generally perform well on each dataset. For example, when the window size is $9 \!\times\! 9$, the result in the lower right corner shows that field A has been over-smoothed and misclassified into two clusters. 

As can be seen from Fig. \ref{fig.elbo_win}, the spatial information helps to accelerate the convergence process, but also makes it difficult for the lower bound to converge to a larger state. This is because although spatial information can help to produce smoother classification maps, it also makes the variational distribution deviates from the true posterior. Moreover, it is also observed that the lower bound achieves the largest convergence value when $win \!=\! 5 \!\times\! 5$. Therefore, $win$ should be set depending on the dataset size and expected richness of land covers. Considering the memory consumption and computational burden brought by large $win$s, the window size is usually taken as $3 \!\times\! 3$ or $5 \!\times\! 5$ in the following experiments. 

\begin{figure*}[htbp]
	\centering
	\subfigtopskip=2pt
	\subfigbottomskip=2pt
	\subfigcapskip=-5pt
	\subfigure[]{
		\label{fig.Foulum_Data}
		\begin{annotate}{\includegraphics[width=.225\textwidth]{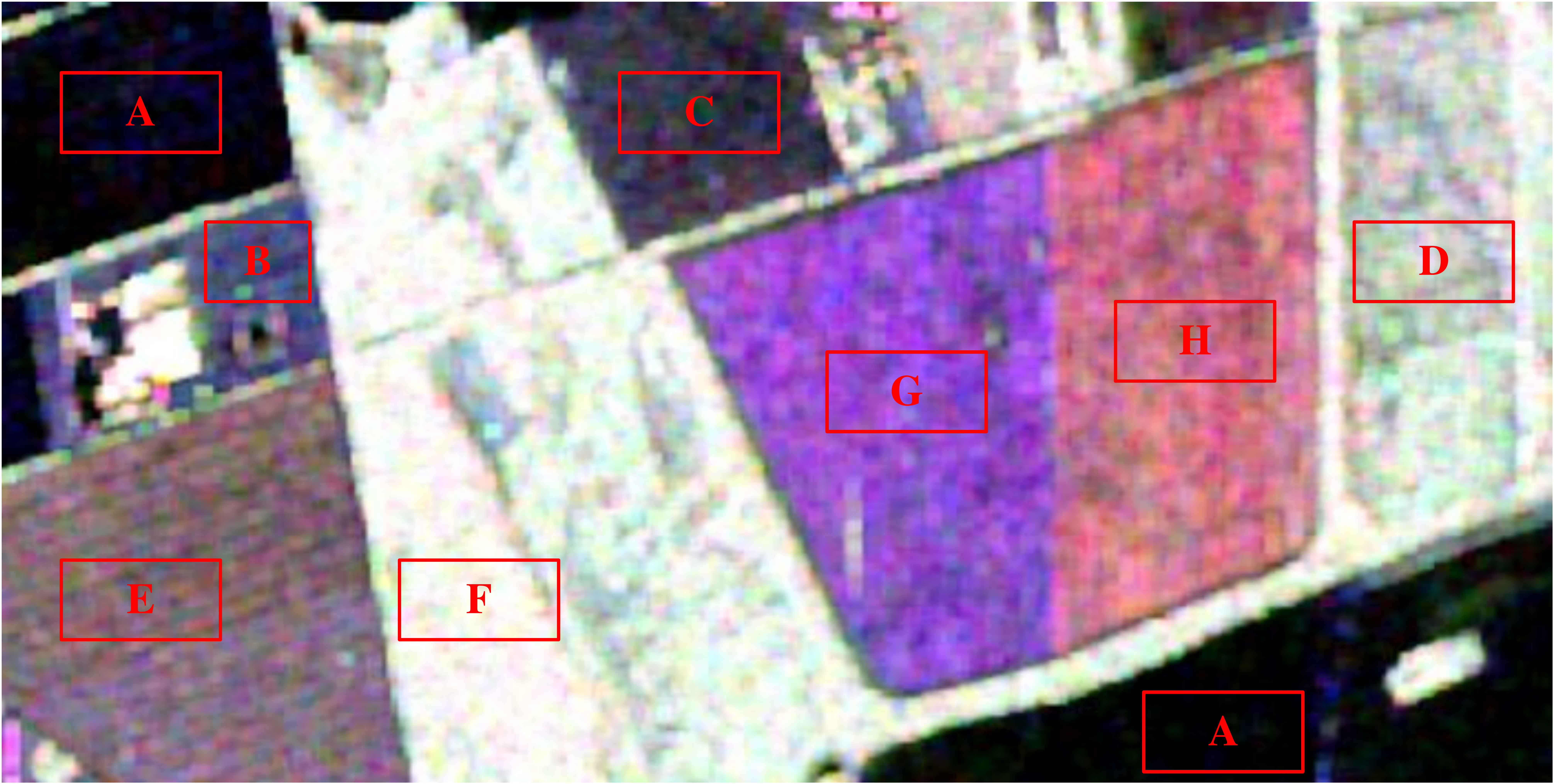}}{0.2}
		\end{annotate}	
	}
	\hspace{-.045\textwidth}
	\subfigure[]{
		\label{fig.Foulum_Ha}
		\begin{annotate}{\includegraphics[width=.225\textwidth]{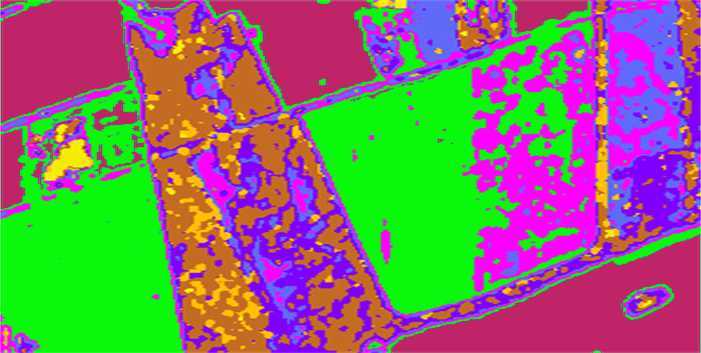}}{0.2}
			\draw[color=white,rotate around={25:(-2.5, 0)}] (-2.5, 0) ellipse (1.5 and 3.9);
			\draw[color=white,rotate around={25:(-5.8, -2.1)}] (-5.8, -2.1) ellipse (1.5 and 1.7);	
			\draw[color=white,rotate around={25:(-6.2, 2.3)}] (-6.2, 2.5) ellipse (1.35 and 1.1);
			\draw[color=white] (6.5, 1.35) ellipse (1.0 and 2.35);
			\draw[color=white] (5.45, -2.8) ellipse (0.5 and 0.9);
			\draw[color=white] (4, 0.5) ellipse (1.25 and 2.5);			
			\draw[color=white,rotate around={25:(3.3, -2.55)}] (3.3, -2.55) ellipse (2.25 and 0.2);		
		\end{annotate}	
	}
	\hspace{-.045\textwidth}
	\subfigure[]{
		\label{fig.Foulum_standard}
		\begin{annotate}{\includegraphics[width=.225\textwidth]{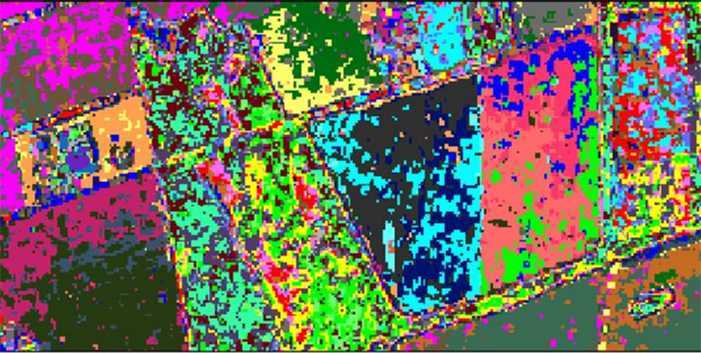}}{0.2}
			\draw[color=white,rotate around={25:(-2.5, 0)}] (-2.5, 0) ellipse (1.5 and 3.9);
			\draw[color=white,rotate around={25:(-5.8, -2.1)}] (-5.8, -2.1) ellipse (1.5 and 1.7);	
			\draw[color=white,rotate around={25:(-6.2, 2.3)}] (-6.2, 2.5) ellipse (1.35 and 1.1);
			\draw[color=white] (6.5, 1.35) ellipse (1.0 and 2.35);
			\draw[color=white] (5.45, -2.8) ellipse (0.5 and 0.9);
			\draw[color=white] (4, 0.5) ellipse (1.25 and 2.5);			
			\draw[color=white,rotate around={25:(3.3, -2.55)}] (3.3, -2.55) ellipse (2.25 and 0.2);		
		\end{annotate}	
	}
	\hspace{-.045\textwidth}
	\subfigure[]{
		\label{fig.Foulum_relaxed}
		\begin{annotate}{\includegraphics[width=.225\textwidth]{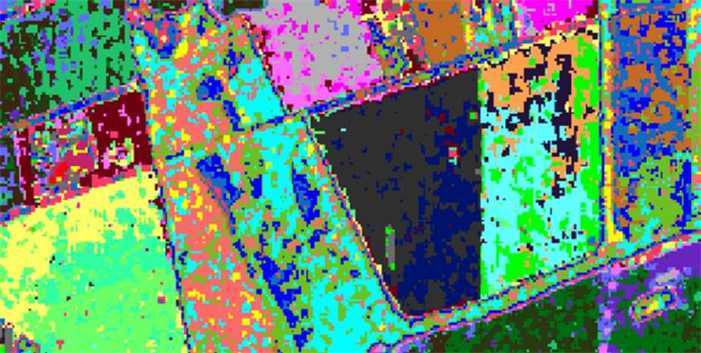}}{0.2}
			\draw[color=white,rotate around={25:(-2.5, 0)}] (-2.5, 0) ellipse (1.5 and 3.9);
			\draw[color=white,rotate around={25:(-5.8, -2.1)}] (-5.8, -2.1) ellipse (1.5 and 1.7);	
			\draw[color=white,rotate around={25:(-6.2, 2.3)}] (-6.2, 2.5) ellipse (1.35 and 1.1);
			\draw[color=white] (6.5, 1.35) ellipse (1.0 and 2.35);
			\draw[color=white] (5.45, -2.8) ellipse (0.5 and 0.9);
			\draw[color=white] (4, 0.5) ellipse (1.25 and 2.5);			
			\draw[color=white,rotate around={25:(3.3, -2.55)}] (3.3, -2.55) ellipse (2.25 and 0.2);		
		\end{annotate}	
	}
	\quad
	\subfigure[]{
		\label{fig.Foulum_WMM}
		\begin{annotate}{\includegraphics[width=.225\textwidth]{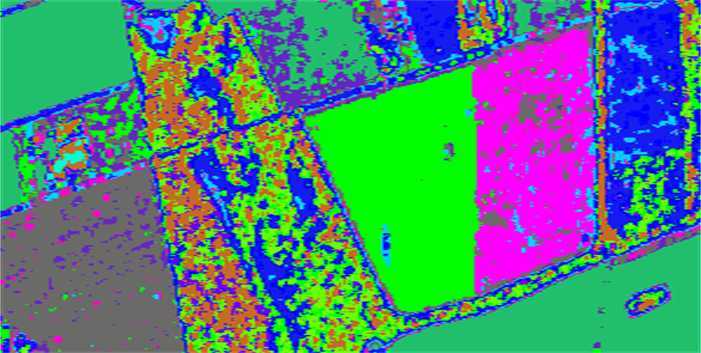}}{0.2}
			\draw[color=white,rotate around={25:(-2.5, 0)}] (-2.5, 0) ellipse (1.5 and 3.9);
			\draw[color=white,rotate around={25:(-5.8, -2.1)}] (-5.8, -2.1) ellipse (1.5 and 1.7);	
			\draw[color=white,rotate around={25:(-6.2, 2.3)}] (-6.2, 2.5) ellipse (1.35 and 1.1);
			\draw[color=white] (6.5, 1.35) ellipse (1.0 and 2.35);
			\draw[color=white] (5.45, -2.8) ellipse (0.5 and 0.9);
			\draw[color=white] (4, 0.5) ellipse (1.25 and 2.5);			
			\draw[color=white,rotate around={25:(3.3, -2.55)}] (3.3, -2.55) ellipse (2.25 and 0.2);		
		\end{annotate}	
	}
	\hspace{-.045\textwidth}
	\subfigure[]{
		\label{fig.Foulum_SVWMM}
		\begin{annotate}{\includegraphics[width=.225\textwidth]{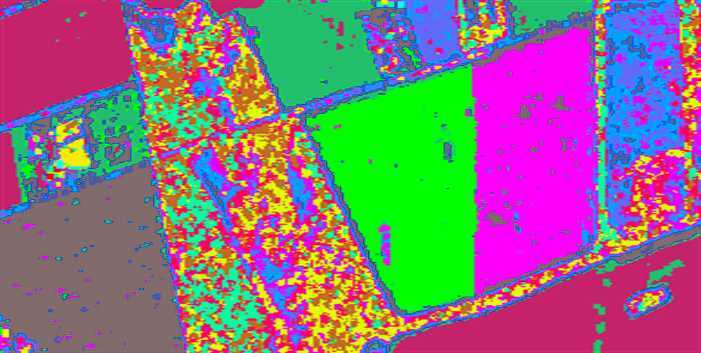}}{0.2}
			\draw[color=white,rotate around={25:(-2.5, 0)}] (-2.5, 0) ellipse (1.5 and 3.9);
			\draw[color=white,rotate around={25:(-5.8, -2.1)}] (-5.8, -2.1) ellipse (1.5 and 1.7);	
			\draw[color=white,rotate around={25:(-6.2, 2.3)}] (-6.2, 2.5) ellipse (1.35 and 1.1);
			\draw[color=white] (6.5, 1.35) ellipse (1.0 and 2.35);
			\draw[color=white] (5.45, -2.8) ellipse (0.5 and 0.9);
			\draw[color=white] (4, 0.5) ellipse (1.25 and 2.5);			
			\draw[color=white,rotate around={25:(3.3, -2.55)}] (3.3, -2.55) ellipse (2.25 and 0.2);		
		\end{annotate}	
	}
	\hspace{-.045\textwidth}
	\subfigure[]{
		\label{fig.Foulum_VBWMM}
		\begin{annotate}{\includegraphics[width=.225\textwidth]{Foulum_VBWMM.png}}{0.2}
			\draw[color=white,rotate around={25:(-2.5, 0)}] (-2.5, 0) ellipse (1.5 and 3.9);
			\draw[color=white,rotate around={25:(-5.8, -2.1)}] (-5.8, -2.1) ellipse (1.5 and 1.7);	
			\draw[color=white,rotate around={25:(-6.2, 2.3)}] (-6.2, 2.5) ellipse (1.35 and 1.1);
			\draw[color=white] (6.5, 1.35) ellipse (1.0 and 2.35);
			\draw[color=white] (5.45, -2.8) ellipse (0.5 and 0.9);
			\draw[color=white] (4, 0.5) ellipse (1.25 and 2.5);			
			\draw[color=white,rotate around={25:(3.3, -2.55)}] (3.3, -2.55) ellipse (2.25 and 0.2);		
		\end{annotate}	
	}
	\hspace{-.045\textwidth}
	\subfigure[]{
		\label{fig.Foulum_SVVBWMM}
		\begin{annotate}{\includegraphics[width=.225\textwidth]{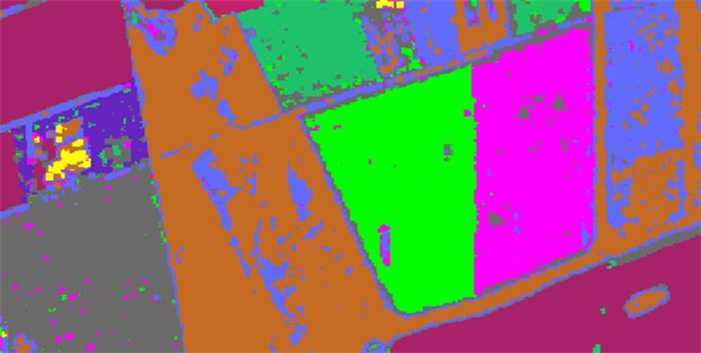}}{0.2}
			\draw[color=white,rotate around={25:(-2.5, 0)}] (-2.5, 0) ellipse (1.5 and 3.9);
			\draw[color=white,rotate around={25:(-5.8, -2.1)}] (-5.8, -2.1) ellipse (1.5 and 1.7);	
			\draw[color=white,rotate around={25:(-6.2, 2.3)}] (-6.2, 2.5) ellipse (1.35 and 1.1);
			\draw[color=white] (6.5, 1.35) ellipse (1.0 and 2.35);
			\draw[color=white] (5.45, -2.8) ellipse (0.5 and 0.9);
			\draw[color=white] (4, 0.5) ellipse (1.25 and 2.5);			
			\draw[color=white,rotate around={25:(3.3, -2.55)}] (3.3, -2.55) ellipse (2.25 and 0.2);		
		\end{annotate}	
	}
	\caption{(a) Pauli RGB image and classification results of Foulum dataset using (b) $H/\alpha$-Wishart \cite{HalphaWis}, (c) standard-Wishart \cite{APD11}, (d) relaxed-Wishart \cite{APD11}, (e) WMM \cite{SVWMM}, (f) SVWMM \cite{SVWMM}, (g) proposed VBWMM, (h) proposed SVVBWMM.}
	\label{fig.Exp_Foulum}
\end{figure*}

\begin{table}[ht]
	\centering
	\caption{ENL Estimation on Foulum Dataset}
	\begin{tabular}{p{4.055em}ccccc}
		\toprule
		\toprule
		\multicolumn{1}{c}{\textit{Class}} & \multicolumn{1}{c}{\textit{ML}} & \multicolumn{1}{c}{\textit{TM}} & \multicolumn{1}{c}{\textit{WMLC}} & \multicolumn{1}{c}{\textit{VB}} & \multicolumn{1}{c}{\textit{\textbf{SVVB}}} \\			
		\midrule
		\rowcolor[rgb]{ .769,  .137,  .42} Field A & \cellcolor[rgb]{ 1,  1,  1}22.04  & \cellcolor[rgb]{ 1,  1,  1}18.27  & \cellcolor[rgb]{ 1,  1,  1}18.41  & \cellcolor[rgb]{ 1,  1,  1}18.88  & \cellcolor[rgb]{ 1,  1,  1}18.77  \\
		\rowcolor[rgb]{ .396,  .153,  .757} Field B & \cellcolor[rgb]{ 1,  1,  1}21.62  & \cellcolor[rgb]{ 1,  1,  1}21.05  & \cellcolor[rgb]{ 1,  1,  1}16.24  & \cellcolor[rgb]{ 1,  1,  1}19.11  & \cellcolor[rgb]{ 1,  1,  1}19.17  \\
		\rowcolor[rgb]{ .129,  .749,  .424} Field C & \cellcolor[rgb]{ 1,  1,  1}24.15  & \cellcolor[rgb]{ 1,  1,  1}21.22  & \cellcolor[rgb]{ 1,  1,  1}19.22  & \cellcolor[rgb]{ 1,  1,  1}19.57  & \cellcolor[rgb]{ 1,  1,  1}18.86  \\
		\rowcolor[rgb]{ .38,  .412,  1} Forest D & \cellcolor[rgb]{ 1,  1,  1}18.11  & \cellcolor[rgb]{ 1,  1,  1}16.70  & \cellcolor[rgb]{ 1,  1,  1}16.33  & \cellcolor[rgb]{ 1,  1,  1}18.39  & \cellcolor[rgb]{ 1,  1,  1}18.40  \\
		\rowcolor[rgb]{ .416,  .416,  .416} Field E & \cellcolor[rgb]{ 1,  1,  1}23.18  & \cellcolor[rgb]{ 1,  1,  1}18.93  & \cellcolor[rgb]{ 1,  1,  1}19.03  & \cellcolor[rgb]{ 1,  1,  1}19.07  & \cellcolor[rgb]{ 1,  1,  1}19.03  \\
		\rowcolor[rgb]{ .769,  .42,  .137} Forest F & \cellcolor[rgb]{ 1,  1,  1}20.41  & \cellcolor[rgb]{ 1,  1,  1}17.81  & \cellcolor[rgb]{ 1,  1,  1}17.45  & \cellcolor[rgb]{ 1,  1,  1}18.40  & \cellcolor[rgb]{ 1,  1,  1}18.50  \\
		\rowcolor[rgb]{ 0,  1,  0} Field G & \cellcolor[rgb]{ 1,  1,  1}23.76  & \cellcolor[rgb]{ 1,  1,  1}22.13  & \cellcolor[rgb]{ 1,  1,  1}15.41  & \cellcolor[rgb]{ 1,  1,  1}19.39  & \cellcolor[rgb]{ 1,  1,  1}18.87  \\
		\rowcolor[rgb]{ 1,  0,  1} Field H & \cellcolor[rgb]{ 1,  1,  1}19.91  & \cellcolor[rgb]{ 1,  1,  1}16.39  & \cellcolor[rgb]{ 1,  1,  1}15.26  & \cellcolor[rgb]{ 1,  1,  1}18.31  & \cellcolor[rgb]{ 1,  1,  1}17.99  \\
		\bottomrule
		\bottomrule
	\end{tabular}%
	\label{tab.ENL_FoC_Slice}%
\end{table}%
 
Finally, the experimental results of proposed and comparative methods have been illustrated in Fig. \ref{fig.Exp_Foulum}. Fig. \ref{fig.Foulum_Ha} shows that the baseline $H/\alpha$-Wishart method fails to tell the differences between two pairs of land covers, namely field A and field C, together with field B and forest E. The `fine' results of standard-Wishart in Fig. \ref{fig.Foulum_standard} and relaxed-Wishart in Fig. \ref{fig.Foulum_relaxed} are divided into 44 and 30 clusters respectively, where both sub-sampling rates are set to $1/4$ \cite{APD11}. For these methods, the automatic determination of cluster number comes at the price of sub-sampling, which reduces the number of components and speeds up the training process, but fails to utilize the spatial information in neighboring pixels. The best results of WMM and SVWMM \cite{SVWMM} are given in Fig. \ref{fig.Foulum_WMM} and Fig. \ref{fig.Foulum_SVWMM}. To preserve more edge and texture details, the spatial window size $win$ of SVWMM and proposed SVVBWMM is chosen as $5 \!\times\! 5$. The hyperparameters of proposed VBWMM, namely $\alpha ^{(0)}$,  $\beta ^{(0)}$, $b^{(0)}$ and $c^{(0)}$, are set to $1\!\times\!10^{-1}$, $5 \!\times\! 10^{2}$, 5 and 5, respectively. 

Compared with the results from Fig. \ref{fig.Foulum_Ha} to Fig. \ref{fig.Foulum_SVWMM}, there are less miscellaneous pixels in the white eclipses of Fig. \ref{fig.Foulum_VBWMM} and \ref{fig.Foulum_SVVBWMM}, revealing that the proposed VBWMM and SVVBWMM can provide better clustering results than the above four approaches. After integrating spatial information from the nearby pixels, the speckles in forest F have been eliminated. Besides, the proposed SVVBWMM also exhibits better interpretation in the bottom right field A, and there is no remarkable loss in texture details. 

%
%
Fig. \ref{fig.ENL_Distributions} illustrates the ENL distribution given by ML estimator. Apart from field B, it is observed that the peak values have shifted to the left of mean values, because the estimators tend to produce low ENL estimates in heterogeneous windows. This also implies the necessity to find a new distribution to model the ENL of each cluster. The results in Table \ref{tab.ENL_FoC_Slice} show that the ENL estimation expectation of VBWMM is much closer to the reference value 18 than those of ML, TM and WMLC. Also, the SVVBWMM can give a more reasonable estimation of ENL, which indicates that spatial information can help to correct possible bias of ENL to some extent.
\subsection{Experimental Results on San Francisco Dataset}
\label{subsec:ExpSanP}
\begin{figure*}[htbp]
	\centering
	\subfigtopskip=2pt
	\subfigbottomskip=2pt
	\subfigcapskip=-5pt
	\subfigure[]{
		\label{fig.SanF_Data}
		\includegraphics[width=.225\textwidth]{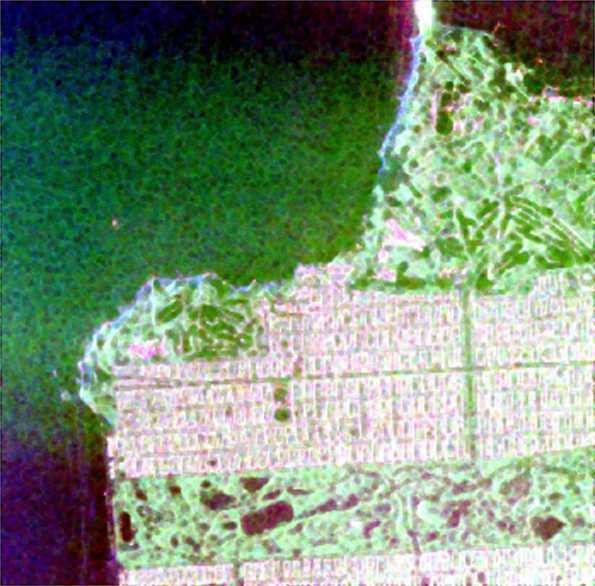}
	}
	\subfigure[]{
		\label{fig.SanF_Ha}
		\includegraphics[width=.225\textwidth]{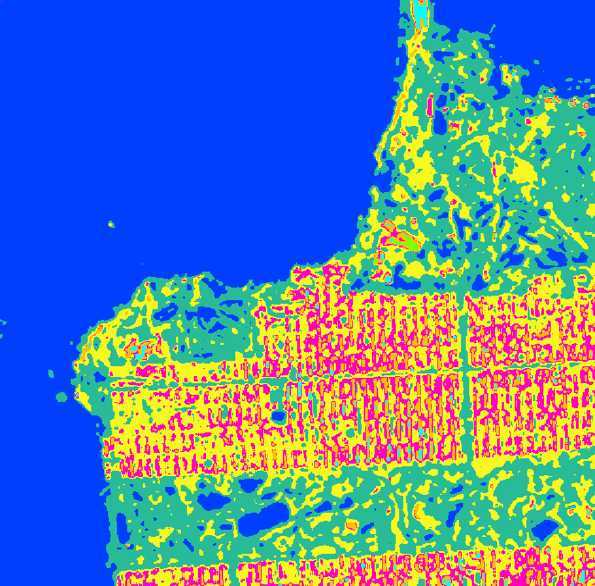}
	}
	\subfigure[]{
		\label{fig.SanF_standard}
		\includegraphics[width=.225\textwidth]{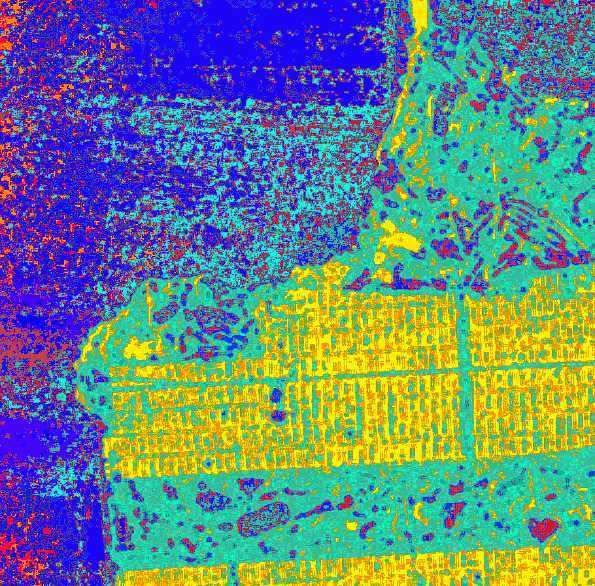}
	}
	\subfigure[]{
		\label{fig.SanF_relaxed}
		\includegraphics[width=.225\textwidth]{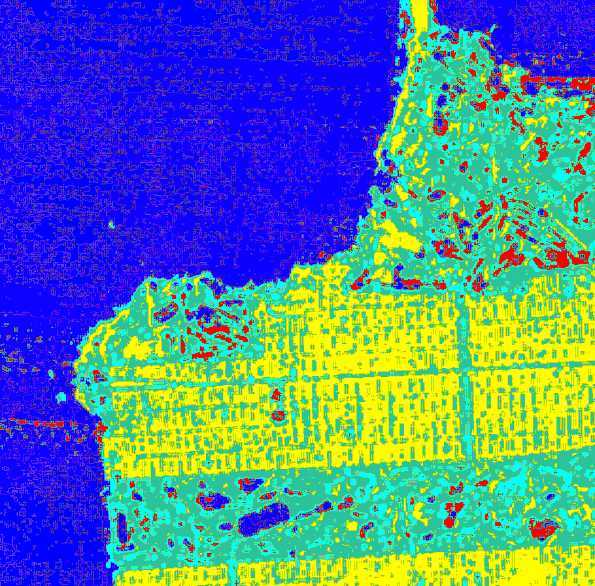}
	}
	\quad
	\subfigure[]{
		\label{fig.SanF_WMM}
		\includegraphics[width=.225\textwidth]{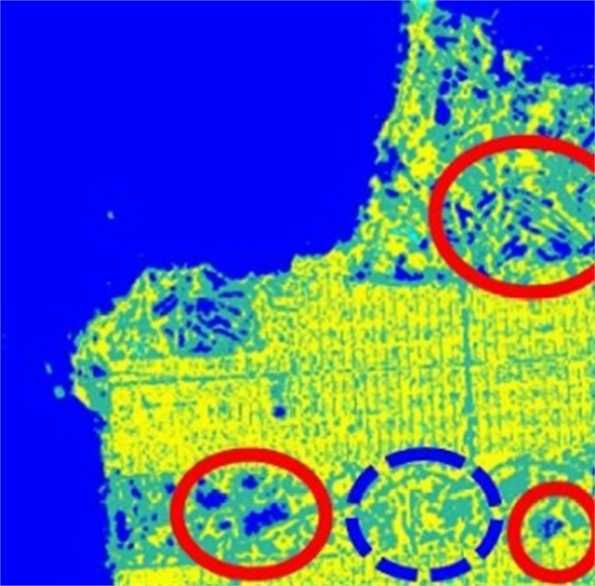}
	}
	\subfigure[]{
		\label{fig.SanF_SVWMM}
		\includegraphics[width=.225\textwidth]{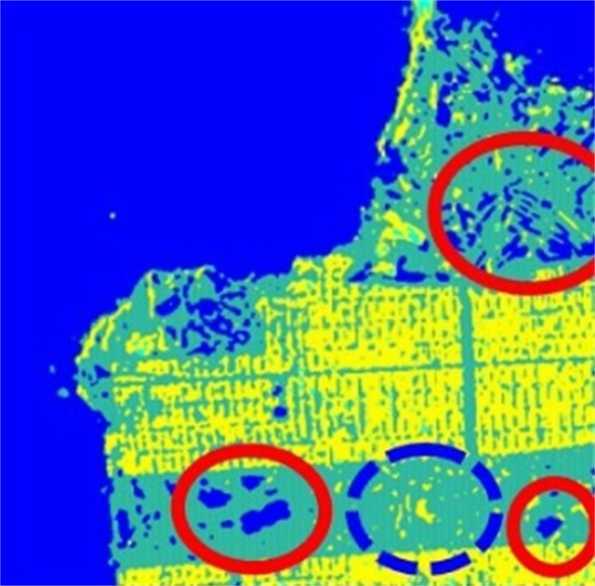}
	}
	\subfigure[]{
		\label{fig.SanF_VBWMM}
		\includegraphics[width=.225\textwidth]{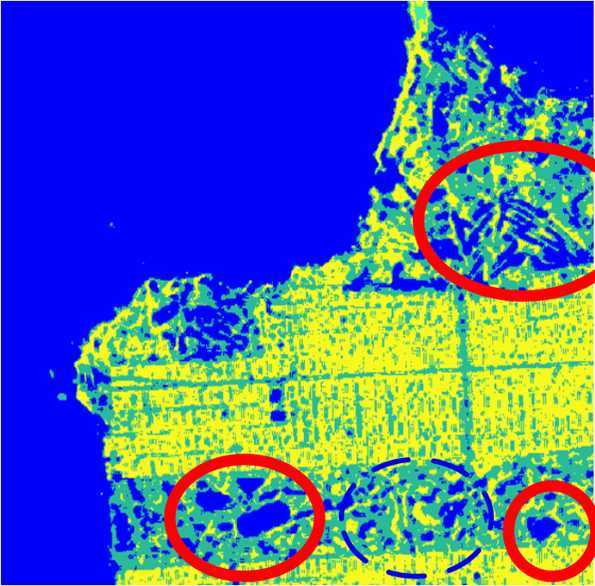}
	}
	\subfigure[]{
		\label{fig.SanF_SVVBWMM}
		\includegraphics[width=.225\textwidth]{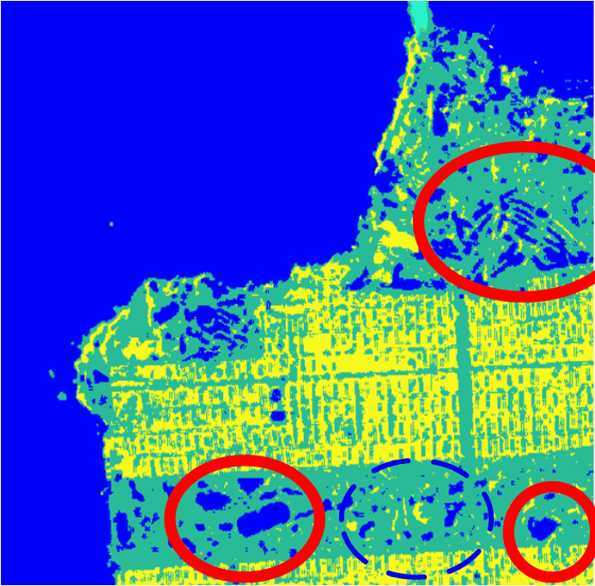}
	}
	\caption{(a) Pauli RGB image and classification results of San Francisco dataset using (b) $H/\alpha$-Wishart \cite{HalphaWis}, (c) standard-Wishart \cite{APD11}, (d) relaxed-Wishart \cite{APD11}, (e) WMM \cite{SVWMM}, (f) SVWMM \cite{SVWMM}, (g) proposed VBWMM, (h) proposed SVVBWMM.}
	\label{fig.Exp_SanF}
\end{figure*}

To validate the performance of proposed method in urban areas, the following experiments are conducted with PolSAR data provided by AIRSAR over San Francisco in the P-band, as shown in Fig. \ref{fig.Exp_SanF}. The imaging scene is $586 \time 595$ in size and contains a variety of land covers, such as urban, ocean, forests, mountains, polo fields and the Golden Gate Bridge. 

In Fig. \ref{fig.Exp_SanF}, it is observed that the $H/\alpha$-Wishart method has classified the land covers into 5 classes, namely water, vegetation, bridge and two kinds of urban areas. The clustering result of standard-Wishart in Fig. \ref{fig.SanF_standard} performs worse than that of relaxed-Wishart in Fig. \ref{fig.SanF_relaxed}, especially in the sea area that has been misclassified into many clusters, which indicates the necessity of different ENL values for each cluster.

Compared to WMM \cite{SVWMM} in Fig. \ref{fig.SanF_WMM}, which confuses mountains with forests and urban in the top red eclipses, the proposed VBWMM can retain more textures details. From Fig. \ref{fig.SanF_SVWMM}, it can be seen that SVWMM fails to tell urban from the dark areas in the eclipses below, while the proposed SVVBWMM can capture their difference more accurately.
From another point of view, it is true that spatial information helps to improving the interpretability of clustering results, but also suffers the risk of over-smoothing and deviating from the scattering properties. Therefore, whether to introduce spatial information is also a trade-off between image interpretability and texture details. Table \ref{tab.ENL_SanP} illustrates that the ENL estimations of proposed VB approaches are comparable to those of ML \cite{MLwin}, TM \cite{EstENL} and WMLC \cite{GoF_Mellin}.

\begin{table}[ht]
	\centering
	\caption{ENL Estimation on San Francisco Dataset}
	\begin{tabular}{p{4.055em}ccccc}
		\toprule
		\toprule
		\multicolumn{1}{c}{\textit{Class}} & \multicolumn{1}{c}{\textit{ML}} & \multicolumn{1}{c}{\textit{TM}} & \multicolumn{1}{c}{\textit{WMLC}} & \multicolumn{1}{c}{\textit{VB}} & \multicolumn{1}{c}{\textit{\textbf{SVVB}}} \\			
		\midrule		
		\rowcolor[rgb]{ 0,  0,  1} Water & \cellcolor[rgb]{ 1,  1,  1}35.95  & \cellcolor[rgb]{ 1,  1,  1}40.42  & \cellcolor[rgb]{ 1,  1,  1}35.79  & \cellcolor[rgb]{ 1,  1,  1}31.51  & \cellcolor[rgb]{ 1,  1,  1}39.52  \\
		\rowcolor[rgb]{ .192,  .729,  .651} Vegetation & \cellcolor[rgb]{ 1,  1,  1}48.11  & \cellcolor[rgb]{ 1,  1,  1}46.34  & \cellcolor[rgb]{ 1,  1,  1}46.94  & \cellcolor[rgb]{ 1,  1,  1}40.09  & \cellcolor[rgb]{ 1,  1,  1}43.71  \\
		\rowcolor[rgb]{ 1,  1,  0} Urban & \cellcolor[rgb]{ 1,  1,  1}43.96  & \cellcolor[rgb]{ 1,  1,  1}43.82  & \cellcolor[rgb]{ 1,  1,  1}40.55  & \cellcolor[rgb]{ 1,  1,  1}40.98  & \cellcolor[rgb]{ 1,  1,  1}41.66  \\
		\rowcolor[rgb]{ 0,  1,  1} Bridge & \cellcolor[rgb]{ 1,  1,  1}41.48  & \cellcolor[rgb]{ 1,  1,  1}48.33  & \cellcolor[rgb]{ 1,  1,  1}39.87  & \cellcolor[rgb]{ 1,  1,  1}40.76  & \cellcolor[rgb]{ 1,  1,  1}41.15  \\
		\bottomrule
		\bottomrule
	\end{tabular}%
	\label{tab.ENL_SanP}%
\end{table}%

\subsection{Experimental Results on Flevoland Dataset}
\label{subsec:ExpAIRFlev}
\begin{figure*}[htbp]
	\centering
	\subfigtopskip=2pt
	\subfigbottomskip=2pt
	\subfigcapskip=-5pt
	\subfigure[]{
		\label{fig.Flevoland_Data}
		\begin{annotate}{\includegraphics[width=.175\textwidth]{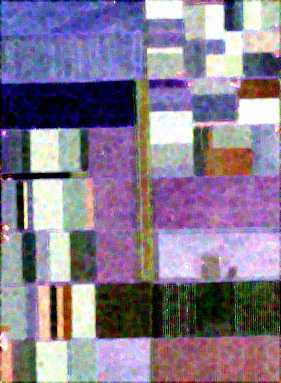}}{0.2}
		\end{annotate}	
	}
	\hspace{-.045\textwidth}
	\subfigure[]{
		\label{fig.Flevoland_GND}
		\begin{annotate}{\includegraphics[width=.175\textwidth]{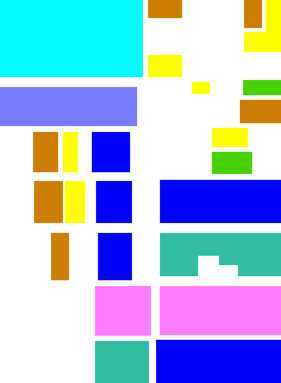}}{0.2}
		\end{annotate}	
	}
	\hspace{-.045\textwidth}
	\subfigure[]{
		\label{fig.Flevoland_Legend}
		\begin{annotate}{\includegraphics[width=.175\textwidth]{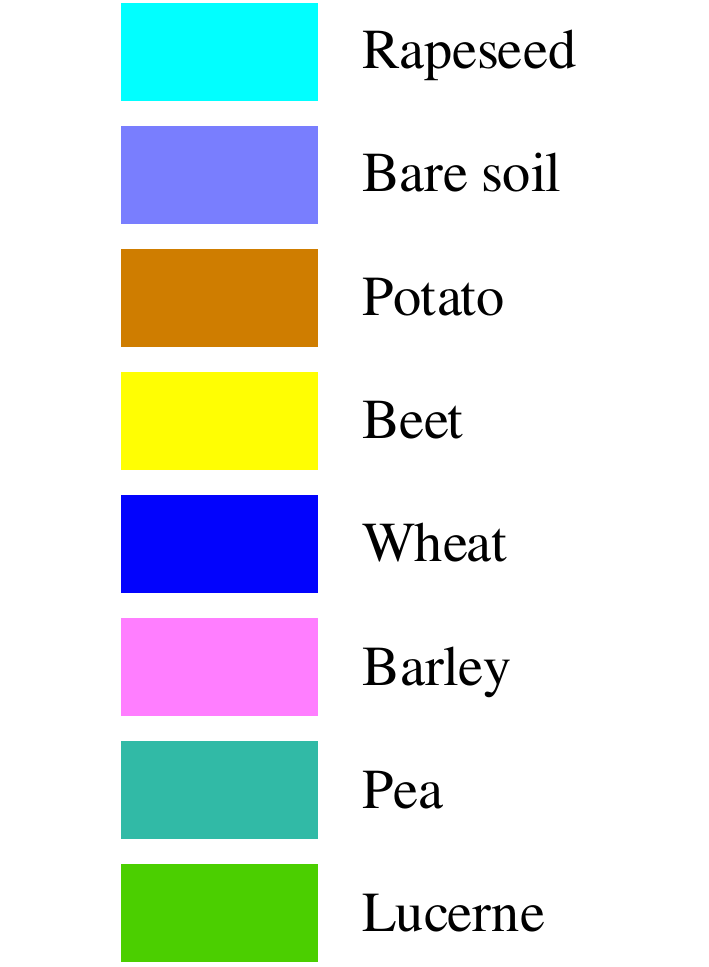}}{0.2}
		\end{annotate}	
	}
	\hspace{-.045\textwidth}
	\subfigure[]{
		\label{fig.Flevoland_Ha}
		\begin{annotate}{\includegraphics[width=.175\textwidth]{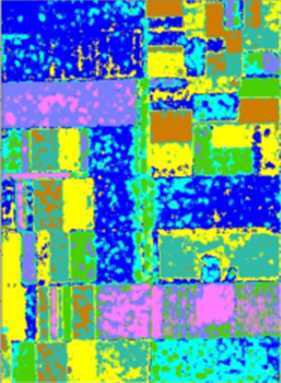}}{0.2}
			\draw[color=white] (-1, -0.5) ellipse (1.0 and 3.15);		
			\draw[color=white] (-3, 6.3) ellipse (2.85 and 1.65);
			\draw[color=white] (-3, 3.65) ellipse (2.85 and 0.85);
			\draw[color=white,rotate around={5:(-3.8, -0.55)}] (-3.8, -0.55) ellipse (0.65 and 3.15);	
			\draw[color=white,rotate around={-25:(-3.1, -6.05)}] (-3.1, -6.05) ellipse (0.65 and 2.15);	
			\draw[color=white] (3.25, -7.05) ellipse (2.6 and 0.85);	
			\draw[color=black] (2.05, -5.05) ellipse (3.75 and 1.05);	
			\draw[color=white] (3.25, -2.70) ellipse (2.6 and 1.05);
			\draw[color=white] (-0.75, -7.05) ellipse (1.1 and 0.925);					
		\end{annotate}	
	}
	\hspace{-.045\textwidth}
	\subfigure[]{
		\label{fig.Flevoland_standard}
		\begin{annotate}{\includegraphics[width=.175\textwidth]{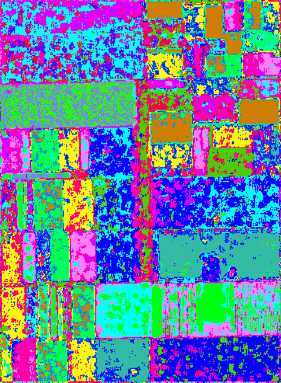}}{0.2}
			\draw[color=white] (-1, -0.5) ellipse (1.0 and 3.15);		
			\draw[color=white] (-3, 6.3) ellipse (2.85 and 1.65);
			\draw[color=white] (-3, 3.65) ellipse (2.85 and 0.85);
			\draw[color=white,rotate around={5:(-3.8, -0.55)}] (-3.8, -0.55) ellipse (0.65 and 3.15);	
			\draw[color=white,rotate around={-25:(-3.1, -6.05)}] (-3.1, -6.05) ellipse (0.65 and 2.15);	
			\draw[color=white] (3.25, -7.05) ellipse (2.6 and 0.85);	
			\draw[color=black] (2.05, -5.05) ellipse (3.75 and 1.05);	
			\draw[color=white] (3.25, -2.70) ellipse (2.6 and 1.05);
			\draw[color=white] (-0.9, -7.05) ellipse (1.0 and 0.925);		
		\end{annotate}	
	}
	\quad
	\subfigure[]{
		\label{fig.Flevoland_relaxed}
		\begin{annotate}{\includegraphics[width=.175\textwidth]{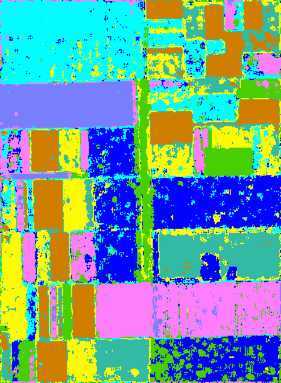}}{0.2}
			\draw[color=white] (-1, -0.5) ellipse (1.0 and 3.15);		
			\draw[color=white] (-3, 6.3) ellipse (2.85 and 1.65);
			\draw[color=white] (-3, 3.65) ellipse (2.85 and 0.85);
			\draw[color=white,rotate around={5:(-3.8, -0.55)}] (-3.8, -0.55) ellipse (0.65 and 3.15);	
			\draw[color=white,rotate around={-25:(-3.1, -6.05)}] (-3.1, -6.05) ellipse (0.65 and 2.15);	
			\draw[color=white] (3.25, -7.05) ellipse (2.6 and 0.85);	
			\draw[color=black] (2.05, -5.05) ellipse (3.75 and 1.05);	
			\draw[color=white] (3.25, -2.70) ellipse (2.6 and 1.05);
			\draw[color=white] (-0.9, -7.05) ellipse (1.0 and 0.925);			
		\end{annotate}	
	}
	\hspace{-.045\textwidth}
	\subfigure[]{
		\label{fig.Flevoland_WMM}
		\begin{annotate}{\includegraphics[width=.175\textwidth]{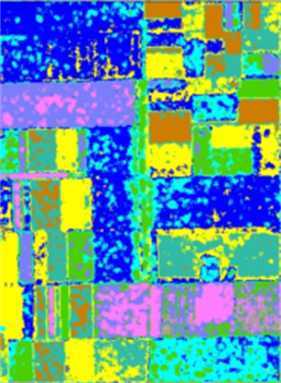}}{0.2}
			\draw[color=white] (-1, -0.5) ellipse (1.0 and 3.15);		
			\draw[color=white] (-3, 6.3) ellipse (2.85 and 1.65);
			\draw[color=white] (-3, 3.65) ellipse (2.85 and 0.85);
			\draw[color=white,rotate around={5:(-3.8, -0.55)}] (-3.8, -0.55) ellipse (0.65 and 3.15);	
			\draw[color=white,rotate around={-25:(-3.1, -6.05)}] (-3.1, -6.05) ellipse (0.65 and 2.15);	
			\draw[color=white] (3.25, -7.05) ellipse (2.6 and 0.85);	
			\draw[color=black] (2.05, -5.05) ellipse (3.75 and 1.05);	
			\draw[color=white] (3.25, -2.70) ellipse (2.6 and 1.05);
			\draw[color=white] (-0.9, -7.05) ellipse (1.0 and 0.925);		
		\end{annotate}	
	}
	\hspace{-.045\textwidth}
	\subfigure[]{
		\label{fig.Flevoland_SVWMM}
		\begin{annotate}{\includegraphics[width=.175\textwidth]{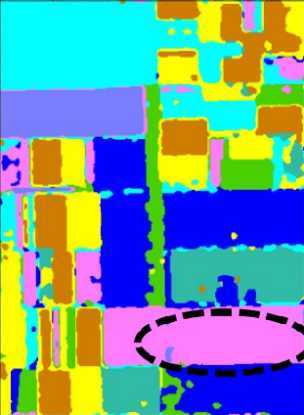}}{0.2}
			\draw[color=white] (-1, -0.5) ellipse (1.0 and 3.15);		
			\draw[color=white] (-3, 6.3) ellipse (2.85 and 1.65);
			\draw[color=white] (-3, 3.65) ellipse (2.85 and 0.85);
			\draw[color=white,rotate around={5:(-3.8, -0.55)}] (-3.8, -0.55) ellipse (0.65 and 3.15);	
			\draw[color=white,rotate around={-25:(-3.1, -6.05)}] (-3.1, -6.05) ellipse (0.65 and 2.15);	
			\draw[color=white] (3.25, -7.05) ellipse (2.6 and 0.85);	
			\draw[color=white] (3.25, -2.70) ellipse (2.6 and 1.05);
			\draw[color=white] (-0.9, -7.05) ellipse (1.0 and 0.925);				
		\end{annotate}	
	}
	\hspace{-.045\textwidth}
	\subfigure[]{
		\label{fig.Flevoland_VBWMM}
		\begin{annotate}{\includegraphics[width=.175\textwidth]{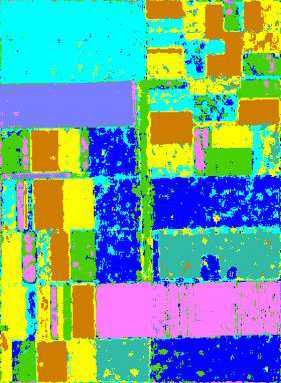}}{0.2}
			\draw[color=white] (-1, -0.5) ellipse (1.0 and 3.15);		
			\draw[color=white] (-3, 6.3) ellipse (2.85 and 1.65);
			\draw[color=white] (-3, 3.65) ellipse (2.85 and 0.85);
			\draw[color=white,rotate around={5:(-3.8, -0.55)}] (-3.8, -0.55) ellipse (0.65 and 3.15);	
			\draw[color=white,rotate around={-25:(-3.1, -6.05)}] (-3.1, -6.05) ellipse (0.65 and 2.15);	
			\draw[color=white] (3.25, -7.05) ellipse (2.6 and 0.85);	
			\draw[color=black] (2.05, -5.05) ellipse (3.75 and 1.05);	
			\draw[color=white] (3.25, -2.70) ellipse (2.6 and 1.05);
			\draw[color=white] (-0.9, -7.05) ellipse (1.0 and 0.925);			
		\end{annotate}	
	}
	\hspace{-.045\textwidth}
	\subfigure[]{
		\label{fig.Flevoland_SVVBWMM}
		\begin{annotate}{\includegraphics[width=.175\textwidth]{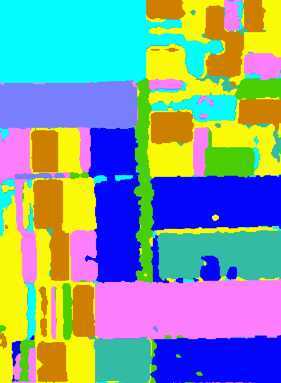}}{0.2}
			\draw[color=white] (-1, -0.5) ellipse (1.0 and 3.15);		
			\draw[color=white] (-3, 6.3) ellipse (2.85 and 1.65);
			\draw[color=white] (-3, 3.65) ellipse (2.85 and 0.85);
			\draw[color=white,rotate around={5:(-3.8, -0.55)}] (-3.8, -0.55) ellipse (0.65 and 3.15);	
			\draw[color=white,rotate around={-25:(-3.1, -6.05)}] (-3.1, -6.05) ellipse (0.65 and 2.15);	
			\draw[color=white] (3.25, -7.05) ellipse (2.6 and 0.85);	
			\draw[color=black] (2.05, -5.05) ellipse (3.75 and 1.05);	
			\draw[color=white] (3.25, -2.70) ellipse (2.6 and 1.05);
			\draw[color=white] (-0.9, -7.05) ellipse (1.0 and 0.925);		
		\end{annotate}	
	}
	\caption{(a) Pauli RGB image, (b) ground truth, (c) legend, and classification results of Flevoland dataset using (d) $H/\alpha$-Wishart \cite{HalphaWis}, (e) standard-Wishart \cite{APD11}, (f) relaxed-Wishart \cite{APD11}, (g) WMM \cite{SVWMM}, (h) SVWMM \cite{SVWMM}, (i) proposed VBWMM, (j) proposed SVVBWMM.}
	\label{fig.Exp_Flevoland}
\end{figure*}
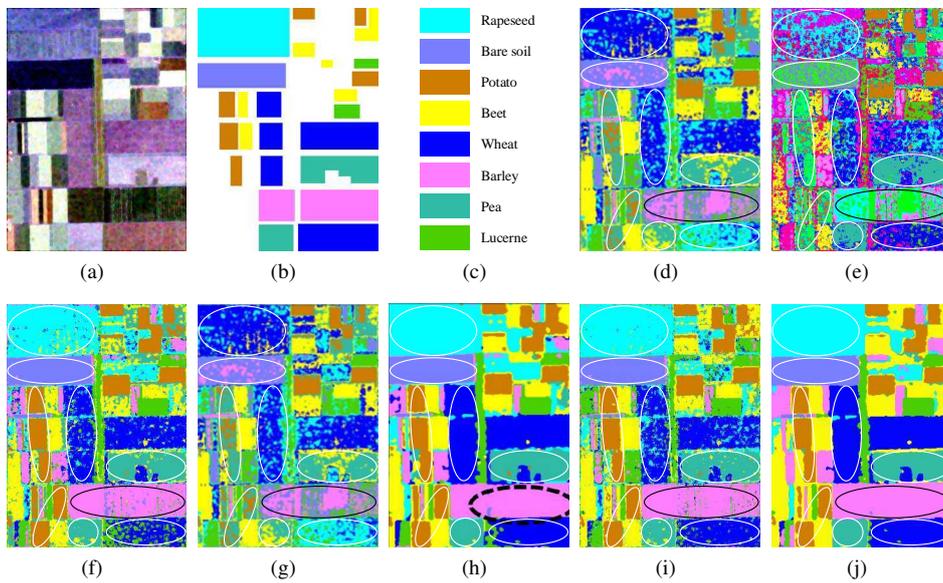
The experimental tests of this section aim at quantitatively evaluating the performance of proposed method on agricultural areas. This L-band AIRSAR dataset is acquired by NASA/JPL over Flevoland in 2010, which is preprocessed with refined Lee filter to suppress speckle noise. The selected area is $383 \!\times\! 281$ in size, which consists of eight classes, with its Pauli RGB image, ground truth and legend shown in Fig. \ref{fig.Flevoland_Data}, Fig. \ref{fig.Flevoland_GND} and Fig. \ref{fig.Flevoland_Legend}, respectively. In the following experiments, overall accuracy (OA) and kappa coefficients $\kappa$ are chosen as the criteria, which have been summarized in Table \ref{tab.Flevoland_Acc} for quantitative evaluation. Moreover, the ENL estimations are demonstrated in Table \ref{tab.ENL_Flevoland}.

In Fig. \ref{fig.Flevoland_Ha}, the result shows that $H/\alpha$-Wishart fails to distinguish between rapeseed and wheat, nor can it tell potato from pea. From Fig. \ref{fig.Flevoland_standard} and Fig. \ref{fig.Flevoland_relaxed}, it is observed that each of the 8 classes has been found out by relaxed-Wishart, while the clustering result of standard-Wishart is far from satisfactory. 
In \ref{fig.Flevoland_WMM}, it is observed that WMM's incapability to tell rapeseed from wheat in the red eclipses has deteriorated the OA. What's more, it is shown that WMM also confuses potato with pea, while the proposed VBWMM performs better in such areas, as highlighted in Fig. \ref{fig.Flevoland_VBWMM}. Also, the clustering result in black eclipses is smoother than that of WMM, which clearly demonstrates the advantage of proposed VBWMM in describing the scattering characteristics of land covers. As a result, the accuracy of proposed VBWMM method is 91.85\%, which increases by 19.05\% in comparison with WMM. The spatial window size of SVWMM and proposed SVVBWMM is set to $5 \!\times\! 5$. By comparing their clustering results in Fig. \ref{fig.Flevoland_SVWMM} and \ref{fig.Flevoland_SVVBWMM}, it is known that the proposed SVVBWMM performs better than SVWMM, with the accuracy improved from 96.5\% to 97.49\%, which is 0.99\% higher than that of SVWMM and verifies the effectiveness of spatial similarities. 

According to Table \ref{tab.Flevoland_Acc}, the accuracy values, OA and $\kappa$ of proposed VBWMM and SVVBWMM are generally larger than the other methods, implying that the proposed method performs better on most clusters. The ENL estimation results in Table \ref{tab.ENL_Flevoland} have verified the effectiveness of proposed IGG prior in characterizing the variance of \textit{L} between clusters, while making the clustering results more accurate and interpretable. 

\begin{table*}[ht]
	\centering
	\caption{Accuracy Evaluation on Flevoland Dataset~(\%)}
	\begin{tabular}{p{4.000em}<{\centering}cccccccc}
		\toprule
		\toprule
		\multicolumn{1}{c}{\textit{Class}} & \multicolumn{1}{c}{\textit{$H/\alpha$-Wishart}} & \multicolumn{1}{c}{\textit{Standard}} & \multicolumn{1}{c}{\textit{Relaxed}} & \multicolumn{1}{c}{\textit{WMM}} & \multicolumn{1}{c}{\textit{\textbf{VBWMM}}} &
		\multicolumn{1}{c}{\textit{SVWMM}} & \multicolumn{1}{c}{\textit{\textbf{SVVBWMM}}}\\		
		\midrule		
		\rowcolor[rgb]{ 0,  1,  1} Rapeseed & \cellcolor[rgb]{ 1,  1,  1}77.25  & \cellcolor[rgb]{ 1,  1,  1}41.16  & \cellcolor[rgb]{ 1,  1,  1}61.40  & \cellcolor[rgb]{ 1,  1,  1}37.3 & \cellcolor[rgb]{ 1,  1,  1}\textbf{91.93 } & \cellcolor[rgb]{ 1,  1,  1}98.3 & \cellcolor[rgb]{ 1,  1,  1}\textbf{99.23} \\
		\rowcolor[rgb]{ .475,  .475,  1} {Bare soil} & \cellcolor[rgb]{ 1,  1,  1}97.96  & \cellcolor[rgb]{ 1,  1,  1}47.65  & \cellcolor[rgb]{ 1,  1,  1}78.90  & \cellcolor[rgb]{ 1,  1,  1}96.1 & \cellcolor[rgb]{ 1,  1,  1}\textbf{96.66 } & \cellcolor[rgb]{ 1,  1,  1}97.3 & \cellcolor[rgb]{ 1,  1,  1}\textbf{97.58} \\
		\rowcolor[rgb]{ .812,  .412,  0} {Potato} & \cellcolor[rgb]{ 1,  1,  1}40.80  & \cellcolor[rgb]{ 1,  1,  1}42.83  & \cellcolor[rgb]{ 1,  1,  1}96.87  & \cellcolor[rgb]{ 1,  1,  1}\textbf{97.2} & \cellcolor[rgb]{ 1,  1,  1}88.65  & \cellcolor[rgb]{ 1,  1,  1}97.4 & \cellcolor[rgb]{ 1,  1,  1}\textbf{97.56} \\
		\rowcolor[rgb]{ 1,  1,  0} {Beet} & \cellcolor[rgb]{ 1,  1,  1}74.09  & \cellcolor[rgb]{ 1,  1,  1}54.20  & \cellcolor[rgb]{ 1,  1,  1}92.27  & \cellcolor[rgb]{ 1,  1,  1}82.8 & \cellcolor[rgb]{ 1,  1,  1}\textbf{94.29 } & \cellcolor[rgb]{ 1,  1,  1}86.0 & \cellcolor[rgb]{ 1,  1,  1}\textbf{88.64} \\
		\rowcolor[rgb]{ 0,  0,  1} {Wheat} & \cellcolor[rgb]{ 1,  1,  1}53.78  & \cellcolor[rgb]{ 1,  1,  1}58.11  & \cellcolor[rgb]{ 1,  1,  1}90.13  & \cellcolor[rgb]{ 1,  1,  1}58.2 & \cellcolor[rgb]{ 1,  1,  1}\textbf{96.59 } & \cellcolor[rgb]{ 1,  1,  1}96.2 & \cellcolor[rgb]{ 1,  1,  1}\textbf{97.45} \\
		\rowcolor[rgb]{ 1,  .494,  1} {Barley} & \cellcolor[rgb]{ 1,  1,  1}68.27  & \cellcolor[rgb]{ 1,  1,  1}36.74  & \cellcolor[rgb]{ 1,  1,  1}95.92  & \cellcolor[rgb]{ 1,  1,  1}91.3 & \cellcolor[rgb]{ 1,  1,  1}\textbf{96.14 } & \cellcolor[rgb]{ 1,  1,  1}98.5 & \cellcolor[rgb]{ 1,  1,  1}\textbf{99.51} \\
		\rowcolor[rgb]{ .192,  .729,  .651} {Pea} & \cellcolor[rgb]{ 1,  1,  1}59.16  & \cellcolor[rgb]{ 1,  1,  1}83.16  & \cellcolor[rgb]{ 1,  1,  1}93.79  & \cellcolor[rgb]{ 1,  1,  1}92.2 & \cellcolor[rgb]{ 1,  1,  1}\textbf{95.03 } & \cellcolor[rgb]{ 1,  1,  1}97.3 & \cellcolor[rgb]{ 1,  1,  1}\textbf{99.34} \\
		\rowcolor[rgb]{ .294,  .812,  0} {Lucerne} & \cellcolor[rgb]{ 1,  1,  1}- & \cellcolor[rgb]{ 1,  1,  1}49.38  & \cellcolor[rgb]{ 1,  1,  1}90.51  & \cellcolor[rgb]{ 1,  1,  1}\textbf{90.8} & \cellcolor[rgb]{ 1,  1,  1}84.44  & \cellcolor[rgb]{ 1,  1,  1}\textbf{96.4} & \cellcolor[rgb]{ 1,  1,  1}96.32 \\
		OA    & 71.9  & 58.56  & 86.64  & 72.8  & \textbf{91.85 } & 96.5  & \textbf{97.49} \\
		$\kappa$     & 65.5  & 53.61  & 84.39  & 67.9  & \textbf{90.33 } & 95.9  & \textbf{96.95 } \\
		\bottomrule
		\bottomrule
	\end{tabular}%
	\label{tab.Flevoland_Acc}%
\end{table*}%

\begin{table}[htbp]
	\centering
	\caption{ENL Estimation on Flevoland Dataset}
	\begin{tabular}{p{4.055em}ccccc}
		\toprule
		\toprule
		\multicolumn{1}{c}{\textit{Class}} & \multicolumn{1}{c}{\textit{ML}} & \multicolumn{1}{c}{\textit{TM}} & \multicolumn{1}{c}{\textit{WMLC}} & \multicolumn{1}{c}{\textit{VB}} & \multicolumn{1}{c}{\textit{\textbf{SVVB}}} \\			
		\midrule
		\rowcolor[rgb]{ 0,  1,  1} Rapeseed  & \cellcolor[rgb]{ 1,  1,  1}32.28  & \cellcolor[rgb]{ 1,  1,  1}31.89  & \cellcolor[rgb]{ 1,  1,  1}35.79  & \cellcolor[rgb]{ 1,  1,  1}30.94  & \cellcolor[rgb]{ 1,  1,  1}27.06  \\
		\rowcolor[rgb]{ .475,  .475,  1} Bare soil & \cellcolor[rgb]{ 1,  1,  1}40.62  & \cellcolor[rgb]{ 1,  1,  1}40.45  & \cellcolor[rgb]{ 1,  1,  1}36.94  & \cellcolor[rgb]{ 1,  1,  1}31.43  & \cellcolor[rgb]{ 1,  1,  1}30.30  \\
		\rowcolor[rgb]{ .812,  .412,  0} Potato & \cellcolor[rgb]{ 1,  1,  1}43.69  & \cellcolor[rgb]{ 1,  1,  1}38.63  & \cellcolor[rgb]{ 1,  1,  1}40.55  & \cellcolor[rgb]{ 1,  1,  1}31.00  & \cellcolor[rgb]{ 1,  1,  1}34.96  \\
		\rowcolor[rgb]{ 1,  1,  0} Beet  & \cellcolor[rgb]{ 1,  1,  1}44.54  & \cellcolor[rgb]{ 1,  1,  1}42.86  & \cellcolor[rgb]{ 1,  1,  1}39.87  & \cellcolor[rgb]{ 1,  1,  1}31.66  & \cellcolor[rgb]{ 1,  1,  1}32.44  \\
		\rowcolor[rgb]{ 0,  0,  1} Wheat & \cellcolor[rgb]{ 1,  1,  1}43.36  & \cellcolor[rgb]{ 1,  1,  1}39.59  & \cellcolor[rgb]{ 1,  1,  1}39.18  & \cellcolor[rgb]{ 1,  1,  1}30.68  & \cellcolor[rgb]{ 1,  1,  1}31.44  \\
		\rowcolor[rgb]{ 1,  .494,  1} Barley & \cellcolor[rgb]{ 1,  1,  1}21.15  & \cellcolor[rgb]{ 1,  1,  1}23.21  & \cellcolor[rgb]{ 1,  1,  1}29.11  & \cellcolor[rgb]{ 1,  1,  1}30.98  & \cellcolor[rgb]{ 1,  1,  1}29.77  \\
		\rowcolor[rgb]{ .192,  .729,  .651} Pea   & \cellcolor[rgb]{ 1,  1,  1}31.75  & \cellcolor[rgb]{ 1,  1,  1}27.90  & \cellcolor[rgb]{ 1,  1,  1}35.63  & \cellcolor[rgb]{ 1,  1,  1}31.50  & \cellcolor[rgb]{ 1,  1,  1}31.83  \\
		\rowcolor[rgb]{ .294,  .812,  0} Lucerne & \cellcolor[rgb]{ 1,  1,  1}25.18  & \cellcolor[rgb]{ 1,  1,  1}25.17  & \cellcolor[rgb]{ 1,  1,  1}32.89  & \cellcolor[rgb]{ 1,  1,  1}31.48  & \cellcolor[rgb]{ 1,  1,  1}34.95  \\
		\bottomrule
		\bottomrule
	\end{tabular}%
	\label{tab.ENL_Flevoland}%
\end{table}%

\subsection{Experimental Results on Quebec Dataset}
\label{subsec:ExpQuebec}
\begin{figure*}[htbp]
	\centering
	\subfigtopskip=2pt
	\subfigbottomskip=2pt
	\subfigcapskip=-5pt
	\subfigure[]{
		\label{fig.Quebec_Data}
		\begin{annotate}{\includegraphics[width=.225\textwidth]{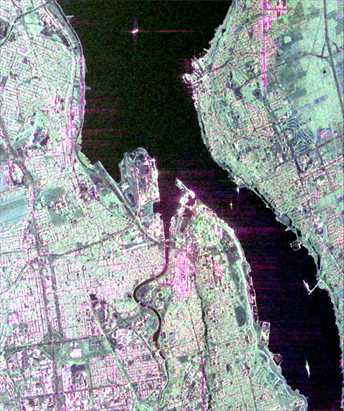}}{0.2}
			\draw[color=red] (0.1, 7.7) ellipse (1.1 and 0.35);
			\draw[color=red] (-3.1, 4.2) ellipse (1.0 and 0.35);
			\draw[color=red] (-3.1, 3.3) ellipse (1.0 and 0.35);
			\draw[color=red] (0.3, 1.5) ellipse (1.0 and 0.35);
			\draw[color=red] (3.8, -1.1) ellipse (1.0 and 0.35);
			\draw[color=white,rotate around={25:(-1.0, -4.7)}] (-1.0, -4.7) ellipse (0.55 and 1.5);
			\draw[color=white,rotate around={-30:(1.0, 7.5)}] (1.0, 7.5) ellipse (0.55 and 1.5);
			\draw[color=white,rotate around={5:(6.5, 5.8)}] (6.5, 5.8) ellipse (0.55 and 2.5);
			\draw[color=white,rotate around={5:(6.8, -5.75)}] (6.8, -5.75) ellipse (0.55 and 2.35);
		\end{annotate}	
	}
	\hspace{-.045\textwidth}
	\subfigure[]{
		\label{fig.Quebec_Ha}
		\begin{annotate}{\includegraphics[width=.225\textwidth]{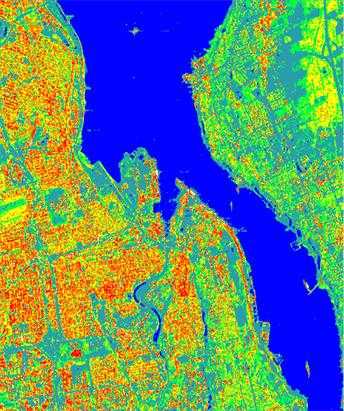}}{0.2}
			\draw[color=red] (0.1, 7.7) ellipse (1.1 and 0.35);
			\draw[color=red] (-3.1, 4.2) ellipse (1.0 and 0.35);
			\draw[color=red] (-3.1, 3.3) ellipse (1.0 and 0.35);
			\draw[color=red] (0.3, 1.5) ellipse (1.0 and 0.35);
			\draw[color=red] (3.8, -1.1) ellipse (1.0 and 0.35);
			\draw[color=white,rotate around={25:(-1.0, -4.7)}] (-1.0, -4.7) ellipse (0.55 and 1.5);
			\draw[color=white,rotate around={-30:(1.0, 7.5)}] (1.0, 7.5) ellipse (0.55 and 1.5);
			\draw[color=white,rotate around={5:(6.5, 5.8)}] (6.5, 5.8) ellipse (0.55 and 2.5);
			\draw[color=white,rotate around={5:(6.8, -5.75)}] (6.8, -5.75) ellipse (0.55 and 2.35);			
		\end{annotate}	
	}
	\hspace{-.045\textwidth}
	\subfigure[]{
		\label{fig.Quebec_standard}
		\begin{annotate}{\includegraphics[width=.225\textwidth]{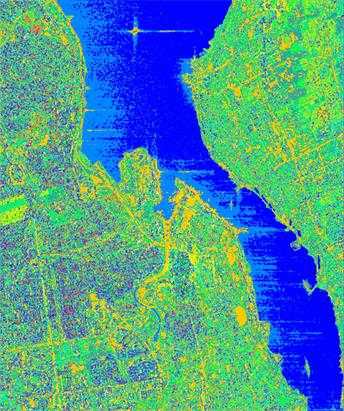}}{0.2}
			\draw[color=red] (0.1, 7.7) ellipse (1.1 and 0.35);
			\draw[color=red] (-3.1, 4.2) ellipse (1.0 and 0.35);
			\draw[color=red] (-3.1, 3.3) ellipse (1.0 and 0.35);
			\draw[color=red] (0.3, 1.5) ellipse (1.0 and 0.35);
			\draw[color=red] (3.8, -1.1) ellipse (1.0 and 0.35);
			\draw[color=white,rotate around={25:(-1.0, -4.7)}] (-1.0, -4.7) ellipse (0.55 and 1.5);
			\draw[color=white,rotate around={-30:(1.0, 7.5)}] (1.0, 7.5) ellipse (0.55 and 1.5);
			\draw[color=white,rotate around={5:(6.5, 5.8)}] (6.5, 5.8) ellipse (0.55 and 2.5);
			\draw[color=white,rotate around={5:(6.8, -5.75)}] (6.8, -5.75) ellipse (0.55 and 2.35);			
		\end{annotate}	
	}
	\hspace{-.045\textwidth}
	\subfigure[]{
		\label{fig.Quebec_relaxed}
		\begin{annotate}{\includegraphics[width=.225\textwidth]{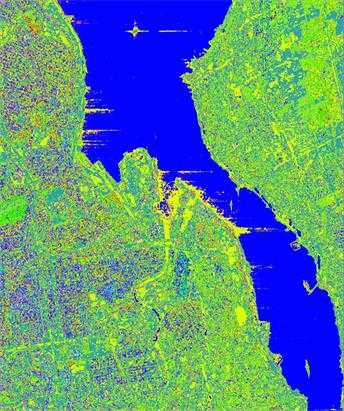}}{0.2}
			\draw[color=red] (0.1, 7.7) ellipse (1.1 and 0.35);
			\draw[color=red] (-3.1, 4.2) ellipse (1.0 and 0.35);
			\draw[color=red] (-3.1, 3.3) ellipse (1.0 and 0.35);
			\draw[color=red] (0.3, 1.5) ellipse (1.0 and 0.35);
			\draw[color=red] (3.8, -1.1) ellipse (1.0 and 0.35);
			\draw[color=white,rotate around={25:(-1.0, -4.7)}] (-1.0, -4.7) ellipse (0.55 and 1.5);
			\draw[color=white,rotate around={-30:(1.0, 7.5)}] (1.0, 7.5) ellipse (0.55 and 1.5);
			\draw[color=white,rotate around={5:(6.5, 5.8)}] (6.5, 5.8) ellipse (0.55 and 2.5);
			\draw[color=white,rotate around={5:(6.8, -5.75)}] (6.8, -5.75) ellipse (0.55 and 2.35);	
		\end{annotate}	
	}
	\quad
	\subfigure[]{
		\label{fig.Quebec_WMM}
		\begin{annotate}{\includegraphics[width=.225\textwidth]{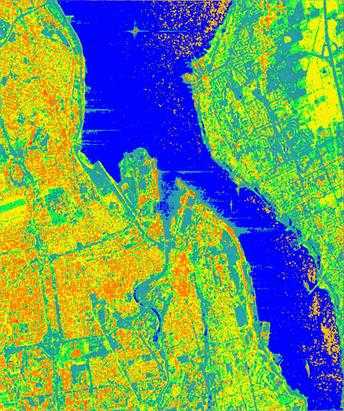}}{0.2}
			\draw[color=red] (0.1, 7.7) ellipse (1.1 and 0.35);
			\draw[color=red] (-3.1, 4.2) ellipse (1.0 and 0.35);
			\draw[color=red] (-3.1, 3.3) ellipse (1.0 and 0.35);
			\draw[color=red] (0.3, 1.5) ellipse (1.0 and 0.35);
			\draw[color=red] (3.8, -1.1) ellipse (1.0 and 0.35);
			\draw[color=white,rotate around={25:(-1.0, -4.7)}] (-1.0, -4.7) ellipse (0.55 and 1.5);
			\draw[color=white,rotate around={-30:(1.0, 7.5)}] (1.0, 7.5) ellipse (0.55 and 1.5);
			\draw[color=white,rotate around={5:(6.5, 5.8)}] (6.5, 5.8) ellipse (0.55 and 2.5);
			\draw[color=white,rotate around={5:(6.8, -5.75)}] (6.8, -5.75) ellipse (0.55 and 2.35);		
		\end{annotate}	
	}
	\hspace{-.045\textwidth}
	\subfigure[]{
		\label{fig.Quebec_SVWMM}
		\begin{annotate}{\includegraphics[width=.225\textwidth]{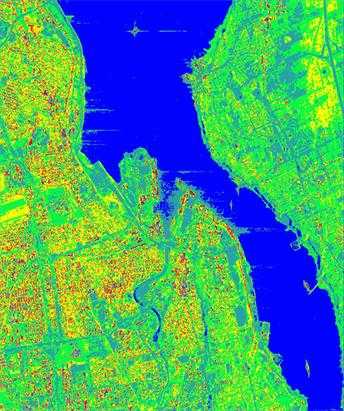}}{0.2}
			\draw[color=red] (0.1, 7.7) ellipse (1.1 and 0.35);
			\draw[color=red] (-3.1, 4.2) ellipse (1.0 and 0.35);
			\draw[color=red] (-3.1, 3.3) ellipse (1.0 and 0.35);
			\draw[color=red] (0.3, 1.5) ellipse (1.0 and 0.35);
			\draw[color=red] (3.8, -1.1) ellipse (1.0 and 0.35);
			\draw[color=white,rotate around={25:(-1.0, -4.7)}] (-1.0, -4.7) ellipse (0.55 and 1.5);
			\draw[color=white,rotate around={-30:(1.0, 7.5)}] (1.0, 7.5) ellipse (0.55 and 1.5);
			\draw[color=white,rotate around={5:(6.5, 5.8)}] (6.5, 5.8) ellipse (0.55 and 2.5);
			\draw[color=white,rotate around={5:(6.8, -5.75)}] (6.8, -5.75) ellipse (0.55 and 2.35);
		\end{annotate}	
	}
	\hspace{-.045\textwidth}
	\subfigure[]{
		\label{fig.Quebec_VBWMM}
		\begin{annotate}{\includegraphics[width=.225\textwidth]{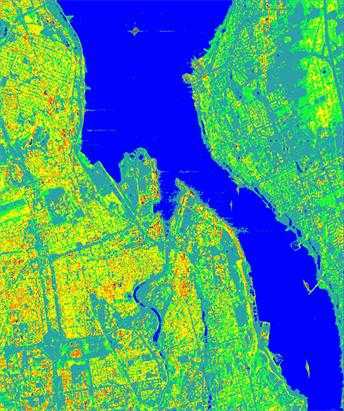}}{0.2}
			\draw[color=red] (0.1, 7.7) ellipse (1.1 and 0.35);
			\draw[color=red] (-3.1, 4.2) ellipse (1.0 and 0.35);
			\draw[color=red] (-3.1, 3.3) ellipse (1.0 and 0.35);
			\draw[color=red] (0.3, 1.5) ellipse (1.0 and 0.35);
			\draw[color=red] (3.8, -1.1) ellipse (1.0 and 0.35);
			\draw[color=white,rotate around={25:(-1.0, -4.7)}] (-1.0, -4.7) ellipse (0.55 and 1.5);
			\draw[color=white,rotate around={-30:(1.0, 7.5)}] (1.0, 7.5) ellipse (0.55 and 1.5);
			\draw[color=white,rotate around={5:(6.5, 5.8)}] (6.5, 5.8) ellipse (0.55 and 2.5);
			\draw[color=white,rotate around={5:(6.8, -5.75)}] (6.8, -5.75) ellipse (0.55 and 2.35);
		\end{annotate}	
	}
	\hspace{-.045\textwidth}
	\subfigure[]{
		\label{fig.Quebec_SVVBWMM}
		\begin{annotate}{\includegraphics[width=.225\textwidth]{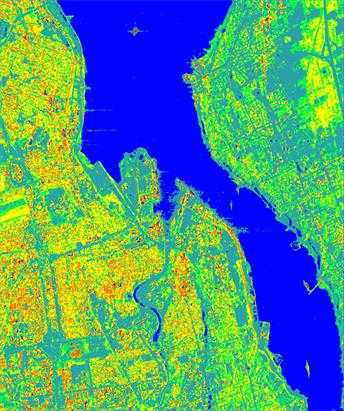}}{0.2}
			\draw[color=red] (0.1, 7.7) ellipse (1.1 and 0.35);
			\draw[color=red] (-3.1, 4.2) ellipse (1.0 and 0.35);
			\draw[color=red] (-3.1, 3.3) ellipse (1.0 and 0.35);
			\draw[color=red] (0.3, 1.5) ellipse (1.0 and 0.35);
			\draw[color=red] (3.8, -1.1) ellipse (1.0 and 0.35);
			\draw[color=white,rotate around={25:(-1.0, -4.7)}] (-1.0, -4.7) ellipse (0.55 and 1.5);
			\draw[color=white,rotate around={-30:(1.0, 7.5)}] (1.0, 7.5) ellipse (0.55 and 1.5);
			\draw[color=white,rotate around={5:(6.5, 5.8)}] (6.5, 5.8) ellipse (0.55 and 2.5);
			\draw[color=white,rotate around={5:(6.8, -5.75)}] (6.8, -5.75) ellipse (0.55 and 2.35);		
		\end{annotate}	
	}
	\caption{(a) Pauli RGB image and classification results of Quebec dataset using (b) $H/\alpha$-Wishart \cite{HalphaWis}, (c) standard-Wishart \cite{APD11}, (d) relaxed-Wishart \cite{APD11}, (e) WMM \cite{SVWMM}, (f) SVWMM \cite{SVWMM}, (g) proposed VBWMM, (h) proposed SVVBWMM.}
	\label{fig.Exp_Quebec}
\end{figure*}
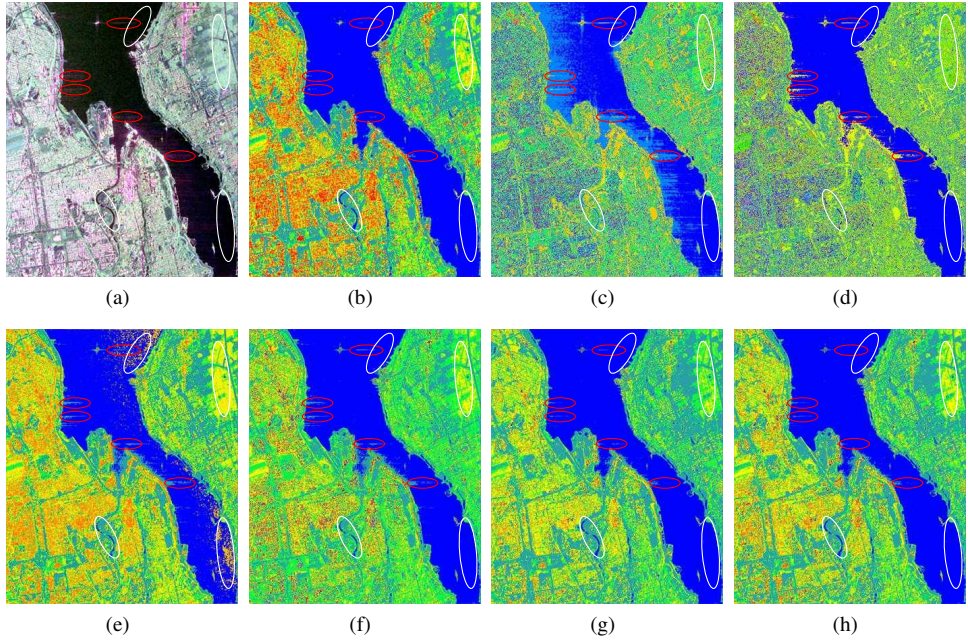

To validate the performance of proposed method on spaceborne datasets, the following experiments are conducted with PolSAR data provided by RADARSAT-2 over Quebec City of Canada in fine beam mode, as shown in Fig. \ref{fig.Exp_Quebec}. The imaging scene is $2055 \time 1720$ in size and contains a variety of land covers, such as urban, vegetation, rivers and harbors. For standard-Wishart and relaxed-Wishart \cite{APD11}, the split-confidence level and sub-sampling rate are set to 95\% and $1/225$, respectively. For the four VB approaches in Fig. \ref{fig.Quebec_WMM} to Fig. \ref{fig.Quebec_SVVBWMM}, the initial value of components $K$ is set to 16. 

The classification maps resulting from all the seven methods are presented in Fig. \ref{fig.Exp_Quebec}, where the ships are correctly identified and rich details are demonstrated by all these methods. As illustrated in Fig. \ref{fig.Quebec_Ha}, one can observe that the urban areas are generally classified into three clusters by $H/\alpha$-Wishart. Nevertheless, from the Pauli RGB image in Fig. \ref{fig.Quebec_Data}, it is observed that Quebec Dataset exhibits complex structures such as the horizontal lines in the red eclipses are not completely assigned to an appropriate cluster, which brings about the challenge for its interpretation. Fig. \ref{fig.Quebec_standard} and Fig. \ref{fig.Quebec_relaxed} show that standard-Wishart and relaxed-Wishart methods perform poorly on large datasets, where many more unexpected clusters are found. Still, relaxed-Wishart performs better than standard-Wishart, where the horizontal lines and water areas have been correctly identified into the same cluster. 

For this scenario, all four VB approaches have found 6 clusters, where the urban areas are generally clustered into two groups, namely low- and high-density urban, as represented in yellow and light orange colors from Fig. \ref{fig.Quebec_WMM} to Fig. \ref{fig.Quebec_SVVBWMM}. The clustering results of water areas indicate that WMM also cannot accurately capture these structures, leading to the miscellaneous pixels in the red eclipses in Fig. \ref{fig.Quebec_WMM}. In contrast, SVWMM performs better than WMM, where the water area is treated as one cluster, implying that the incorporation of spatial information is helpful in facilitating the visual interpretation of PolSAR images, as exhibited in Fig. \ref{fig.Quebec_SVWMM}. However, although spatial context is quite useful in providing correct guidance on appropriate clustering of these pixels, many details have been over-smoothed, such as the river area in bottom white eclipse. The highlighted white eclipses in Fig. \ref{fig.Quebec_VBWMM} reveals that the proposed VBWMM demonstrates competitive performance in both homogeneous and heterogeneous areas. Meanwhile, the proposed SVVBWMM allows for better preservation of structures, while reducing speckle noise and improving the interpretability of PolSAR image, especially in the dash-dotted black eclipses. Besides, there is an evidence that the proposed VBWMM and SVVBWMM can provide appropriate ENL expectation values, which is confirmed by the estimation results of top 5 classes with the largest proportions, as illustrated in Table \ref{tab.ENL_Quebec}.

\begin{table}[htbp]
	\centering
	\caption{ENL Estimation on Quebec Dataset}
	\begin{tabular}{p{4.055em}ccccc}
		\toprule
		\toprule
		\multicolumn{1}{c}{\textit{Class}} & \multicolumn{1}{c}{\textit{ML}} & \multicolumn{1}{c}{\textit{TM}} & \multicolumn{1}{c}{\textit{WMLC}} & \multicolumn{1}{c}{\textit{VB}} & \multicolumn{1}{c}{\textit{\textbf{SVVB}}} \\			
		\midrule		
		\rowcolor[rgb]{ 0,  0,  1} Water & \cellcolor[rgb]{ 1,  1,  1}29.80  & \cellcolor[rgb]{ 1,  1,  1}29.99  & \cellcolor[rgb]{ 1,  1,  1}29.14  & \cellcolor[rgb]{ 1,  1,  1}25.53  & \cellcolor[rgb]{ 1,  1,  1}28.31  \\
		\rowcolor[rgb]{ .161,  .631,  .263} Vegetation & \cellcolor[rgb]{ 1,  1,  1}24.69  & \cellcolor[rgb]{ 1,  1,  1}21.06  & \cellcolor[rgb]{ 1,  1,  1}21.69  & \cellcolor[rgb]{ 1,  1,  1}24.56  & \cellcolor[rgb]{ 1,  1,  1}25.29  \\
		\rowcolor[rgb]{ .047,  .965,  .267} Field  & \cellcolor[rgb]{ 1,  1,  1}25.79  & \cellcolor[rgb]{ 1,  1,  1}26.53  & \cellcolor[rgb]{ 1,  1,  1}25.45  & \cellcolor[rgb]{ 1,  1,  1}27.39  & \cellcolor[rgb]{ 1,  1,  1}26.43  \\
		\rowcolor[rgb]{ 1,  1,  0} Low & \cellcolor[rgb]{ 1,  1,  1}19.29  & \cellcolor[rgb]{ 1,  1,  1}22.24  & \cellcolor[rgb]{ 1,  1,  1}21.86  & \cellcolor[rgb]{ 1,  1,  1}18.01  & \cellcolor[rgb]{ 1,  1,  1}20.94  \\
		\rowcolor[rgb]{ .98,  .522,  .016} High & \cellcolor[rgb]{ 1,  1,  1}13.15  & \cellcolor[rgb]{ 1,  1,  1}10.78  & \cellcolor[rgb]{ 1,  1,  1}13.11  & \cellcolor[rgb]{ 1,  1,  1}12.49  & \cellcolor[rgb]{ 1,  1,  1}14.76  \\
		\bottomrule
		\bottomrule
	\end{tabular}%
	\label{tab.ENL_Quebec}%
\end{table}%

\section{Conclusion and Future Work}
\label{sec:Conclusion}
In this paper, a variational Bayesian Wishart mixture model (VBWMM) has been proposed to realize unsupervised classification of PolSAR data, which maximizes the closed-form lower bound and estimates the expectation value of ENL for each cluster through a newly derived inverse gamma-gamma (IGG) prior distribution. Besides, covariance matrix similarity and geometric similarity of neighboring pixels are incorporated to exploit the spatial information of PolSAR images. And most importantly, another advantage of proposed model lies in the closed-form lower bound used to evaluate model convergence, which has not been achieved by other Bayesian approaches, e.g. SVWMM \cite{SVWMM}, as it is not practical to manually pick the best result after a certain number of iterations in real application. Experiments on airborne and spaceborne PolSAR images are performed to validate the superiority of proposed method in automated clustering and ENL estimation towards other iterative Bayesian approaches. 

The PolSAR images used in the experiments have been preprocessed with multi-look boxcar or refined Lee filter, which offers increased interpretability at the cost of lower resolution. However, complex Wishart distribution assumes that the number of looks is larger than 2, which does not hold true for high-resolution PolSAR images. Therefore, our future work focuses on variational Bayesian techniques with more complex matrix variate distributions, such as K-Wishart distribution and $\mathcal{G}_{d}^{0}$-distribution, so as to better interpret the true distribution of high-resolution PolSAR images, especially in heterogeneous areas. 

The proposed model is an attempt for the unsupervised classification of PolSAR images based on variational Bayesian with matrix variate mixture models and a variable number of looks. Recently, deep learning has emerged as a promising alternative to classification tasks, which updates its parameters via back-propagation instead of deriving closed-form solutions. Therefore, a key question for further research is to derive effective architectures that can represent the coherent and covariance matrices with variable texture and number of looks, which is expected to increase the convergence speed, greatly reduce the inference time and generalize well to large-scale PolSAR images. Another possible direction of our future research is to take advantage of both deep learning and Bayesian theory, with an aim to more effectively extract the textural information, statistical characteristics and spatial information of land covers.
\section{Acknowledgment}
The authors would like to thank the anonymous reviewers for their useful comments and suggestions, which were of great help in improving this paper. The authors also would like to thank JPL/NASA and CSA for providing free download links for the PolSAR data used in the experiments. Finally, the author would like to thank C. Liu \cite{SVWMM} for providing the ground truth of Flevoland dataset and advice on the initialization skills, which are quite useful in improving model robustness and obtaining better clustering results.

\appendix

\section{Bounds for Ratios of Modified Bessel Functions}
Segura \cite{BoundsBes} proved the bounds for ratio of modified Bessel functions, where the double inequality in \eqref{eq.seineq1} holds for any $v \geq 0$ and $x > 0$.
\begin{equation}
\label{eq.seineq1}
\frac{{v \!+\! \sqrt {{x^2} \!+\! {v^2}} }}{x} \!<\! \frac{{{K_{v \!+\! 1}}(x)}}{{{K_v}(x)}} \!<\! \frac{{v \!+\! \frac{1}{2} \!+\! \sqrt {{x^2} \!+\! {{\left( {v \!+\! \frac{1}{2}} \right)}^2}} }}{x}
\end{equation}

Therefore, it is easy to obtain the iterative formula of $K_a(2\!\sqrt {{b}{c}})$ by letting $v \!=\! a$ and $x \!=\! 2\!\sqrt {{b}{c}}$, as is shown in \eqref{eq.seineq2}.
\begin{equation}
\label{eq.seineq2}
\frac{{a + \sqrt {4{b}{c} + {a^2}} }}{{2\!\sqrt {{b}{c}} }} \!<\! \frac{{{K_{a  + 1}}(2\!\sqrt {{b}{c}} )}}{{{K_a }(2\!\sqrt {{b}{c}} )}} \!<\! \frac{{a + {1 \mathord{\left/
				{\vphantom {1 2}} \right.
				\kern-\nulldelimiterspace} 2} + \sqrt {4{b}{c} + {{\left( {a + {1 \mathord{\left/
								{\vphantom {1 2}} \right.
								\kern-\nulldelimiterspace} 2}} \right)}^2}} }}{{2\!\sqrt {{b}{c}} }}
\end{equation}

The Tur{\'a}n-type inequalities \cite{BoundsBes,IneqBes,ApproBes,ProdBes} are commonly used and well suited to approximate the iterative formulas related to $K'_v\left( \cdot \right)$ and $K_v\left( \cdot \right)$. For all $u \!>\! 0$ and $v \!>\! 1$, the following inequalities hold
\begin{equation}
\label{eq.Tur_1}
- \sqrt {\frac{v}{{v \!-\! 1}}{u^2} + {v^2}}  \!<\! \frac{{u{{K'}_v}\left( u \right)}}{{{K_v}\left( u \right)}} \!<\!  \!-\! \sqrt {{u^2} + {v^2}} 
\end{equation}
Moreover, the right-hand side of \eqref{eq.Tur_1} holds true for all $v \in {\cal R}$.

\begin{equation}
\label{eq.Tur_2}
\frac{{{K_{v + 1}}\left( u \right)}}{{{K_v}\left( u \right)}}  \!= \!  \!-\! \frac{{{{K'}_v}\left( u \right)}}{{{K_v}\left( u \right)}} + \frac{v}{u}
\end{equation}

Based on \eqref{eq.seineq1}\verb|-|\eqref{eq.Tur_2}, the following inequality in \eqref{eq.KdrK} is obtained by assuming $u \!=\! 2\!\sqrt {{b}{c}}$ and $v \!=\! a$.
\begin{equation}
\label{eq.KdrK}
- \sqrt {\frac{a}{{a \!-\! 1}} + \frac{{{a^2}}}{{4{b}{c}}}}  \!<\! \frac{{{{K'}_a}\left( {2\!\sqrt {{b}{c}} } \right)}}{{{K_a}\left( {2\!\sqrt {{b}{c}} } \right)}} \!<\!  \!-\! \sqrt {1 + \frac{{{a^2}}}{{4{b}{c}}}}
\end{equation}

For the double inequalities in \eqref{eq.seineq2} and \eqref{eq.KdrK}, the mean value of lower and upper bounds is used as an approximation, and thereby resulting in the closed-form solutions in \eqref{eq.clo_1} and \eqref{eq.clo_2}.
\begin{equation}
\label{eq.clo_1}
\frac{{{K_{a + 1}}\left( {2\!\sqrt {{b}{c}} } \right)}}{{{K_a}\left( {2\!\sqrt {{b}{c}} } \right)}} = \frac{{\sqrt {4{b}{c} + {a^2}}  + \sqrt {4{b}{c} + {{\left( {a + {1 \mathord{\left/
								{\vphantom {1 2}} \right.
								\kern-\nulldelimiterspace} 2}} \right)}^2}} {\text{ + }}2a + {1 \mathord{\left/
				{\vphantom {1 2}} \right.
				\kern-\nulldelimiterspace} 2}}}{{4\!\sqrt {{b}{c}} }}
\end{equation}
\begin{equation}
\label{eq.clo_2}
\frac{{{{K'}_a }\left( {2\!\sqrt {{b}{c}} } \right)}}{{{K_a }\left( {2\!\sqrt {{b}{c}} } \right)}} =  \!-\! \frac{{\sqrt {\frac{a}{{a \!-\! 1}} + \frac{{{a^2}}}{{4{b}{c}}}}  + \sqrt {1 + \frac{{{a^2}}}{{4{b}{c}}}} }}{2}
\end{equation}

\section{Computing the Variational Lower Bound}
Based on the approximation for $\ln {p_L}\left( l \right)$ in Section \ref{subsubsec:AppLogPL}, the various terms in ELBO are easily evaluated to give the following results, as shown in \eqref{eq.ElogpC}\verb|-|\eqref{eq.logpL}.
\begin{equation}
\label{eq.ElogpC}
\begin{aligned}
\langle\ln p(C \mid \Sigma, L, z)\rangle &=\sum_{n=1}^{N} \sum_{k=1}^{K} r_{n,k}\left\{\left(\left\langle L_{k}\right\rangle-3\right) \ln \left|C_{n}\right|+\left(3+\left\langle\ln \left|\Sigma_{k}^{-1}\right|\right\rangle-\operatorname{tr}\left(\Omega_{k} C_{n}\right)\right)\left\langle L_{k}\right\rangle\right.\\
&\left.+\frac{9}{2}\left\langle\ln L_{k}\right\rangle-4\left\langle\frac{1}{L_{k}}\right\rangle-3 \ln \pi-\frac{3}{2} \ln 2 \pi\right\}
\end{aligned}
\end{equation}

\begin{equation}
	\label{eq.ElogpSigmaL}
	\begin{aligned}
		\langle\ln p(\Sigma, L)\rangle &=\sum_{k=1}^{K}\left\{\left\langle\ln p\left(\Sigma_{k} \mid L_{k}\right)\right\rangle+\left\langle\ln p\left(L_{k}\right)\right\rangle\right\} \\
		&=\sum_{k=1}^{K}\left\{\left\langle\ln W\left(\Sigma_{k}^{-1} \mid \beta^{(0)} L_{k}, \Omega^{(0)}\right)\right\rangle+\left\langle\ln I G G\left(L_{k} \mid a^{(0)}, b^{(0)}, c^{(0)}\right)\right\rangle\right\} \\
		&=\sum_{k=1}^{K}\left\{\left(a^{(0)}+\frac{7}{2}\right)\left\langle\ln L_{k}\right\rangle+\left[\beta^{(0)}\left(3-\ln \left|\Omega^{(0)}\right|-\operatorname{tr}\left(\left(\Omega^{(0)}\right)^{-1} \Omega_{k}\right)\right)-b^{(0)}\right]\left\langle L_{k}\right\rangle-\left(\frac{4}{\beta^{(0)}}+c^{(0)}\right)\left\langle\frac{1}{L_{k}}\right\rangle\right.\\
		&\left.+\left(\beta^{(0)}\left\langle L_{k}\right\rangle-d\right)\left\langle\ln \left|\Sigma_{k}^{-1}\right|\right\rangle\right\}+K\left[\frac{9}{2} \ln \beta^{(0)}+a^{(0)} \ln b^{(0)}-\ln \Gamma\left(a^{(0)}\right)-\frac{3}{2} \ln 2 \pi\right]
	\end{aligned}
\end{equation}

\begin{equation}
	\label{eq.ElogqSigmaL}
	\begin{aligned}
		\langle\ln q(\Sigma, L)\rangle &=\sum_{k=1}^{K}\left\{\left\langle\ln q\left(\Sigma_{k} \mid L_{k}\right)\right\rangle+\left\langle\ln q\left(L_{k}\right)\right\rangle\right\} \\
		&=\sum_{k=1}^{K}\left\{\left\langle\ln W\left(\Sigma_{k}^{-1} \mid \beta_{k} L_{k}, \Omega_{k}\right)\right\rangle+\left\langle\ln G I\left(L_{k} \mid a_{k}, b_{k}, c_{k}\right)\right\rangle\right\} \\
		&=\sum_{k=1}^{K}\left\{\left(a_{k}+\frac{7}{2}\right)\left\langle\ln L_{k}\right\rangle-\left(\beta_{k} \ln \left|\Omega_{k}\right|+b_{k}\right)\left\langle L_{k}\right\rangle-\left(\frac{4}{\beta_{k}}+c_{k}\right)\left\langle\frac{1}{L_{k}}\right\rangle\right.\\
		&\left.+\left(\beta_{k}\left\langle L_{k}\right\rangle-d\right)\left\langle\ln \left|\Sigma_{k}^{-1}\right|\right\rangle-\ln \Gamma\left(a_{k}\right)+a_{k} \ln b_{k}+\frac{9}{2} \ln \beta_{k}-\frac{3}{2} \ln 2 \pi\right\}
	\end{aligned}
\end{equation}
where the log-gamma function originates from the logarithm of Bessel function, which is typically approximated with \eqref{eq.lnK_a} and \eqref{eq.logpL}.

%

%

\bibliographystyle{elsarticle-num-names}      
\bibliography{REF_WMMVBEM}  

%
%
%
\end{document}